\PassOptionsToPackage{unicode}{hyperref}
\PassOptionsToPackage{hyphens}{url}
\PassOptionsToPackage{dvipsnames,svgnames*,x11names*}{xcolor}
\documentclass[
  11pt,
  british,
  a4paper,
  twoside,openright]{scrreprt}
\usepackage{lmodern}
\usepackage{amssymb,amsmath}
\usepackage{ifxetex,ifluatex}
\ifnum 0\ifxetex 1\fi\ifluatex 1\fi=0 
  \usepackage[T1]{fontenc}
  \usepackage[utf8]{inputenc}
  \usepackage{textcomp} 
\else 
  \usepackage{unicode-math}
  \defaultfontfeatures{Scale=MatchLowercase}
  \defaultfontfeatures[\rmfamily]{Ligatures=TeX,Scale=1}
\fi
\IfFileExists{upquote.sty}{\usepackage{upquote}}{}
\IfFileExists{microtype.sty}{
  \usepackage[]{microtype}
  \UseMicrotypeSet[protrusion]{basicmath} 
}{}
\usepackage{xcolor}
\IfFileExists{xurl.sty}{\usepackage{xurl}}{} 
\IfFileExists{bookmark.sty}{\usepackage{bookmark}}{\usepackage{hyperref}}
\hypersetup{
  pdftitle={Proof of All},
  pdfauthor={Mario A. Barbara},
  pdfsubject={topic},
  pdfkeywords={keyword},
  colorlinks=true,
  linkcolor=purple,
  filecolor=Maroon,
  citecolor=purple,
  urlcolor=cyan,
  pdfcreator={LaTeX via pandoc}}
\usepackage{longtable,booktabs}
\usepackage{etoolbox}
\makeatletter
\patchcmd\longtable{\par}{\if@noskipsec\mbox{}\fi\par}{}{}
\makeatother
\IfFileExists{footnotehyper.sty}{\usepackage{footnotehyper}}{\usepackage{footnote}}
\makesavenoteenv{longtable}
\providecommand{\tightlist}{%
  \setlength{\itemsep}{0pt}\setlength{\parskip}{0pt}}
\usepackage{centernot}

\usepackage[utf8]{inputenc}
\usepackage{extarrows}
\usepackage{comment}
\excludecomment{verbatim}
\def\[#1\]{\begin{gather*}#1\end{gather*}}
\usepackage{fouriernc}
\usepackage{mathrsfs}
\usepackage{tikz}
\makeatletter
\@ifpackageloaded{subfig}{}{\usepackage{subfig}}
\@ifpackageloaded{caption}{}{\usepackage{caption}}
\AtBeginDocument{%

}
\AtBeginDocument{%

}
\@ifpackageloaded{float}{}{\usepackage{float}}
\floatstyle{ruled}
\@ifundefined{c@chapter}{\newfloat{codelisting}{h}{lop}}{\newfloat{codelisting}{h}{lop}[chapter]}
\floatname{codelisting}{Listing}

\makeatother
\ifxetex
  \usepackage{polyglossia}
  \setmainlanguage[variant=british]{english}
\else
  \usepackage[shorthands=off,main=british]{babel}
\fi

\title{Proof of All\thanks{Eötvös Loránd University, University of Twente, European Institute of
Technology, E-Group}}
\usepackage{etoolbox}
\makeatletter
\providecommand{\subtitle}[1]{
  \apptocmd{\@title}{\par {\large #1 \par}}{}{}
}
\makeatother
\subtitle{\emph{Verifiable Computation in a Nutshell}}
\author{Mario A. Barbara}
\date{Budapest, 2019}

\begin{document}
	\begin{titlepage}
		\scshape
		\thispagestyle{empty}
		
		\begin{minipage}{0.3\textwidth}\flushleft
			\includegraphics[scale=0.25]{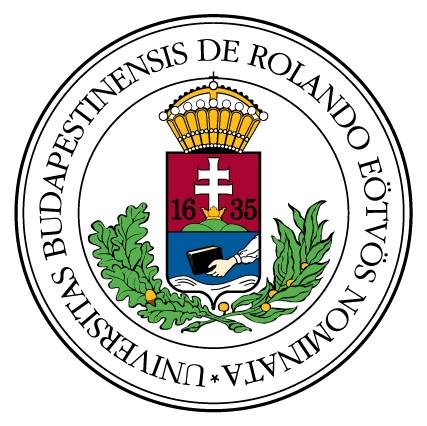}
		\end{minipage}\hfill\begin{minipage}{0.7\textwidth}\flushright
			\LARGE Eötvös Loránd University\\
			\Large Faculty of Informatics
		\end{minipage}
		
		\vspace{3cm}
		\centering
		\Huge Proof of All\\ 
		\LARGE \emph{Verifiable Computation in a Nutshell}
		
		\vspace{5cm}
		\begin{minipage}{0.5\textwidth}\flushleft
			\Large Dr.~Peter Ligeti\\
			\normalsize ELTE\\\ \\
			\Large Dr.~Andreas Peter\\
			\normalsize University of Twente\\\ \\
			\Large Mr.~Áron Szabó\\
			\normalsize E-Group
		\end{minipage}\hfill\begin{minipage}{0.5\textwidth}\flushright
			\LARGE Mario A. Barbara\\
			\large MSc. Computer Science
		\end{minipage}
		
		\centering\vfill
		\Large Budapest, 2019	
	\end{titlepage}
\begin{abstract}
\textbf{\emph{Abstract}} Recent advances in the cryptographic field of
``Zero-Knowledge Proofs'' have sparked a new wave of research, giving
birth to many exciting theoretical approaches in the last few years.
Such research has often overlapped with the need for private and
scalable solutions of Blockchain-based communities, resulting in the
first practical implementations of such systems. Many of these
innovative constructions have developed in parallel, using different
terminologies and evolving into a fragmented ecosystem, calling for
their consolidation into the more stable domain of ``Verifiable
Computation''. In this master thesis I propose a unifying Verifiable
Computation model for the simplification and efficient comparison of all
cryptographic proof systems. I take advantage of this model to analyse
innovative technologies (Homomorphic Authenticators, Verifiable Delay
Functions) which developed into their own specialised domains, and I
attempt to make them more accessible for newcomers to the field.
Furthermore, I expand on the future of Verifiable Computation, Universal
proof compilers and ``Proofs of All'', by approaching the
state-of-the-art zk-STARK construction from a more accessible and
informal design perspective. \newline \newline \textbf{\emph{Thanks}} To
those who supported me in times of need, and to those who gave me
technical advice. In particular: Mamma e Papà, prof. Ligeti, Mathijs,
Eszter, István. \vfill \emph{Let's shoot for the moon. If we miss, we'll
still land amongst the stars.}
\end{abstract}

{
\hypersetup{linkcolor=purple}
\setcounter{tocdepth}{2}
\tableofcontents
}
\hypertarget{objectives}{%
\chapter*{Objectives}\label{objectives}}
\addcontentsline{toc}{chapter}{Objectives}

The objective of this MSc thesis is to tackle innovative technologies
from a unifying perspective, prioritising simplicity and understanding
over obscure constructions, and benefitting further popularisation of
the cryptographic proofs domain. My main incentive for writing this
thesis is the conflict of interest that exists in the security field
between the desire to exploit innovations in cryptography for building
disruptive technologies (e.g.~the ``blockchain revolution''\footnote{for
  a list of companies currently investing in blockchain, see
  {[}\protect\hyperlink{ref-blockchain-forbes}{1}{]}}), and the
complexity and knowledge barrier required to understand them, which
leads to consistent misinformation in the market (e.g.~the ``Bitconnect
scandal'' {[}\protect\hyperlink{ref-bitconnect}{2}{]}) as well as in
developer channels, and fragmentation within the research community.

The domain I target is that of cryptographic proof systems, and,
specifically, I gather them under the umbrella term \emph{``Verifiable
Computation''}. There are \textbf{3 thesis objectives} which this work
hopes to achieve:

\begin{enumerate}
\def\labelenumi{\arabic{enumi}.}
\item
  \emph{A Unifying Model}

  for the cryptographic Verifiable Computation domain. The idea is to
  select and define the most important and comprehensive properties that
  have been spread out over various VC technologies in the course of
  more than 3 decades, sometimes using different names and definitions,
  By using a standardised model for defining protocols, researchers can
  attempt to merge the fragmented domain of cryptographic proofs, and
  thus unite their efforts under a single research domain.
\item
  \emph{Technical Analysis}

  of VC technologies, revisiting exciting and correlated protocols using
  the unifying model. While the focus is kept only on more favoured
  constructions of recent years, I wish to help new researchers get
  quickly acquainted with the VC landscape, hopefully leading to further
  popularisation and systemisation of the domain.
\item
  \emph{A Simplified Guide}

  to understanding the VC domain and its prominent technologies. This
  ``layman's view'' of core cryptographic properties is achieved through
  uncompromisingly logical and verbose debates on each technical design,
  where no assumptions are left to the imagination. While there are many
  formal definitions used to organise technical details, they are always
  accompanied by informal descriptions. I believe in simplifying
  technical constructions as much as possible, as it is paramount to
  their implementation and diffusion in the engineering field; it also
  stimulates further research, as the cryptocurrency community has
  proven{[}\protect\hyperlink{ref-ethresearch}{3}{]}.
\end{enumerate}

\begin{verbatim}
 4. **TBD, probably not enough time** _A Practical Use-Case_ 

    is demonstrated through the design of a business application protocol using VC technologies. The scenario deals with the Open-Banking PSD2 standardisation of the European Union, and its appeal to innovative FinTech companies. I provide scientific and engineering solutions, which can later be exploited Open-Banking product implementations. Context, requirements, and research data is provided by E-Group, EIT partner and my internship provider. 
\end{verbatim}

This thesis is divided into 3 chapters, which reflect the objectives set
forth during the development of this work. \textbf{Chapter I,
\emph{''Introduction and VC Model''}}, discusses the history behind
proof protocols while giving a gentle introduction to the topic,
progressively presenting my systemising model for understanding and
analysing VC protocols and properties. \textbf{Chapter II,
\emph{''Non-universal VC Protocols''}}, introduces and gathers currently
expanding and innovative VC technologies (i.e.~HAUTHs, VDFs) which have
previously been considered within separate domains, offering a
comprehensive analysis under the model provided in the first chapter.
\textbf{Chapter III, \emph{''Universal VC Compilers''}}, introduces to
our model and breaks down state-of-art prominent VC technology
(i.e.~STARKs) which has yielded groundbreaking results, with the
potential to revolutionise the cryptographic community as well as
disrupt the cryptocurrency market itself.

\begin{verbatim}
Finally, **Chapter IV, _”An Open-Banking Use-Case”_**, provides practical insight into the potential of VC technologies through the exhibition of a simple but innovative protocol capable of appealing to the EU's novel Open-Banking ("PSD2" standardised) market, as well as my EIT scholarship[@eit-webpage] and its partner, and my internship provider, E-Group[@egroup-webpage].
\end{verbatim}

\hypertarget{introduction-and-vc-model}{%
\chapter{Introduction and VC Model}\label{introduction-and-vc-model}}

\pagenumbering{arabic}

Throughout the past few decades, our society has put a great deal of
effort into developing technologies upon which to build \emph{trusted
platforms} and services. Along with this explosion of services, the
Internet has brought digital freedom into our daily lives. The latest
example of this is distributed networking (e.g.~Bitcoin, Ethereum),
which aims to replace important societal functions. This latest trend
marks an important milestone of our globalised society: the time has
come to build \emph{trustless platforms}, built upon technologies we can
all indiscriminately trust.

The services we are speaking of take electronic form and exist only in
the realm of Computer Science, which imposes restrictions on what
security and trust really mean, expressed fundamentally in the form of
Cryptography. Cryptography the art, Cryptography the science, has been
developing at an accelerated rate of research ever since human
conflict,\footnote{it appears that Caesar was also a big fan of
  cryptography :)} and the need for trusted communication, has existed.
What we deal with in this thesis is ``how to trust \emph{someone} doing
\emph{something} with \emph{some secret data}''. That phrase might seem
a little vague, but I promise: it encompasses so many notions of
computer science and cryptography that it extends to virtually any
computation on paper or silicon. To define this domain, I'll use the
term ``Verifiable Computation'' (or VC).

In order to explain what it means to \emph{prove computation}, I would
like to start by taking a brief look at the most common form of provably
secure computation since the birth of the Web: Authentication (Section
\ref{sec:auth}). Afterwards, I'll move on to define formally (and
informally) what generic cryptographic proofs entail (Sections
\ref{sec:theorems}, \ref{sec:types}); then how to perform
privacy-friendly computations of hidden variables using Zero-Knowledge
Proofs (Section \ref{sec:zkip}); then, how to do this without multiple
rounds of communication (Section \ref{sec:fs}), more efficiently and
with less communication overheads (Section \ref{sec:scal}), as well as
other VC properties and open questions (Section \ref{sec:othervc}). For
a discussion on how to put it all together in a single convenient
package, please check Section \ref{sec:ci} in the chapter on Universal
Verifiable Computation.

\hypertarget{sec:auth}{%
\section{Identification Schemes and Authentication}\label{sec:auth}}

Before we talk about Verifiable Computation, let's scale down a bit and
talk about a simpler concept: Identification Schemes. The process of
authenticating a user, which simply defines a Prover showing to a
Verifier that he knows a specific non-deterministic secret relating to
his publicly known identifier, has roughly evolved under the three
following cryptographic constructions.

\hypertarget{simple-password-based-authentication}{%
\subsection{Simple Password-based
Authentication}\label{simple-password-based-authentication}}

In this naïve approach the Verifier doesn't trust the Prover, so he asks
him to send over the secret password (i.e.~``knowledge witness'') so
that he may verify it.
\[Prover \overset{password}{\rightarrow} Verifier\] This is an extremely
flawed protocol because:

\begin{enumerate}
\def\labelenumi{\arabic{enumi}.}
\tightlist
\item
  the whole secret is revealed to the Verifier (if the Prover actually
  knows the password, that is);
\item
  and anybody else looking at this conversation;
\item
  the secret password needs to be well protected and stored by both
  parties.
\end{enumerate}

The following solutions have been devised to improve this method:

\begin{enumerate}
\def\labelenumi{\arabic{enumi}.}
\tightlist
\item
  N/A (it is a requirement of the protocol);
\item
  secure unicast communication channels, e.g.~HTTPs;
\item
  the Server stores the password in Hash+Salt format, or Encrypted
  format. This helps take stress off hacked servers whose database is
  compromised, as long as the hack is detected (otherwise, the webpage
  can be modified to redirect login attempts). Another solution is to
  have an auxiliary check (called Two Factor Authentication, or 2FA)
  using one-time tokens sent via SMS (or an app); unfortunately this is
  mainly a convenient hack invented by the industry to patch up the
  inherent weaknesses of this authentication approach, and it is as
  secure as the device and communication used to receive the token, as
  well as the token generation process itself.
\end{enumerate}

Even though Simple Password-based authentication is a very
straightforward interaction (the user types into a text box) it still
conveys a false sense of security, leads to many failed login attempts
(due to typing mystakes) as well as poor password generation habits by
lazy or misinformed users.

\hypertarget{challresponse}{%
\subsection{Challenge-Response Authentication
Protocols}\label{challresponse}}

In this approach neither party can trust the other, or the communication
channel may be unsafe, so they take advantage of any ``hard
cryptographic problem'' to challenge knowledge of the secret.
Essentially, Asymmetric Encryption and Signature schemes help provers
avoid revealing their secret to malicious verifiers.
\[Prover \overset{challenge}{\leftarrow} Verifier
\\Prover \overset{response}{\rightarrow} Verifier\] \emph{(where the
challenge is a random number chosen by the Verifier, which the Prover
must use to generate as response a unique signature or reveal a random
message previously encrypted by the Verifier.)}

This method is already a huge improvement over the previous one, and yet
it has received very little adoption amongst the most popular Web
services, even though it could easily be implemented through browser
plugins and software wallets. In fact, its most widespread adoption
seems to be physical authentication cards, used for traditional banking
transactions at ATMs or for authorising entry to company offices.

There is still one small issue: Challenge-Response protocols still
reveal some information, such as unique signatures or decrypted
cipher-texts. While Encryption and Signature schemes are chosen to leak
as little information as possible (e.g.~Computationally
Indistinguishable from random values), there is still something to be
learned from selective forgery attacks (for signatures) and chosen
cipher-text attacks (for encryption); should the underlying
cryptographic scheme be broken, the credentials and privacy of the users
might be compromised.

\hypertarget{zero-knowledge-identification-protocols}{%
\subsection{Zero-Knowledge Identification
Protocols}\label{zero-knowledge-identification-protocols}}

Neither party trusts the other. With this technique exactly
\underline{nothing about the secret is revealed} to the Verifier, except
that it is valid. The secret to achieving this marvellous result lies
within Interactive Proof Systems and their properties, which we will
discuss in the rest of this chapter. A common approach to such protocols
is through one or more rounds of interactivity:
\[\textit{commit step}\begin{cases}Prover \overset{statement}{\rightarrow} Verifier\end{cases}
\\\ \textit{round}\ i \begin{cases}
Prover \overset{challenge}{\leftarrow} Verifier
\\Prover \overset{proof}{\rightarrow} Verifier\end{cases}\]

Alternatively, there is a field of protocols which performs transparent
preprocessing and then sends off a single large proof to be
probabilistically checked offline:
\[Prover \overset{large\ proof}{\rightarrow} Verifier\]

\hypertarget{sec:theorems}{%
\section{Theorem Proving and Interactive Proofs}\label{sec:theorems}}

The roots of Verifiable Computation extend all the way to Theorem
Proving, when mathematicians still wrote their proofs on paper. If we
wish to convert mathematical theorems to the domain of computer science,
we should take a look at the well fleshed out theories of NP complexity
classes; here is an informal definition of NP Theorem Proving:

\begin{description}
\item[NP Languages]
\(Th \in NP \iff \exists \textit{ "witness" } w : \textit{ Th is "easy" to verify using } w\)

\emph{Note: You should look at the witness as a sequence of logical
deductions which start from truthful statements and lead all the way to
the theorem claim}:
\[\textit{axiom(s)} \implies \overbrace{... \implies ... \implies ... }^\textit{witness} \implies Th\]
\emph{If one part of the sequence is already known to the Prover, the
witness represents the part which is not known.}
\end{description}

This definition was extended in 1985 by
{[}\protect\hyperlink{ref-GMR85}{4}{]} to represent an Interactive Proof
System IP:

\begin{description}
\tightlist
\item[IP Languages]
\[\textit{Given }
\begin{cases}
 P_\text{UNBOUNDED}, V_\text{POLY} \in ITM \textit{ (Interactive Turing Machine)} \\
 L \subseteq \{0,1\}^* \in NP-lang \\
 n \textit{ input size}, c \textit{ large constant} \\
 w \textit{ secret witness of P} \phantom{(??) define as set of implications…} \end{cases}
 \\
 \textit{Then } X \in L \iff
 \\\land\begin{cases}
 \textbf{Completeness} \iff \forall \textit { input } X \in L \textit{ to }(P, V): Pr[V \textit{ accepts } X] \geq 1 - \frac{1}{n^c} \\
 \textbf{Soundness}\iff \forall P’_{POLY} \in ITM \land X \notin L \textit{ input to } (P', V): Pr[V \textit{ accepts } X] \leq \frac{1}{n^c} \\
 \qquad\textit{or} \iff \forall P'_{POLY} \in ITM \land X \notin L \textit{ input to } (P', V):
 \\\quad\qquad \Big(Pr[V \textit{ accepts } X] \geq 1 - \frac{1}{n^c} \implies 
 \\\qquad\qquad\exists \textit{ "Extractor" } E_{POLY} \in ITM : \exists R \subseteq \{0,1\}^*: E(X) = R(w)\Big)
 \end{cases}\] \emph{(please note that we've defined the language as NP,
but IP protocols have been shown to support even more expressive spaces
such as PSPACE or even NEXP)}
\end{description}

These systems are often called ``Proofs (or Protocols) of Knowledge'',
because the Completeness property defines a protocol (i.e.~set of rules)
to follow in order to accept a given statement, and the Soundness
property implies instead the existence of some sort of ``knowledge''
(also known as ``witness''), needed to distinguish right from wrong. The
alternative definition of Soundness, which makes use of an Extractor
machine, is typically used to single out unique Knowledge which is
possessed by the Prover, and is useful for understanding
``Zero-Knowledge Proofs of Knowledge''. Here is a more intuitive
definition of those properties:

\begin{description}
\item[Completeness]
if the statement \(X\) is valid, then there is an ``easy'' way to prove
it using the protocol. The Verifier will be able to efficiently check
this in polynomial time. In order words: \emph{all valid statements are
always accepted.}
\item[Soundness]
if the statement \(X\) is false, then there is ``almost'' no way to
prove it. The Verifier only needs to trust its own knowledge and
randomness to disprove false proofs from an all-powerful Prover. In
other words: \emph{all invalid statements are always rejected.}

\emph{(When discussing the Extractor, the key to understanding the
definition is that it shouldn't be possible to accept false statements,
unless the illegitimate Prover was somehow capable of extracting the
witness, or a relationship on the witness, from the statement \(X\)
itself in order to use it.)}
\end{description}

An important security observation to make, especially when considering
the second definition for Soundness, is that there is no restriction to
the amount of ``knowledge leaked'' by the execution of an IP instance.
This essentially means that the Prover could naïvely just send his
secret password (i.e.~witness) over, and the protocol might still be
valid. Restrictions on such flaws, as well as the importance of
Interactivity, will be added by the Zero-Knowledge definition.

A second important security observation is that the Soundness property
is so vague that it does not really provide any security guarantee that
the statement \(X \in L\) will be hard to prove for cheaters, only that
it should not be possible to prove \(X \notin L\). In fact, if the
language \(L\) trivial enough, it might even be possible to randomly
choose any \(X \in L\) and extract the witness required to prove it.
Thus, the security of our proof lies entirely within the chosen language
\(L\), which is typically based on some hard cryptographic problem such
as finding the prime factors of a large number.

\emph{NOTE: While we've defined soundness based on negligible
probability, practical constructions only require \(Pr < \frac{1}{2}\),
and repetition is employed to achieve the definition above. Also, almost
all practical systems have perfect completeness (\(Pr = 1\)).}

\hypertarget{sec:pcp}{%
\subsection{Probabilistically Checkable Proofs}\label{sec:pcp}}

There is an alternative field of cryptographic proof systems that is
roughly equivalent to IPs, but uses different constructions:
Probabilistically Checkable Proofs (PCPs). We will not go into detail
regarding PCPs, but suffice to say that they share a lot of similarities
with IPs. The main differences in the definition are minor details
regarding specific cryptographic properties:

\begin{itemize}
\tightlist
\item
  \emph{Soundness}: it is always computational, since the Prover is
  computationally (\(P_{POLY}\)) bounded, just like the Verifier. This
  is due to the fact that PCPs are not technically ``proof'' systems,
  but ``argument''-based systems.
\item
  \emph{Non Interactivity}: such protocols don't require any
  interactivity by default (IPs need the Fiat-Shamir extension discussed
  later); instead, the Prover preprocesses the original language
  statement to generate a (typically large) proof to send off to the
  Verifier for inspection. This notion will be defined in Section
  \ref{sec:fs}.
\item
  \emph{Transparency}: such systems do not employ interactivity because
  everything is prepared in a trustless fashion. The Verifier will be
  able to use (public) randomness to analyse a few elements of the given
  proof. This notion will be defined in Section \ref{sec:othervc}.
\item
  \emph{Verifier Efficiency}: these protocols are required to be
  efficient by default. This is due to the groundbreaking results
  emerging from the PCP Theorem
  {[}\protect\hyperlink{ref-PCPTheorem}{5}{]}--{[}\protect\hyperlink{ref-Babai91second}{8}{]}
  finalised in '98 by Aurora et al., which led to the conferring of the
  Gödel Prize for multiple cryptographers having worked on it throughout
  the 90s. This notion is defined in Section \ref{sec:scal} through
  \emph{proof succinctness} and \emph{verifier scalability}.
\end{itemize}

In practice, PCP systems make heavy use of polynomial arithmetisation,
making them better suited for \emph{Universal} VC systems, as seen in
Chapter 3; IPs, instead, typically focus on constructions based on
specific problem isomorphisms. An extension of PCPs called Interactive
Oracle Proofs (IOPs) {[}\protect\hyperlink{ref-IOP}{9}{]}, which
combines them with IPs, can be found in state-of-the-art Universal VC
systems and I mention it in Section \ref{sec:fri}).

\hypertarget{sec:types}{%
\section{Types of Knowledge}\label{sec:types}}

The issue with most weak cryptographic authentication methods is that
some uniquely identifiable knowledge about the secret is somehow
``leaked'' during the authentication process. Intuitively, we would like
to reduce this ``knowledge leakage'' as much as possible. In order to do
so, we must first understand what possessing ``knowledge'' truly means.
Let us define two major scenarios where knowledge is typically conveyed:

\begin{enumerate}
\def\labelenumi{\arabic{enumi}.}
\item
  \emph{Communication}:

  the Prover has chosen (or is in possession of) some non-deterministic
  private value which the Verifier needs to solve some other publicly
  known problem (likely published by the Prover and verified by a
  Trusted Authority). The only way for this value to be known is through
  the Prover himself.
\item
  \emph{Computation}:

  the Verifier would like to extract some knowledge from a given hard
  problem, but is too computationally bounded to be able to do so. Given
  enough computational power, any Prover would be able to extract the
  required knowledge and convey it to the Verifier.
\end{enumerate}

Knowledge seems to be strictly related to the act of communicating some
value which is the result of a computation that was either too difficult
or even impossible for the Verifier to perform. In other words,
knowledge is transferred between two communicating parties if and only
if the output of their interaction was the result of an infeasible
computation for one or both of the parties.\footnote{An interesting note
  here is that transferring random bits does not typically convey any
  information, since any party can generate randomness by itself (being
  an ITM). This may seem counterintuitive, but those random bits would
  only convey knowledge if related to some pre-defined public statement
  or problem which the Verifier cannot solve by himself.} Here is an
informal definition:

\begin{description}
\tightlist
\item[Knowledge Complexity KC]
\[\textit{Given }
\begin{cases}
P_\text{UNBOUNDED}, V_\text{POLY} \in ITM \\
L \in IP(P,V) \\
f: \mathbb{N} \to \mathbb{N} \land f \textit{non-decreasing} \\
n \textit{ input size}
\end{cases}
\\
\textit{Then } KC_L(f(n)) \iff \vphantom{\textit{ (i.e. the knowledge complexity of L is f(n)) }}
\land \begin{cases}
\textbf{i. } X \in L, X \textit{ \underline{only} input to} (P,V) \\
\textbf{ii. } P \textit{"communicates"} \leq f(n) \textit{ bits of "knowledge"}
\end{cases}\]
\end{description}

Whenever we have \(KC_L(0)\), that means we can only convey one bit of
knowledge with our protocol: \(X \overset{?}{\in} L\).

\hypertarget{sec:zkip}{%
\section{Zero-Knowledge Interactive Proofs}\label{sec:zkip}}

If we embed \(KC_L(0)\) into the notion of IP, we get the following:

\begin{description}
\tightlist
\item[ZKIP Languages]
\[\textit{Given }
 \begin{cases}
 P_\text{UNBOUNDED}, V_\text{POLY} \in ITM \textit{ (Interactive Turing Machine)} \\
 L \subseteq \{0,1\}^* \in NP-lang
 \end{cases}
 \\
 \textit{Then } X \in L \iff
 \\\land\begin{cases}
 \textbf{Completeness}\\
 \textbf{Soundness}\\
 \textbf{Zero-Knowledge} 
 \\\iff \forall V'_\text{POLY} \in ITM: 
 \exists \textit{ "Simulator" } S_\text{POLY} \in ITM: \textit{Tx}(S(V')) \approx \textit{Tx}(P,V) \\
 \quad\implies \textbf{Deniability}
 \end{cases}\]
\end{description}

Or, more intuitively:

\begin{description}
\item[Zero-Knowledge]
The idea is that no extra knowledge can be extracted from a legitimate
valid interaction (i.e.~leading to an accepting state), as long as it is
``indistinguishable'' from a forged valid interaction. In fact, there
should be an efficient Simulator algorithm to simulate a valid
interaction's transaction record \(Tx\) even when the simulating
Verifier \(V'\) doesn't have access to the Prover's real witness. The
Simulator can generate \(Tx(S(V'))\) either by executing many protocol
runs until an accepting state is met, or just by deducing the correct
statement \(X\) starting from any final accepting state (known as
``rewind-ability''). I will elaborate later on what
``indistinguishable'' (i.e.~\(\approx\)) means in Section
\ref{sec:typeszk}.

The reason that this simulator-based definition leads to
privacy-friendly (i.e.~non witness-leaking) protocols is because there
can be no witness Extractor for legitimate transcripts, since they're
indistinguishable from forged transcripts, which are assumed to lack any
witness at all. In other words: \emph{the protocol's soundness needs to
rely entirely on interactivity and randomness!}
\end{description}

Zero-Knowledge also implies Deniability:

\begin{description}
\tightlist
\item[Deniability]
A Transaction record from a valid ZKIP interaction does not constitute
an independent proof of knowledge. No external third parties can watch
(or be given) a valid ZKIP communication and infer that the Prover
really has a witness for \(X\in L\), because the interaction may have
been simulated. Only the original parties of the ZKIP communication can
verify that it is indeed legitimate, because they know that the messages
were not forged when challenging each other.
\end{description}

The trick to actually achieving Zero-Knowledge in a meaningful manner
lies within the combination of \underline{Interactivity and Randomness}.
The two parties cannot use a Challenge-Response protocol, because the
response to the challenged question is rather unique, regardless of the
chosen challenge. However, if the response were to be randomly selected
(i.e.~challenged) out of a random distribution of values selected by the
Prover, it would not contain any meaningful information. In order for
such a protocol to be sound, only a legitimate Prover would be
\textbf{always} able to calculate the required response: a deterministic
relationship (selected using the Verifier's random challenge) on a
statement \(Y\), randomly derived from the original statement
\(X\).\footnote{there can also be multiple \(Y\) statements derived from
  \(X\) at the same time, for efficiency purposes.} Interactivity is
required because, regardless of the chosen challenge, the Prover's set
needs to be random for each protocol execution.

Finally, the same security assumption of IPs apply to ZKIPs: the
difficulty of proving \(X \in L\) lies in the chosen language \(L\) and
its cryptographic hardness assumptions.

\medskip

\emph{NOTE: alternative definitions have been used in the past to
describe zero-knowledge, such as ``witness preservation'' and ``witness
indistinguishability'', but the one given here is the strongest one and
the current standard.}

\medskip

\emph{NOTE2: if you would like an alternative informal explanation of
ZKIP protocols, I highly recommend the beautiful paper by Quisquater et
al.~{[}\protect\hyperlink{ref-Quisquater90}{10}{]} on the metaphor of
the ``Ali Baba Cave''. One important feasibility result for ZKIP proofs,
based on finding Hamiltonian cycles in graphs, was given by Blum in 1986
{[}\protect\hyperlink{ref-Blum86}{11}{]}.}

\hypertarget{zkip-as-a-solution-to-malicious-actors}{%
\subsection{ZKIP as a solution to malicious
actors}\label{zkip-as-a-solution-to-malicious-actors}}

Zero-Knowledge proofs are regarded as being an extremely powerful tool
to convert malicious actors into semi-honest actors. A researcher first
builds a protocol which is shown to be secure when all parties (or
eavesdroppers) are semi-honest (i.e.~they always follow the protocol's
rules); then, any party sending messages is required to provide proofs
that they were generated following protocol requirements. Since each
proof is Zero-Knowledge, the security of the original protocol is not
compromised. Because each message must be accompanied by a proof,
malicious attackers have no choice but follow the rules of the protocol,
or just abort. While there is a computational cost to be paid per proof,
Universal VC systems (discussed in Chapter 3) are a convenient and
efficient solution for adding such capabilities.

\hypertarget{sec:typeszk}{%
\subsection{Types of Zero-Knowledge}\label{sec:typeszk}}

We previously defined a Simulator capable of generating fake valid
protocol which are also indistinguishable from legitimate valid runs:
\(Tx(S(V')) \approx Tx(P,V)\).

There are currently 4 different classifications of indistinguishability
(\(\approx\)):

\begin{enumerate}
\def\labelenumi{\arabic{enumi}.}
\tightlist
\item
  \emph{Perfect}: there exists a Simulator which produces communication
  transcripts \emph{identically distributed} to the legitimate
  distribution of valid transcripts between \((P,V)\).
\item
  \emph{Statistical}: there exists a Simulator which produces
  communication transcripts \emph{identically distributed} to the
  legitimate distribution of valid transcripts between \((P,V)\), except
  for a constant (i.e.~``small'') number of exceptions.
\item
  \emph{Computational} (default): there exists a Simulator which
  produces communication transcripts \emph{not-identically distributed}
  to the legitimate transcripts produced between \((P,V)\), but it is
  believed to be \emph{computationally infeasible} to detect such
  differences.
\item
  \emph{Not Known (No Use)}: there does not exist a Simulator but the
  communication Transactions are still believed to leak nothing about
  the witness.
\end{enumerate}

\emph{NOTE: the ``Not Known'' type of indistinguishability \textbf{does
not} satisfy full Zero-Knowledge requirements, and it also implies
Non-Deniability. See the next section.}

\hypertarget{sec:fs}{%
\section{Non Interactivity and Digital Signature
Algorithms}\label{sec:fs}}

We have covered the basics of Zero-Knowledge Proofs, and we've seen that
two essential aspects are \emph{randomness and interactivity}. Well,
what if the Prover and Verifier's interactivity in the real world is
effectively limited? For example, they may not be online at the same
moment, or the Prover might want to pre-process multiple proofs by
himself. Is it even possible to have ``Non-Interactive'' Zero-Knowledge
Proofs (NIZK)?

In 1987 an article by Israeli researchers Fiat and Shamir
{[}\protect\hyperlink{ref-FS87}{12}{]} proposed a heuristic to solve the
aforementioned problem. The key takeaway here is that, while
Zero-Knowledge is not deemed to exist without Interactivity, we can
adopt the famous Random Oracle Model (ROM)
{[}\protect\hyperlink{ref-ROM}{13}{]} assumption to make use of an
interacting ``oracle'' party which will supply us with ``public-coin''
random challenges for our protocol. If we assume that cryptographic Hash
functions correctly implement a Random Oracle, we can employ them as
universal and passive Verifiers to participate in our proof, thus
obtaining:

\begin{description}
\tightlist
\item[Fiat-Shamir Heuristic]
the Verifier selects a challenge \(e = H(pp)\), where \(H\) is a strong
cryptographic hash function implementing a Public-Coin Random Oracle,
and \(pp\) are the public parameters of the problem and the current
protocol execution (including the Prover's randomness)
\end{description}

As can be easily be understood, everyone with the same Hash function
also has access to the same challenges. Which means that they can
validate the lack of bias within the selection of random challenges,
hence the legitimacy of the proof. Since this check can be performed
independently after the execution of a NIZK, once the recorded
communication trace is given the proof can essentially be verified by
anybody. This is what it means for the protocol to become
``\textbf{Non-Interactive}'', while still retaining Interactivity in the
ROM model.

On a final security note, while the lack of bias in the selection of
challenges is apparent, the challenges are still selected based on the
Prover's random inputs. These can be biased and, since interactivity is
outsourced to the Oracle, an illegitimate Prover can mount an
\textbf{offline attack} to keep simulating protocol runs until he finds
lucky challenges he can satisfy. Therefore, to prevent cheating from the
Prover, we must exponentially decrease the error rate for Soundness to
brute-force levels
(i.e.~\(\epsilon = 2^{-256} \iff Pr[V \textit{ accepts } X] \leq \frac{1}{2^{256}}\)).
This would imply that if a Prover does not have unlimited resources
(like in a real-life scenario, but unlike the formal ZKIP definition),
then he should not be able to come up with a simulated NIZK valid
protocol run.

\hypertarget{flawed-nizk-zero-knowledge-and-non-deniability}{%
\subsection{Flawed NIZK Zero-Knowledge and
Non-Deniability}\label{flawed-nizk-zero-knowledge-and-non-deniability}}

The last remark noted that we're preventing Provers from being able to
simulate protocol runs. Does this also mean that the Zero-Knowledge
property is broken? Well, yes, but actually no. There does not seem to
be a definitive answer in the cryptographic community as to whether
Zero-Knowledge is truly preserved for the Fiat-Shamir heuristic (a good
debate on this can be found in
{[}\protect\hyperlink{ref-fiatshamirisalie}{14}{]}), but \(KC_L(0)\) is
believed to hold as long as the ROM model holds. The Zero-Knowledge
property for a Fiat-Shamir NIZK is currently classified as ``Not Known''
(see the relevant subsection).

As an important consequence of the fact that NIZK proofs can be
validated by anybody with a transcript of the communication, the
Deniability property is broken:
\[NIZK \implies \textbf{Non-Deniability} \centernot\implies\textbf{Deniability}\]
This has the downside that any third parties can detect whether a proof
was legitimate or not. While this may not seem like such a big deal,
uniquely identifying logins (i.e.~Identity Proofs) in censorship states
can pose a real threat to human rights. It is best to use pseudonymous
identities and de-anonymising networks when using NIZK technology
(e.g.~ZCash) under such harsh regimes.

\hypertarget{digital-signature-algorithm-construction}{%
\subsection{Digital Signature Algorithm
construction}\label{digital-signature-algorithm-construction}}

An important upside of NIZKs of Knowledge is that they can be extended
from one-shot Identification Schemes to one-shot Digital Signature
Algorithms!

\begin{description}
\tightlist
\item[DSA Fiat-Shamir Heuristic]
the Verifier selects a challenge \(e = H(pp, m)\), where all parameters
are the same as the standard Fiat-Shamir Heuristic, and \(m\) is the
message that the Prover wants to sign.
\end{description}

The ``signature'' is, of course, actually just a proof of Knowledge
which ties the Identification proof to the presence of a specific
message in the Verifier's challenge. This suffices to show that the
Prover knows the witness for his identity, and that he is committing to
using randomness (i.e.~challenges) derived from a specific message.

\hypertarget{sec:scal}{%
\section{Performance through Scalability}\label{sec:scal}}

Over the years, as cryptographers struggled to develop Zero-Knowledge VC
protocols for practical use cases, a few more properties on
\textbf{performance requirements} were devised\footnote{A result of this
  approach can be seen in the the PCP field of proof protocols.}. These
properties are especially relevant for comparison of recent Universal VC
systems, such as the ones mentioned in Section
\ref{sec:universalconclusion}, which tend to make compromises in the
name of expressiveness.

If outsourcing verifiable computations is to be seen as a commodity,
then they have to be fast to verify. This requirements boils down to two
main properties: there should be an exponential gap between the protocol
execution complexity of the Prover and Verifier (where the Verifier
takes less time), and the proof size should be small enough that the
Verifier can read it. Furthermore, Provers of the past often required
hundreds of gigabytes and weeks just to process simple proofs, we'd like
to avoid that as well. Formally:

\begin{description}
\item[Fully Scalable Proof]
\[Given 
\begin{cases}
P_\text{POLY}, V_\text{POLY} \in ITM \textit{ (Interactive Turing Machine)} \\
(x, y, f) = X \in L,\\
 y = f(x),\ O_y(\Delta)\\
\end{cases}, \\\textit{Then}\ X \in L \implies \land
\begin{cases}
\mathbf{Completeness}\\
\mathbf{Soundness}\\
\mathbf{Prover\ Scalability}\ \iff O_P(\Delta + polylog(\Delta))\\
\mathbf{Verifier\ Scalability}\ \iff  O_V(polylog(\Delta))\\
\mathbf{Proof\ Succinctness}\ \iff \forall \pi = Tx(P, V): O_{|\pi|}(polylog(|x|))
\end{cases}\]

\emph{Intuitively, we want to validate proofs \(\pi\) much faster than
it takes the Verifier to actually check the statement himself, and
without excessive overhead for the Prover. Also, the communication
complexity for such protocols should be always be well within acceptable
standards.}

\emph{It is important to note that most protocols don't achieve such
results, so the actual definitions are typically relaxed based on the
current best solution in that field. For the STARK protocol analysed in
this thesis I'll use as satisfying prover-scalability requirement a
quasilinear Prover \(O_P(\Delta \cdot polylog(\Delta))\), which can
yield acceptable concrete performance results in most cases.}
\end{description}

\emph{NOTE: often verifier-scalability and proof-succinctness are
regarded together as ``verifier efficiency''. For extra confusion,
sometimes researchers also use the term ``succinctness'' to refer to one
or both properties, or just scalability in general. Sometimes a fully
scalable system is called doubly scalable, or just scalable.}

\hypertarget{sec:othervc}{%
\section{Other VC Properties}\label{sec:othervc}}

Finally, some further non-essential but \textbf{highly appreciated
properties} are added, which increase the reliability and flexibility of
a proof system, allowing it to be used in more demanding use-cases:

\begin{description}
\item[Transparency]
\(Tx(P \gets V) \in \textit{public random coins}\) \emph{; i.e.~the
Verifier only ever sends messages taken from a randomness source that is
also available to the Prover.}

This property was first conceived with Arthur-Merlin (AM) protocols
({[}\protect\hyperlink{ref-ArthurMerlin}{15}{]},
{[}\protect\hyperlink{ref-ArthurMerlin2}{16}{]}), which were proven to
be equivalent in expressiveness to IP protocols that had separate
randomness sources for the two parties; it was first called
``transparency'' in {[}\protect\hyperlink{ref-ArthurMerlin3}{17}{]}.
Transparency is typically present in all PCP-family protocols.

The reason that this property is ``transparent'' is because as long as
the Prover and Verifier have access to the same randomness source, there
can be no trusted or trapdoor-derived setup for the underlying
protocol\footnote{all setups are either deterministic, or public-coin
  non-deterministic.}. Because trust is eliminated, the security of the
protocol cannot be compromised as it does not depend on any specific
party, only mathematics. Transparency has also become a matter of
interest lately, due to the increased popularity of zk-SNARK
constructions (see Section \ref{sec:universalconclusion}), which are
infamous for their trusted setups and less suitable for decentralised,
trustless settings.
\item[Universality]
\(L \iff NP-lang\) \emph{; i.e.~the protocol language supports
statements taken from any NP computation.}

This property is extremely useful for implementing basic cryptographic
proving primitives that can be applied to computations of
Turing-complete machines. The utility of such Universal VC protocols
lies with the convenience of being able to freely design an application,
and then automatically generate proofs for the actions performed by said
application. This topic is widely discussed in Chapter 3.
\item[Post-Quantum Safety]
\emph{the protocol makes use of cryptographic assumptions which are not
shown to be compromised by Quantum algorithms.}
\end{description}

Finally, a couple of \textbf{open questions} which have been less (if at
all) studied in popular VC constructions:

\begin{description}
\tightlist
\item[Composition]
\emph{the proofs of different statements can be efficiently combined, or
extended into more complex ones.}
\item[Multi-Party]
\emph{a single proof can be generated using multiple inputs taken from
different Provers.} Achieving such a property for Zero-Knowledge
protocols would be akin to achieving Multi-Party Computation.
\end{description}

\hypertarget{non-universal-vc-protocols}{%
\chapter{Non-universal VC Protocols}\label{non-universal-vc-protocols}}

In the decades leading up to the introduction of practical Universal
VCs, most protocols only dealt with either secret proving or specialised
computational proofs. Common tools to achieve this were either
Interactivity (and isomorphic problems) or Homomorphic Encryption, or
both. In this chapter I will evaluate two \textbf{innovative fields} of
cryptography which have a strong correlation with universal VC
solutions: Homomorphic Authenticators and Verifiable Delay Functions.
While they have mostly been developed under different contexts, they
share many of the fundamental properties introduced in the VC Model
chapter. Homomorphic Authenticators deal with outsourced (homomorphic)
computation and VDFs deal with scalable computation; both yield
interesting protocols which can be adapted, with a little expertise,
into practical ad-hoc applications.

I will carefully evaluate the properties achieved by each construction,
trying to understand the cryptographic design behind it without
sacrificing the simplicity of our VC Model. This aim of this chapter is
to alleviate the fragmentation and complexity of the fields which stand
below Universal VC solutions, showing that they can be useful starting
points for achieving richer VC constructions. Further non-universal VC
protocols will be discussed in the conclusive remarks of this chapter.

\hypertarget{homomorphic-authenticators}{%
\section{Homomorphic Authenticators}\label{homomorphic-authenticators}}

Homomorphic Authenticators (HAUTHs) stem from an interesting and active
area of research, with recent publications being in 2018. The incipit of
this field lies with Homomorphic Signature schemes, which were
originally rejected by cryptographers due to their implicit
susceptibility to forgery attacks. These schemes were brought up again
by Rivest, and formalised by Johnson et al in
2002{[}\protect\hyperlink{ref-Johnson02}{18}{]}, to include better
definitions for security against forger (i.e.~Random Forgery attacks).

Multiple researchers followed down this path, coming up with innovative
constructions for validating \textbf{outsourced computations}
(e.g.~cloud computing). An initially successful design
{[}\protect\hyperlink{ref-Gennaro12}{19}{]} (in terms of VC features)
relied on fully homomorphic MAC constructions using polynomials. These
MACs would then be sent to a server (i.e.~Prover), which would leverage
their homomorphic properties to generate computational proofs on the
given data. This way, homomorphism can be used to yield valid isomorphic
problems, akin to the concepts introduced in Chapter 1. The big
advantages of this technique, compared to IPs, are: outsourced proving,
proof composition, and efficiency for really big problem sizes. In fact,
a core feature of Hom.Authenticators is that a Prover can upload really
large databases to the Verifier, and then delete them; the
Authenticators themselves will be sufficient to verify the validity of
any computation on this data.

The polynomial-based construction was then extended by
{[}\protect\hyperlink{ref-Fiore16}{20}{]} to support inputs for multiple
clients, which would grant a simil multi-party property. This was
achieved by adding a form of homomorphism to the keys themselves, and
then allowing the Verifiers to merge them during the verification phase.
Other forms of publicly verifiable schemes were provided (based on
lattices), but they proved to be very complex and inefficient. In order
to make the construction more practical, Fiore et
al.{[}\protect\hyperlink{ref-Fiore13}{21}{]} managed to achieve verifier
scalability, albeit sacrificing the universality of the protocol. This
modification used a combination of polynomials and additive group
schemes with bilinear pairings for multiplication. Finally, public
verifiability and zero-knowledge were added just now in 2018 by
Schabhüser et al.{[}\protect\hyperlink{ref-SBB18}{22}{]}. Public
verifiability is achieved by building a homomorphic signature scheme out
of the homomorphic MACs; simil ZK was achieved through a property known
as ``context-hiding''.

In this section I will provide the following content: an overview of
Homomorphic Authenticator protocols and commonly used syntax for this
field (Section \ref{sec:hauthsyntax}); the basic homomorphic components
behind them (Section \ref{sec:hauthcomplete}); a basic MAC construction
construction (Section \ref{sec:hauthsound}); an extension to multiple
clients participating in the protocol (Section \ref{sec:hauthmulti});
and finally support for verifier scalability (Section
\ref{sec:hauthscalable}).

\hypertarget{sec:hauthsyntax}{%
\subsection{Protocol Syntax}\label{sec:hauthsyntax}}

Let's consider the protocol as a 3-step proof, which makes it simpler to
compare it with other VC technologies. We wish to prove \(f(x)=y\), the
construction is as follows:
\[client\ Verifier \overset{\sigma_x}\longrightarrow server\ Prover\\
client\ Verifier\ \overset{\sigma_y}\longleftarrow server\ Prover\\
check(\sigma_y) \overset{?}= True\] Here is the sequence of steps which
the parties go through, in a basic construction:

\begin{enumerate}
\def\labelenumi{\arabic{enumi}.}
\tightlist
\item
  \textbf{Preparing the Authenticator:} The Verifier needs to convert
  his message \(x\) into a Homomorphic Authenticator \(\sigma_x\), which
  is a MAC or Signature made using a secret key \(sk\). First, the
  client generates a unique label \(L\) relating to the message \(x\),
  e.g.~``message \#1'' or ``message x on time 10:54''. The label is then
  converted into a random value \(r\), required for the security of the
  scheme, using a keyed one-way PRF\footnote{for example, a seeded PRNG
    constructed from a keyed cryptographic hash function such as
    Keccak256{[}\protect\hyperlink{ref-keccak}{23}{]}}
  \[r \gets PRF_K(L)\] A Homomorphic MAC (HMAC) is typically built using
  polynomial interpolation:
  \[\sigma_x \gets p = (p_0, p_1) = (x, (r-x)/sk) = Interpolate((0, x), (sk, r))
   \\ with\ p(i) = p_0 + p_1 i\] Since it's more common to define a
  function as composition of multiple inputs,
  i.e.~\(f(x_1, x_2, ...) = y\), then this HMAC interpolation process
  can be repeated for each input message, and each message \(x_i\) will
  be associated with a different label \(L_i\):
  \[\sigma_x \gets (\sigma_1, \sigma_2, ...) = (p_1, p_2, ...)\]
\item
  \textbf{Generating an Authenticator-based proof}: The Prover then uses
  the MAC/Signature scheme to convert function \(f\) into a sequence of
  homomorphic operations on \(\sigma_x\). After these operations have
  been performed, they will yield a valid MAC/Signature \(\sigma_y\),
  which is considered as proof for this protocol. First, the server
  converts the function \(f\) into a Turing-complete sequence of HMAC
  operations, e.g.~\(f_+\) or \(f_\times\) for the polynomial
  construction. \[f \implies (f_+, f_\times, ...)\] These operations are
  applied in sequence to \(\sigma_x\), with a resulting Authenticator
  polynomial called \(\sigma_y\):
  \[\sigma_y \gets (f_+, f_\times, ...) (\sigma_x)\]
\item
  \textbf{Verifying the proof's validity}: The Verifier uses the
  protocol's verification function, this is a crucial step in the
  construction of the protocol which establishes our soundness property.
  To verify whether a given \(\sigma_x\) is a valid Authenticator
  polynomial, the client needs to check whether evaluation on the secret
  key \(sk\) yields a value consistent with the input label(s):
  \[\sigma_x(sk) \overset{?}= r\] \emph{(we will discuss why in detail
  later.)} Which means that any homomorphic derivate of \(\sigma_x\)
  will necessarily yield an equivalent homomorphic derivate of \(r\)
  when evaluated on \(sk\): \[\sigma_y(sk) \overset{?}= f( r )\]
\end{enumerate}

For compatibility's sake, and to help newcomers to the field follow the
original papers better, let us also display the \textbf{domain-specific
syntax} for the Homomorphic Authenticators domain\footnote{\(ek\) is a
  protocol-abstracted evaluation key, but it is not typically present in
  most constructions and it can be representative of the public scheme
  parameters; \(\sigma_i\) is computed for each input message \(m_i\).}:
\[(sk, ek) \gets Keygen(\lambda)
\\\sigma_i \gets Auth(sk, m_i, L_i)
\\\sigma \gets Eval(ek, f, \sigma_i ...)
\\\{0,1\} \gets Ver(sk, f, L_i ..., \sigma, m_i ...)\]

In the next subsections, we will see how to enhance this simple protocol
to include multiple VC features.

\hypertarget{sec:hauthcomplete}{%
\subsection{Adding Completeness}\label{sec:hauthcomplete}}

As described in the previous section, we seek full homomorphism in order
to achieve completeness:

\begin{description}
\tightlist
\item[Homomorphism]
a Signature/MAC scheme \(Sig\) is operator \(\odot\) homomorphic
\[\iff \exists \textit{ operator }\otimes: y=Sig(x) \land y'=Sig(x') \implies y \otimes y' = Sig(x \odot x')\]
\item[Full Homomorphism]
a signature/MAC scheme \(Sig\) is fully homomorphic
\[\iff additively\ homomorphic \land multiplicatively\ homomorphic
\\\iff \exists \oplus \textit{ operator }, \odot \textit{ operator}: y = Sig(x) \land y'=Sig(x') \\\implies y \odot y' = Sig(x \cdot x') \land y \oplus y' = Sig(x + x')\]
\end{description}

Finding a fully homomorphic scheme is essential to achieve
\emph{universality}, so the researchers found a mathematical object
(polynomials) which supported both additive and multiplicative
composition, and then built an authentication scheme on top of it. Given
\(c \in \mathbb{Z}\) and polynomials \(p\) and \(q\) such that
\[\begin{cases}
p = (p_0, p_1, ...) \in F[x]\ \ \big(\textit{ with }n = degree(p),\ |p|=n+1\big)  \iff p(x) = \sum_{i=0}^n p_i \cdot
x^i
\\q \in F[x]\ \big(\textit{with }m = degree(q)\big)
\end{cases}\], the following additive and multiplicative polynomial
operators are given: \[\begin{aligned}
p+q &\overset{def}= \forall_{i=0}^{max(n,m)}{p_i + q_i}
\\p+c &\overset{def}= (p_0 + c,\ \ \forall_{i=1}^{n}{p_i})
\\p \times q &\overset{def}= \forall_{i=0}^{n+m}{\sum_{j=0}^{i}{p_j \cdot q_{i-j}}}
\\p \times c &\overset{def}= \forall_{i=0}^n  pi \cdot c
\end{aligned}\] Of course, these two operations are only a minor part
out of all those which have been defined by mathematicians in the coming
ages; however, it is a widely known Computer Science fact that these two
operations suffice to describe any boolean circuit, and thus fully
homomorphic signature schemes are Turing-complete.

Now that we've defined the basic building blocks for our protocol, let's
show that these two operations hold for any polynomial evaluation:

\begin{description}
\tightlist
\item[Completeness]
\[p(x) + q(x) = \sum_{i=0}^{n} {p_i x^i} + \sum_{i=0}^{m}{q_i x^i} = \sum_{0}^{min(n,m)}{p_i x^i} + \sum_{min(n,m)+1}^{max(n,m)}{p_i  x^i} + \sum_0^{min}{q_i  x^i} + \sum_{min+1}^{max}{q_i  x^i}
\\=\sum_0^{min}{p_i x^i + q_i x^i} + \sum_{min+1}^{max}{p_i x^i + q_i x^i} = \sum_0^{max} {p_i x^i + q_i x^i} = \sum_0^{max}{(p_i+q_i) x^i} = (p+q)(x)\]
and
\[p(x) \cdot q(x) = \sum_0^n {p_i x^i} + \sum_0^m {q_i x^i} = \sum_0^n {\sum_0^m {p_i x^i \cdot q_j x^j}} = \sum_0^n {\sum_0^m {p_i q_i x^{i+j}}}
\\=… = \sum_0^{n+m} \Big({x^i \cdot \sum_{0}^i {p_j \cdot q_{i-j}}}\Big) = (pq)(x)\]
\end{description}

\hypertarget{sec:hauthsound}{%
\subsection{Adding Soundness}\label{sec:hauthsound}}

Our interest here lies in two considerations:

\begin{enumerate}
\def\labelenumi{\arabic{enumi}.}
\tightlist
\item
  building a MAC out of polynomials
\item
  making sure that we can take advantage of the operations explained in
  the previous section to achieve a fully homomorphic MAC
\end{enumerate}

Since preserving the full homomorphism is important, we can start from
step (2) and build our way towards step (1). Let's consider the
following relationships on polynomials: \[\begin{aligned}
p(x) + q(x) &= (p+q)(x)\\
p(x) \cdot q(x) &= (pq)(x)
\end{aligned}
\ \iff\
\begin{aligned}
Eval(x, p) + Eval(x, q) &= Eval(x, p+q)\\
Eval(x, p) \cdot Eval(x, q) &= Eval(x, pq)
\end{aligned}\] If we tried to represent this as a more traditional
Encryption/Signature scheme, it might look like this:

\[Sig(sk, p) + Sig(sk, q) = Sig(sk, p+q)\\
Sig(sk, p) \cdot Sig(sk, q) = Sig(sk, pq)\]

We can, therefore, understand that we should use the secret key instead
of the x-coordinate, and the homomorphisms should still hold. Thanks to
the Completeness properties achieved above, operating on a polynomial
has the effect of operating on all its points at the same time; which
means that interpolating two polynomials on the same x-coordinates
allows us to combine them to operate on their y-coordinates:
\[\begin{aligned}
p = Interpolate((0, m_1) (1, m_2) (2, m_3))\\
q = Interpolate((0, m_4) (1, m_5) (2, m_6))\\
(p+q)(0) = m_1 + m_4\\
(p+q)(1) = m_2 + m_5\\
(p \cdot q)(2) = m_3 \cdot m_6
\end{aligned}
\implies
\begin{aligned}
\sigma_p = Interpolate((sk, m_1))\\
\sigma_q = Interpolate((sk, m_2))\\
(\sigma_p+\sigma_q)(sk) = \sigma_p(sk) + \sigma_q(sk) = m_1 + m_2\\
(\sigma_p \cdot \sigma_q)(sk) = \sigma_p(sk) \cdot \sigma_q(sk) = m_1 \cdot m_2
\end{aligned}\] The polynomial \(\sigma_p\) is already very close to a
homomorphic MAC (\(sk\) is the secret key, and \(m_1\) is the message
being signed), but we mustn't disclose the \(sk\) x-coordinate to
anyone. There are two problem:

\begin{enumerate}
\def\labelenumi{\arabic{enumi}.}
\tightlist
\item
  if we disclose the message being signed, something usually allowed by
  signature and MAC schemes, then someone could figure out our secret
  key \(sk\).\footnote{as long as the degree of the polynomial is higher
    than 0, which we will see is a useful thing to have}
\item
  to protect against oracle attacks we should add randomness to the
  scheme (as well as prove that it is secure against Random Forgery).
\end{enumerate}

To address both problems at the same time, we will move the message
\(m\) to a known x-coordinate, while using a random value for our \(sk\)
x-coordinate: \[\sigma = Interpolate((0, m), (sk, r))\] It is paramount
to avoid disclosing \(r\), so it must always be stored privately by the
signer and associated with the message being signed. Since this is often
an inconvenient constraint for the user of MAC, the user instead chooses
a unique label \(L\) associated with the signature on \(m\) at that
specific point in time (e.g.~``m \textbar\textbar{} time''), and
randomises it using a keyed one-way PRF (e.g.~a cryptographic PRNG or a
keyed cryptographic hash function). The following is the \textbf{HMAC
construction} for signing (m, L) using private keys (sk, K):
\[r = PRF_K(L)\\
\sigma = Interpolate((0,m), (sk, r))\] This construction is also shown
in {[}\protect\hyperlink{ref-Gennaro12}{19}{]} to be secure
(\emph{sound}) against Random Forgery attacks, as long as the label L is
never re-used.

Please note that, since our signatures are still just polynomials, our
completeness property from the previous section still holds:
\[{\left.\begin{aligned}
\sigma_1 \gets Interpolate((0, m_1), (sk, r_1)) \land r_1 \gets PRF_K(L_1)\\
\sigma_2 \gets Interpolate((0, m_2), (sk, r_2)) \land r_2 \gets PRF_K(L_2)
\end{aligned}\right\rbrace}
\\\implies
\begin{aligned}
(\sigma_1 + \sigma_2)(0) = \sigma_1(0) + \sigma_2(0) = m_1 + m_2 \\
(\sigma_1 \cdot \sigma_2)(sk) = \sigma_1(sk) + \sigma_2(sk) = r_1 \cdot r_2
\end{aligned}\]

\hypertarget{sec:hauthmulti}{%
\subsection{Adding Multiple Clients}\label{sec:hauthmulti}}

In order to support multiple clients, we will have to change both the
homomorphism and the MAC constructions. For the new homomorphism, we
will take advantage of another property about polynomials: they can be
multi-variate. In fact, a polynomial \(p(x)\) can support
full-homomorphism just as much as \(p(x,y)\) can. This is intuitive if
you remap \(x\) as a composition between two other variables. Given
\(c \in \mathbb{Z}\) and univeriate polynomials \(p\) and \(q\)
\[\begin{cases}
p = (p_0, p_1, ...) \in F[x]\ \ \big(\textit{ with }n = degree(p),\ |p|=n+1\big)  \iff p(x) = \sum_{i=0}^n p_i \cdot
x^i
\\q \in F[y]\ \big(\textit{with }m = degree(q)\big)
\end{cases}\], the following additive and multiplicative polynomial
operators are given: \[\begin{aligned}
p+q &\overset{def}= \forall_{i=0}^{max(n,m)}{p_i + q_i}\ \textit{(for missing values, $p_i=0$ and $q_i=0$)}, 
\\ &\textit{with }max(m,n)=\textit{degree}(p+q),\ |p+q| = 2max(m,n))
\\p+c &\overset{def}= \textit{same as univariate homomorphism}
\\p \times q &\overset{def}= \forall_{i=0}^{n+m} \forall_{j=0}^{i} p_j \cdot q_{i-j}\ \textit{(coefficients for $x^j y^{I-j}$)}, 
\\ &\textit{with }m+n = degree(p \times q), |p \times q| = |m+n|^2
\\p \times c &\overset{def}= \textit{same as univariate homomorphism}
\end{aligned}\] \emph{Note: the size of the multiplicative homomorphism
result can be further compressed down to
\(|p \times q| = \sum_0^{m+n} i + 1 = m^2 + n^2\), and even more using
techniques described in {[}\protect\hyperlink{ref-Fiore16}{20}{]}}

Now that we have modified our polynomials (while still retaining
completeness) we can construct a Multi-Key Fully-Homomorphic MAC out of
different separate keys: \[\sigma_p = Interpolate_X((sk_1, m_1))
\\\sigma_q = Interpolate_Y((sk_2, m_2))
\\(\sigma_p + \sigma_q)(sk_1, sk_2) = \sigma_p(sk_1) + \sigma_q(sk_2) = m_1 + m_2
\\(\sigma_p \sigma_q)(sk_1, sk_2) = \sigma_p(sk_1) * \sigma_q(sk_2) = m_1 \cdot  m_2\]
Clearly both keys are required for the final evaluation step, hence,
verification of any signature requires that the Verifiers share their
secret keys, or perform a MPC computation; Fiore et
al.~{[}\protect\hyperlink{ref-Fiore16}{20}{]} take the simpler approach,
and have the parties share all the secrets. Because of this, scheme is
actually a MAC and not a digital signature, just like the previous
construction. If we group the keys like \(sk = (sk_1, sk_2)\), we can
perform the evaluation on the keys exactly like the main protocol syntax
requires.

Of course, we should harden our primitive MAC using the same
randomisation process as before, revealing only the signed message for
the \(0\) x-coordinate: \[r = PRF_{K_i}(L)
\\\sigma_i = Interpolate((0, m), (sk_i, r))\] If we wish to adjust the
syntax to the final step of the protocol, it'll look like this:
\[sk = (sk_0, sk_1, ..., sk_\textit{last party})\ \textit{(all participants)}
\\r = (r_0, r_1, ..., r_\textit{last message})\ \textit{(all messages)}
\\\sigma_y(sk) \overset{?}= f(r)\] This construction is also shown to be
sound in {[}\protect\hyperlink{ref-Fiore16}{20}{]}, in a way that is
similar to the previous one.

\hypertarget{sec:hauthscalable}{%
\subsection{Adding Verifier Scalability}\label{sec:hauthscalable}}

The scheme obtained so far has a lot of nice properties, such as
outsourced proving and support for large inputs, but it imposes a big
toll on the Verifier: the client must compute the function on an
alternative set of inputs (the labels) each time he wishes to validate a
computation. In short, the scheme is not \emph{verifier scalable}. In
this step, we will essentially change the construction of our Hom.MAC
into {[}\protect\hyperlink{ref-Fiore13}{21}{]}, incorporating
polynomials into additive groups. Before we do that, however, let's
consider what we're going to need: amortisation.

\hypertarget{what-is-amortisation}{%
\subsubsection{What is Amortisation?}\label{what-is-amortisation}}

The final verification step, essentially, requires receiving the
evaluated Authenticator \(\sigma_y\) from the Prover, and then checking
it against a constant \(f(r)\) evaluated by the Verifier. Wouldn't it be
nice to re-use \(f(r)\) for multiple executions of the protocol?

Unfortunately, security assumptions from
{[}\protect\hyperlink{ref-Gennaro12}{19}{]} for the soundness of our
basic HMAC require that \(r\) always be randomly chosen, even for
multiple signatures on the same message --- therefore, \(L\) needs to be
randomly chosen as well. What we can do is split \(L\) into a changing
part \(\Delta\), and a constant part \(l\): \(L = (l, \Delta)\); this is
also called a ``Multi-Label'' by the authors. These multi-labels might
look a little like this:

\begin{itemize}
\tightlist
\item
  \(L = (\textit{”message m”}, \textit{”at time 12:54”})\), so that
  \(f\) can be computed on messages of the same nature (i.e.~index in a
  database), but changing over time; or
\item
  \(L = (\textit{“message m at time 8am”}, \textit{”on day 08/12/2019”})\),
  so that \(f\) can be computed on the same set of messages (i.e.~a
  single row indexed in a database), but changing over dates.
\end{itemize}

While \(f(r) = f(PRF_K(L))\) will still change across multiple execution
runs, we might find a way to precompute \(C=f(PRF_K(l))\), and then
efficiently add the component \(\Delta\) later on:
\[f(r) = Load(C, \Delta)\] Assuming the function \(Load\) has an
exponentially lower complexity than \(f\), the check should also be
\emph{verifier scalable}.

In order to actually build the \(Load\) function, we'll have to somehow
pull the \(\Delta\) out of \(f\):
\[f(r) = f(PRF_K(L)) = f(PRF_K((l, \Delta))) \implies \exists f’:f’(PRF_K(l), \Delta) = Load(f(PRF_K(l), \Delta)\]
This act of ``pulling out'' a value is exactly what full homomorphism
allows us to achieve for a function \(g\):
\[\exists g’: E(g(x, y)) = g'(E(x), E(y))\] Unfortunately, while \(f\)
may operate on polynomials, \(L\) is not one. In fact, even \(PRF\)
operates on specific values (you may think of them as numbers, but a
string is also valid input to a hash function), and returns a value as
well. In order to ``pull \(\Delta\) out'', we will perform two tricks:

\begin{enumerate}
\def\labelenumi{\arabic{enumi}.}
\tightlist
\item
  transform \(l\) into a 1st degree polynomial, whose variable
  represents \(\Delta\)
\item
  convert \(PRF\) into its equivalent sequence of operators \(PRF'\) for
  the homomorphic signature scheme
\end{enumerate}

We can then evaluate this polynomial on \(\Delta\):
\[Load(f(PRF_K(l), \Delta) \overset{def}= f(PRF’_K(l))(\Delta)\] While
this approach certainly works, \(PRF'\) would probably be cumbersome to
evaluate on polynomials, especially when \(PRF\) is actually a keyed
hash-function such as Keccak256{[}\protect\hyperlink{ref-keccak}{23}{]}.
Instead, the researchers came up with a more efficient construction,
which manages to first evaluate the \(PRF\) on simple values, and then
convert it into a polynomial:

\begin{enumerate}
\def\labelenumi{\arabic{enumi}.}
\tightlist
\item
  \[r_1 = PRF1_K(l)
      \\r_2 = PRF2_K(\Delta)
      \\PRF_K((l, \Delta)) \overset{def}= r_1 \oplus r_2\]\footnote{the
    actual operation to merge \(PRF1\) and \(PRF2\) is not really a XOR,
    but another trickery defined on top of additive groups. The cost for
    \(PRF’\) is \(O_{f\ amortised}(|r_1|)\), so \(O(1)\) for the 2nd
    degree restriction that was added by the authors, as we will see
    later.} \(PRF1\) and \(PRF2\) are defined similarly to the original
  PRF
\item
  transform \(r_1\) into a 1st degree polynomial whose variable
  represents \(\Delta\)
\item
  convert \(PRF\) into its equivalent sequence of operators \(PRF’\) for
  the homomorphic signature scheme
\item
  The new check becomes (after amortisation):
  \[r_1 \overset{\textit{amortised}}= PRF1_K(l)
  \\r_2 = PRF2_K(\Delta)
  \\f(r) = Load(r_1, r_2) = PRF’(r_1, r_2)
  \\\sigma_y(sk) \overset{?}= f(r)\]
\end{enumerate}

The construction provided by the authors for \(PRF'\) requires the use
of additive groups, therefore we will adapt the rest of our homomorphism
construction to this requirement.

\hypertarget{amortized-completeness}{%
\subsubsection{Amortized Completeness}\label{amortized-completeness}}

Now that we have obtained amortisation, we just need to move our
previous HMAC, based on polynomials, to an additive group
\(\mathbb{G}\):
\[\sigma \overset{def}= Interpolate((0, m) (sk, r)) = p = (p_0, p_1) = (m,  (r - m)/sk)
\\\iff \sigma_\mathbb{G} \overset{def}= Interpolate_\mathbb{G}((0, m) (sk, r)) = p_\mathbb{G} = (g^{p_0}, g^{p_1}) = (g^m, g^{(r-m)/sk})\]
As can be seen, we just simply move all the polynomial coefficients into
the group generator's exponent. All polynomial homomorphisms only need
to work on the exponents; given \(c \in \mathbb{Z}\) and polynomials
\(p\) and \(q\): \[\begin{cases}
p_\mathbb{G} = (g^{p_0}, g^{p_1}, ...) \in G[x]\ \ \big(\textit{with }n=deg(p_\mathbb{G}),\ |p_\mathbb{G}| = n + 1\big)
\\q_\mathbb{G} \in G[x]\ \big(\textit{with }m= deg(q_\mathbb{G})\big)
\end{cases}\] the following additive and multiplicative group-polynomial
operators are given: \[\begin{aligned}
p+q &\overset{def}= \forall_{i=0}^{max(n,m)}{p_i + q_i} 
\\p+c &\overset{def}= (p_0 + c,\ \ \forall_{i=1}^{n}{p_i}) 
\\p \cdot q &\overset{def}= \forall_{i=0}^{n+m} {\sum_{j=0}^{i}{p_j \cdot q_{i-j}}} 
\\p \times c &\overset{def}= \forall_{i=0}^n  pi \cdot c 
\end{aligned}\implies
\begin{aligned}
p_\mathbb{G} + q_\mathbb{G} &\overset{def}= \forall_{i=0}^{max(n,m)} g^{p_i+q_i} = \forall_{i=0}^{max(n,m)} {(g^{p_i} \cdot g^{q_i})} 
\\p_\mathbb{G}+c &\overset{def}= (g^{p_0+c}, \forall_{i=1}^{n} g^{p_i}) = ((g^{p_o})^c, \forall_{i=1}^{n} g^{p_i})
\\p_\mathbb{G} \cdot q_\mathbb{G} &\overset{def}= \forall_{i=0}^{n+m} \sum_{j=0}^{i} g^{p_j \cdot q_{i-j}} = \forall_{i=0}^{n+m} \sum_{j=0}^{i} (g^{p_j})^{q_{i-j}} 
\\&= \forall_{i=0}^{n+m} \sum_{j=0}^{i} (g^{p_i})^{dlog(g^{q_{i-j}})}
\\p_\mathbb{G} \times c &\overset{def}= \forall_{i=0}^n g^{p_i \cdot c} = \forall_{i=0}^n (g^{p_i})^c
\end{aligned}\] Completeness is straightforward and leverages the same
concepts mentioned previously. Evaluation is also pretty simple:
\[p_\mathbb{G} = (g^{p_0}, g^{p_1}, g^{p_2}) \in G[x]
\\p_\mathbb{G}(x) = g^{p_0 + p_1 x + p_2 x^2} = g^{\sum_0^n {p_i x^i}} = \prod_0^n g^{p_i \cdot x^i} = \prod_0^n (g^{p_i})^{x^i} = g^{p_0} \cdot (g^{p_1})^x \cdot (g^{p_2})^{x^2}\]
As can be noticed, the multiplicative homomorphism requires that we use
\(dlog\) to compute the multiplication between any two elements of
\(\mathbb{G}\). However, for security purposes, the authors decided to
integrate our polynomial-based fully-homomorphic scheme into groups
where the Discrete Logarithm Problem would hold. The alternative is to
apply a bilinear mapping in order to simulate (and obtain) up to one
multiplicatively homomorphic operation:
\[e: \mathbb{G} \times \mathbb{G} \to \mathbb{G}_T,\ e(g^a, g^b) = e(g,g)^{ab} = g_t^{ab},\ g_t = e(g,g),
\\\langle g \rangle = \mathbb{G},\ \langle g_t \rangle = \mathbb{G}_T\]

In particular, two choices were made:

\begin{enumerate}
\def\labelenumi{\arabic{enumi}.}
\tightlist
\item
  Use an additive group with just one bilinear mapping. This effectively
  limits \(f\) to only functions of 2nd degree, thus also eliminating
  the scheme's previous \emph{universality} claim.
\item
  There are a couple of small changes to the group-polynomials'
  definition when applied to our HMAC scheme:

  \begin{itemize}
  \item
    \(r\) is actually calculated a little differently, as its \(dlog\)
    (calculated by the client) is used instead; it should still hold as
    valid entropy source, see the paper for more
    details{[}\protect\hyperlink{ref-Fiore13}{21}{]}.
  \item
    the very fist coefficient of any polynomial \(p_\mathbb{G}\)
    (i.e.~\(g^{p_0}\)), is actually set to \(p_0\). This makes
    multiplication for two polynomials of first degree a little more
    efficient, because \[p_\mathbb{G} \overset{def}= (g^{p_0},\ g^{p_1})
      \\q_\mathbb{G} \overset{def}= (g^{q_0},\ g^{q_1})
      \\p_\mathbb{G} \times q_\mathbb{G} = (g^{p_0 q_0},\ g^{p_1 q_0 + q_1 p_0},\ g^{p_1 q_1}) = ((g^{p_0})^{q_0},\ (g^{p_1})^{q_0}  \cdot (g^{q_1})^{p_0},\ (g^{p_1})^{q_1}) 
      \\\qquad\qquad= ((g^{p_0})^{dlog(g^{q_0})},\ (g^{p_1})^{dlog(g^{q_0})} \cdot (g^{q_1})^{dloq(g^{q_0})},\ (g^{p_1})^{dlog(g^{q_1})}) 
      \\\qquad\qquad= (e(g^{p_0},\ g^{q_0}),\ e(g^{p_1},\ g^{q_0}) \cdot e(g^{q_1},\ g^{q_0}),\ e(g^{p_1},\ g^{q_1}))\]
    becomes
    \[p_\mathbb{G} \times q_\mathbb{G} = (p_0 q_0,\ g^{p_1 q_0 + q_1 p_0},\ g^{p_1 q_1}) = ((p_0 q_0,\ (g^{p_1})^{q_0} \cdot (g^{q_1})^{p_0},\ (g^{p_1})^{q_1}) 
      \\\qquad\qquad= ((p_0 q_0,\ (g^{p_1})^{q_0} \cdot (g^{q_1})^{p_0},\ (g^{p_1})^{dlog(g^{q_1})}) 
      \\\qquad\qquad= ((p_0 q_0,\ (g^{p_1})^{q_0} \cdot (g^{q_1})^{p_0},\ e(g^{p_1},\ g^{q_1}))\]
  \end{itemize}
\end{enumerate}

\hypertarget{amortized-soundness-and-scalability}{%
\subsubsection{Amortized Soundness and
Scalability}\label{amortized-soundness-and-scalability}}

The construction of the HMAC follows the same idea as in the previous
ones, so I will be brief. \[\begin{aligned}
r &= PRF_K(L)\ \quad\textit{($L$ is a full multi-label)}
\\\sigma_x  &= Interpolate_\mathbb{G}((0, m) (sk, r))
\end{aligned}\] Then, \(f(x)\) gets evaluated by mapping \(f\) to its
counterpart \(f’\) using the group homomorphisms:
\(\sigma_y = f'(\sigma_x)\). And, finally, the check is the same because
it leverages polynomial evaluation within the additive
group-polynomials: \[\sigma_y(sk) \overset{?}= f(r)\]

\begin{description}
\tightlist
\item[Scalable Verifier]
\emph{in the multi-client or the multi-message construction, the idea is
that all \(L_i\) have the same \(\Delta\), the \(Load\) function only
takes 1 value, so its complexity is \(O_V(1)\)}
\end{description}

\newpage

\hypertarget{verifiable-delay-functions}{%
\section{Verifiable Delay Functions}\label{verifiable-delay-functions}}

Verifiable Delay Functions (VDFs) are currently a very active research
area in the cryptocurrency community, but they have actually been around
for a long time, with a formal definition given only in 2018 by
{[}\protect\hyperlink{ref-BBBF18}{24}{]}. Until recently, researchers
had been toying with many different constructions, trying to find
adequate ``time-lock puzzles''. In 1996 Rivest et
al.~{[}\protect\hyperlink{ref-RSW96}{25}{]} introduced a mathematical
problem which seemed to exhibit interesting properties, with relation to
time delaying functionality, previously only briefly considered in naïve
PoW-like schemes by researchers such as Merkle
{[}\protect\hyperlink{ref-Merkle78}{26}{]}.

The main objective is to come up with a cryptographic proof of elapsed
time, i.e.~a delay. Researchers figured that a universal reference for
measuring the passage of time could be represented by the maximum speed
at which a single operation can be processed on a circuit (of any kind),
so they set out to find ``sequential functions'' -- i.e.~which could
only be computed on a single cpu core one operation at a time. This idea
can be seen as a PoSW, ``Proof of Sequential Work''; we will discuss
later the implications for this construction.

Once such a ``time-lock puzzle'' (or PoSW) was found, the need emerged
for an \emph{efficient verification} mechanism, to relieve the Verifier
from the burden of wasting the same amount of time as the Prover just to
check that he did indeed compute the right result. This would allow for
efficient outsourcing of elapsed time, which may sound like a useless
tool, but it can lead to surprisingly innovative solutions in the
time-agnostic world of computer science. Attempts to find such
\emph{verifier scalable} PoSWs lasted for years, with some improved but
incomplete results in 2015 {[}\protect\hyperlink{ref-LW15}{27}{]} and
2018 {[}\protect\hyperlink{ref-BBBF18}{24}{]}, culminating with two
complete solutions that same year by Wesolowski
{[}\protect\hyperlink{ref-Wes18}{28}{]} and Pietrzak
{[}\protect\hyperlink{ref-Piet18}{29}{]}.

The recent rush of new research in this field is probably due to the
increased popularity of Blockchain technology (see use-cases in Section
\ref{sec:vcusecases}), and a formal definition for these systems was
finally given by {[}\protect\hyperlink{ref-BBBF18}{24}{]} under the new
name ``Verifiable Delay Functions''. We will be mainly considering
Wesolowski's scheme in this section, with some references to Pietrzak's.
A good comparison of the two schemes is also provided in
{[}\protect\hyperlink{ref-BBF18}{30}{]}.

\hypertarget{utility}{%
\subsection{Utility}\label{utility}}

The issue of time synchronisation has long plagued electronic computers
which interact on the Internet. To solve synchronisation between honest
parties, a hardware clock (or constant delay networks) might suffice,
but malicious parties would still be able to report incorrect
timestamps. Most importantly, the issue of time synchronisation also
extends to time delay proving. There are two main ways to detect such
malicious attempts:

\begin{enumerate}
\def\labelenumi{\arabic{enumi}.}
\item
  \emph{Distributed consensus mechanism.}

  This idea basically adapts the concept of a trusted third party to a
  scenario where no such party exists (or at least it is not recognised
  as such by all honest parties).Trust is distributed amongst all the
  parties (according to some satisfactory proportion or relation), and
  the validity of a claim is based on whether it is the most supported
  one by the network.

  In order for this system to work there needs to be a majority of
  parties incentivised to act honestly, which is commonly achieved by
  distributing trust amongst a large number of independently motivated
  parties, all interested in using the same protocol (e.g.~pseudonymous
  Bitcoin users participating from all around the world). Also, the
  network itself needs to always be available to all parties
  (i.e.~censorship resistant), otherwise honest parties might be unable
  to stave off false claims by supporting only the correct ones.
\item
  \emph{Use a universal time delay measurement reference.}

  This would be some sort of event occurring in our world which can be
  universally verified just based on the laws of physics. The Prover
  would perform some sort of action or operation over a period of time,
  and it would automatically reflect on some object in the universe in
  such a way that it would be infeasible to replicate the same exact
  object without the same period of time having elapsed.
\end{enumerate}

Of course, these two systems can be combined into a single solution --
Bitcoin makes both use of the PoW system as well as the distributed
consensus model -- VDFs, instead, take the second approach and try to
find a trustless and convenient solution to measure the passage of time.

A commonly proposed universal time delay source is ``maximum computation
speed'', as in the fastest way that a specific computation can be
performed in any implementation of any computational model in the world.
This is deemed as ``universal'' because, if \emph{no participant in the
world} can perform a certain computation faster than the expected amount
of time, it can be used as a universal time delay reference for humans.

\hypertarget{sec:vcusecases}{%
\subsection{Use-Cases}\label{sec:vcusecases}}

VDF schemes are particularly interesting because reliable and efficient
time delay outsourcing leads to innovative computer science
applications. The original applications were of cryptographic value:

\begin{itemize}
\tightlist
\item
  \textbf{timed encryption}: also known as ``time capsules'', they would
  allow for self-decrypting messages through a ``timed key escrow''
  ({[}\protect\hyperlink{ref-timedkeyescrow1}{31}{]},
  {[}\protect\hyperlink{ref-timedkeyescrow2}{32}{]}) mechanism, where a
  Trapdoor-VDF would reveal the key after some elapsed time. Timed
  encryption can be leveraged to build \textbf{scheduled payments}: one
  could prepare multiple transactions in advance, and they would
  self-decrypt in due time. At any moment prior to the deadline, the
  owner can invalidate the payments. Timed encryption can be used for
  many scenarios, such as \textbf{timed top secret archives}, in order
  to guarantee security and transparency for a country's intelligence
  services.
\item
  \textbf{timed commitments}: using timed encryption as a building
  block, one can build self-revealing commitment
  schemes{[}\protect\hyperlink{ref-timedcommitments}{33}{]}, which can
  be used for lots of protocols, including \textbf{auction bidding}:
  everyone commits during the first phase, and in the second phase the
  bids self-reveal. Timed commitments can also be used for other
  \textbf{voting protocols}, where the vote is revealed after the voting
  has taken place.
\item
  \textbf{slow-timed hash functions}: delay functions are interesting
  alternatives to classic iterated hashing techniques and Key Derivation
  Functions, with the advantage of being \emph{scalable} and
  \emph{sequential}. They can be used for \textbf{password storage},
  when the password are generated by humans, in order to stave off
  brute-force pre-image attacks. Compared to classic techniques (such as
  \emph{scrypt}), VDFs do not leverage memory, instead relying on
  sequentiality. Initial slow-timed hash function constructions were the
  precursors to what eventually became ``Verifiable Delay Functions'' in
  {[}\protect\hyperlink{ref-BBBF18}{24}{]}.
\end{itemize}

In practice, the ability for VDFs to generate public random numbers
(when given a biased entropy source) can be the basis for achieving
Transparency in many other protocols, such as a lottery. Over the last
few years, there has also been increased interest in adapting VDFs to
cryptocurrencies, where the lack of a trusted third party is a common
assumption:

\begin{itemize}
\item
  \textbf{transparent public PRNG beacon}: The main properties of random
  numbers is that they're both \emph{unpredictable} and their generation
  is \emph{unbiased}. Classic solutions to generating public random
  numbers on the blockchain have been to either: take block hashes, or
  use MPC computations {[}\protect\hyperlink{ref-BGB17}{34}{]}. The
  problem is that repeating MPC computations is highly inconvenient (all
  parties must be online at the same time and perform hefty
  computations), and that block hashes are subject to the biased
  selection of transactions by PoW miners.

  Using VDFs as ``slow-timed hash functions'', we can generate random
  numbers on the blockchain which remain secret for a short period of
  time. The main idea is that we can use transaction history as an
  (unpredictable) entropy source\footnote{the entropy for a Bitcoin
    block hash (approx 10 minutes of transaction time) was estimated to
    be \(\approx 70\) bits in 2015 by
    {[}\protect\hyperlink{ref-BCG15}{35}{]}, and it is based directly on
    the difficulty of the mining problem. On Ethereum, blocks are
    published \(\approx 40\) times more frequently (i.e.~around \(15\)
    seconds per transaction) {[}\protect\hyperlink{ref-BGB17}{34}{]},
    which entails lower entropy for each block hash.}, and then remove
  the bias introduced by miners by using VDFs. The VDF inputs are block
  hashes, and the outputs are our random numbers: the miners can bias
  the inputs only up to the block's confirmation time (not just its
  publication), after which they cannot be changed. If the VDF delay is
  longer than an input block's confirmation time, then the outputs will
  become unbiased because the miners won't be able to evaluate and
  change them at the same time. Of course, it is important to accurately
  measure the maximum block confirmation time for the blockchain at
  hand, after which any miner attack becomes infeasible, and set it to
  be smaller than the VDF delay.
\item
  \textbf{transparent lottery systems}: much like Randomness Beacons,
  lottery systems require the selection of a random number only after
  all relevant actions (i.e.~the betting) have taken place. The trick is
  the same, and players are given less time to bet than it takes to
  figure out the random number, fixed at the beginning of the
  computation by using block hashes. I've implemented a prototype
  trapdoor version of such a protocol myself, on Ethereum
  Kovan{[}\protect\hyperlink{ref-traplottery}{36}{]}. Of course, the
  lottery game could take advantage of a Randomness Beacon and just give
  players \(n\) blocks' time to bet, taking as winning number the
  Beacon's output of the \(n\)-th block's hash after the start of the
  lottery.
\item
  \textbf{improved blockchain efficiency}: arguably one of the biggest
  issues cryptocurrencies have right now is the incredible waste of
  resources used for mining in PoW-based blockchains. There are entire
  mining farms which combined consume as much energy as a small country,
  all for the purpose of making Bitcoin run. The Ethereum2.0 research
  team is experimenting with PoS (Proof of Stake) consensus protocols,
  where a new leader is randomly chosen to publish each block, without
  the need for wasteful mining. VDFs can be used for this purpose
  because their output is deterministic, but can still be used to choose
  leaders in a fair (pseudo-random) fashion.

  Comparing the Nakamoto hash inversion puzzle (used in Bitcoin,
  BitGold, and others) with VDFs, leader selection would be akin to
  fixing a PoW output from the start, and then running many parallel
  processes to brute force the input space. The advantage for PoW
  schemes is that they are \emph{verifier scalable}, because other users
  can quickly check that the correct pre-image was found. However, the
  price to pay for fair currency distribution starting from a given
  biased state (i.e.~the previous block's hash) is a non-deterministic
  search by exhaustion, which results in huge energy consumptions. VDFs
  can remove the same bias present in block hashes, while still being
  \emph{verifier scalable} \emph{\textbf{and deterministic}}. That's
  because we fix the input instead of the output, and then proceed
  sequentially with the computation: only one person needs to calculate
  and publish the proof. This leads to a drastic reduction in resource
  wastage, and is a highly anticipated feature of the Ethereum
  blockchain.\footnote{of course, the consensus protocol also requires
    incentivising users towards a unified blockchain state. In PoW
    consensus protocols, miners are incentivised to mine for the longest
    chain, or they risk wasting time and money; but in PoS consensus
    protocols, leaders can generate multiple chains without wasting any
    time. One of the suggested solutions is to force cheating leaders to
    lose money, but the details are still being fleshed out for
    Ethereum2.0.}
\end{itemize}

Other interesting applications of VDFs have been identified for more
ambitious scenarios, such as Web3 and SWARM-like
{[}\protect\hyperlink{ref-SWARM}{37}{]} solutions. An example is
\textbf{proof of ``age''}, where the minimum age of a given file can be
proven to show that some information was indeed known ahead of
time\footnote{this can be regarded as the opposite of showing that a
  given file is recent, such as what what abductors used to do when
  taking a photo of their captives along with the daily newspaper.}.

\hypertarget{a-language-for-time-delays}{%
\subsection{A Language for Time
(Delays)}\label{a-language-for-time-delays}}

But how to measure time in the time-agnostic world of computers? We
could try to equate cpu cycles and operations to the flow of time,
measuring them with a \(\textit{ real-time}\to\mu\textit{-time }\)
formula. There are a couple of \textbf{issues} with this approach:

\begin{enumerate}
\def\labelenumi{\arabic{enumi}.}
\item
  \(\mu\textit{-time}\neq\textit{ real-time}\)

  Algorithms are not an immediately useful tool for measuring time,
  since time runs by itself and they don't. Users might use our protocol
  for measuring time, but we cannot ask them to run it indefinitely.
  This means that we need to adapt our language to measuring time
  delays, and not time; as long as the algorithm runs, a delay will be
  measured! We won't be able to prove something like ``this message at
  13:54 on 1/1/2019'', but we might be able to prove something like
  ``this message took one week to process''. As long as the message is
  unique, we can also prove ``this message is from more than a week
  ago''.
\item
  \(\mu\textit{-time}\neq\textit{universal}\)

  The flow of time has the nice property of being the same for everyone:
  nobody can speed it up or slow it down! However, this does not apply
  to computers -- anyone with more money can buy more processors, and
  then use them to parallelise and speedup most computations.

  For this reason, we aim to find sequential computations which cannot
  be parallelised, such that money will not be a factor when measuring
  the flow of time; thus making our protocol fair and transparent for
  all users.
\end{enumerate}

Now that we've identified the main issues, let's discuss the
\textbf{main properties} that we want for a statement \(X \in L\):

\begin{enumerate}
\def\labelenumi{\arabic{enumi}.}
\item
  \emph{Sequential}

  In order to build a reliable language for measuring time delays, we
  wish to base it on sequential computations. This is because the time
  spent computing paralelisable algorithms varies wildly according to
  the amount of money invested: a poor individual with a single 10€
  processor will run a Bitcoin mining algorithm orders of magnitude
  slower than a rich company with 1000x times the same amount of
  processors; this makes for unreliable time delay measurement, hence
  unfair VDF protocols.The same does not occur when comparing processor
  frequencies: average processors on the market lie at around 1GhZ
  speeds, while the fastest ones in the world at 9GhZ -- just a factor
  of ten! As long as we account for maximum 10GhZ speeds in our
  sequential computation, our delay measurements should apply to
  everyone: nobody will be able to complete the VDF faster than the
  expected amount of time, although some might take a little longer.

  This solution, however, is not without flaws. An alternative to
  sequentiality, frequently used in Cryptography for KDFs and password
  storage, has been to employ algorithms which require using large
  amounts of memory in order to greatly increase the cost for achieving
  parallelised computation. Two successful examples of this are
  \emph{scrypt}{[}\protect\hyperlink{ref-scrypt}{38}{]}, commonly used
  for password-based key-derivation-functions, and
  \emph{Ethash}{[}\protect\hyperlink{ref-ethash}{39}{]}, used for the
  Ethereum cryptocurrency's PoW. Another issue is with the assumption
  that the difference between the world's average processor speed and
  the world's fastest one is ``small'', and that specialised hardware
  implementations (e.g.~ASICs) cannot improve this margin by a
  substantial amount. These assumptions are currently being researched
  by the Chia Foundation and the Ethereum foundation
  ({[}\protect\hyperlink{ref-Wes18}{28}{]},
  {[}\protect\hyperlink{ref-ethereumvdfmpc}{40}{]}).
\item
  \emph{Deterministic}

  Since we're trying to measure effective time delays, and not average
  ones, our scheme cannot rely on well studied PoW protocols. The issue
  being that they're typically probabilistic (as well as paralelisable):
  a problem with an estimated difficulty of 1 hour might end up taking 1
  second, just out of sheer luck! A deterministic computation would give
  us a guarantee as to the number of performed computations, hence, the
  minimum elapsed time.
\end{enumerate}

\hypertarget{building-a-spow-protocol}{%
\subsection{Building a SPoW Protocol}\label{building-a-spow-protocol}}

The major idea behind the success of current VDFs is the specific
protocol language designed by Rivest et al.~in
{[}\protect\hyperlink{ref-RSW96}{25}{]}, based on repeated squaring in
RSA groups. This time-delay language will become the basis for improved
VDF protocols, here is its definition:

\begin{description}
\tightlist
\item[\(\textit{Time-Lock Puzzle } TL(\Delta, \lambda, \mu)\)]
\[\Big\{ (x,y) \mid y \gets \overbrace{(\mu \circ \mu \circ … \circ \mu)}^\textit{T times} (x), T \gets \Delta \cdot \frac{sec}{\Omega_{\mu_\lambda}},\ \Omega_y(\Delta), \ 
\\T \in \mathbb{Z}, \Delta \in \textit{seconds}, \mu: D \to C, x \in D_\lambda,\ y \in C_\lambda\Big\}\]\footnote{in
  the practical scenarios, \(T\) is typically determined heuristically
  according to the specific implementation, or based on concrete metrics
  of the basic \(\mu\) operation. A typical example provided by most
  researchers in their academic articles is \(T \gets 2^{40}\), however,
  concrete time measurements are not typically discussed.}
\end{description}

Rivest et al.~{[}\protect\hyperlink{ref-RSW96}{25}{]} believed their
language contained intrinsic sequentiality properties, and based their
``time-lock'' protocol on it. Given the difficulty of estimating a
function \((\Delta, \lambda) \to T\)\footnote{since there can be many
  other costs associated with usage of SPoWs (such as network
  transmission), they are not well suited for precise time measurements.
  It's best to choose delays which range from a few minutes to hours or
  days.}, the puzzle was simply based on any \(T\) directly:

\begin{description}
\tightlist
\item[RSW96 \(TL(T, \lambda, \mu)\)]
\[\Big\{ (x, y, N) \mid y \gets x^{2^T} \pmod N,\ \mu = x \mapsto x^2 \pmod N,\ \Omega_y(\Delta),\ 
 \\T \in \mathbb{Z},\  N \in_R RSA_\lambda \textit{ modulus},\ x \in \mathbb{Z}_N^*,\ \lambda_{RSA} \textit{ derived from }\lambda\Big\}\]\footnote{\(\lambda\)
  is the security parameter in bits for the RSA group. From \(\lambda\)
  we typically derive \(\lambda_{RSA}\), according to conventions based
  on statistical brute-force attacks shared by the cryptographic
  community. Today it is believed that
  \(\lambda=100 \implies \lambda_{RSA} = 2048\), but this assumption may
  change in the future, or have changed already.}
\end{description}

Clearly, calculating a power \(2^T\) which is huge (e.g.~\(T=2^{40}\))
is not feasible, so we will not be able to employ classic modular
exponentiation techniques; two known techniques are shown in the
\emph{completeness} proof. Here is the protocol we can derive from the
language:

\begin{description}
\tightlist
\item[\(SPoW\)]
\[\textit{Given }
    \begin{cases}
    L \equiv TL(T, \lambda, \mu) \subseteq \{0,1\}^* \in NP \\
    T \in \mathbb{Z} \textit{ timing parameter}\\
    \lambda \textit{ security parameter in bits}\\
    \mu: \textit{squaring in $RSA_\lambda$}\\
    N \textit{ RSA modulus}
    \end{cases}\] \[\textit{And }X \in L \iff 
    \land\begin{cases}
    \textbf{Complete }\\
    \textbf{Sound }
\end{cases}\]
\end{description}

The protocol is clearly complete, since the repeated \(\mu\) operation
does yield a correct \(y\):

\begin{description}
\tightlist
\item[Completeness]
\[\forall X \in L,\ (x, y, N) = X: Pr[y = x^{2^T} \pmod N] = 1\] with
the algorithm for computing \(y\) being: \[\begin{cases}
x \overbrace{\to x^2 \to x^{2^2} \to x^{2^3} \to … \to}^\textit{$T$ group squarings} x^{2^T} \pmod N \textit{ (order is unknown)}
\\
e \gets 2^T \pmod{\phi(N)} \land  x^e = x^{2^T} \pmod N \textit{ (order is known)}
\end{cases}\]\footnote{in a typical SPoW scenario, the RSA group order
  \(\phi(N)\) is not known to the Prover.}\\
\emph{(i.e.~\(X \in L\) means that \((x,y)\) are correct, \textbf{and}
that \(\Omega(T)\) time was spent. The second algorithm for calculating
\(y\) is particularly important, since it implies knowledge of the RSA
trapdoor and private key).}
\end{description}

The actual soundness for this language's claim to being a universal
time-delay reference (i.e.~sequential and deterministic computation) is
not proven, but it does rely on two assumptions:

\begin{description}
\item[Soundness]
\[\forall X \notin TL(x): Pr[y = x^{2^T}] = \textit{negl}(\lambda)\]
\emph{if cracking RSA is hard \textbf{and} all TL puzzles can only be
solved in minimum T time without \(\phi(N)\) \textbf{and}
\(X \notin TL\) but the puzzle solved means that it took less than T
time, \textbf{then} the puzzle was solved with \(\phi(N)\),
\textbf{then} the prover had to have cracked RSA, which has negligible
probability!} In order words: \[{\left.\begin{aligned}
Pr[\exists \textit{"extractor"}\ E_{POLY}: \phi(N) \gets E(N)] = negl(\lambda) 
\\ \forall x \in \mathbb{Z}_N^*: \Omega_{x^{2^T}\ w/out\ \phi(N)}(T \cdot \mu_\lambda) 
\\ \Big(X \notin TL \land y = x^{2^T} \implies \Omega_{y}(< T \cdot \mu_\lambda)\Big) 
\end{aligned}\right\rbrace}\land\implies
\\\Omega_y = \Omega_{x^{2^T}\ w/\ \phi(N)}(< T \cdot \mu_\lambda) \implies \exists \textit{"extractor"}\ E_{POLY} \iff
\\negl(\lambda) = Pr[\exists E: \phi(N) \gets E(N)] =  Pr[y = x^{2^T} \land w/out\ \phi(N)] = Pr[y = x^{2^T} \land X \notin L]\]

The soundness of the protocol relies on the usage of groups of unknown
order, and the inability to reverse a cryptographic one-way
function.\footnote{we will discuss another groups of unknown order
  \(\mathbb{G}(\sqrt{q})\) in the subsection on transparency Section
  \ref{sec:vdftransparency}.} In particular, two assumptions are
required for this protocol to work:

\begin{enumerate}
\def\labelenumi{\arabic{enumi}.}
\item
  Cracking RSA is hard (i.e.~extracting \(\phi(N)\) from \(N\))

  \emph{This assumption has been upheld by the cryptographic community
  for decades, only being proven invalid within the still developing
  context of Quantum Computers.}
\item
  There is no faster way to solve the puzzle w/out \(\phi(N)\) than with
  \(\Omega(T)\) sequential \(RSA_\lambda\) group squarings
  (i.e.~\(\mu_\lambda\))

  \emph{This was not proven in th original '96 paper by Rivest et
  al.{[}\protect\hyperlink{ref-RSW96}{25}{]}, but it is believed to be
  true by all subsequent authors (inluding
  {[}\protect\hyperlink{ref-LW15}{27}{]},
  {[}\protect\hyperlink{ref-BBBF18}{24}{]},
  {[}\protect\hyperlink{ref-Piet18}{29}{]},
  {[}\protect\hyperlink{ref-Wes18}{28}{]} and others).}
\end{enumerate}
\end{description}

No specific construction for the protocol is provided\footnote{actually,
  {[}\protect\hyperlink{ref-RSW96}{25}{]} only states the language
  \(TL\), and assumes that protocols based on its time-delaying
  sequentiality will be sound. A construction is only provided for a
  \emph{timed encryption} use case.}, but you could think of it as
something similar to Homomorphic Authenticators:
\[client\ Verifier \overset{\sigma_x}\longrightarrow server\ Prover\\
client\ Verifier\ \overset{\sigma_y}\longleftarrow server\ Prover\\
check(\sigma_y) \overset{?}= True
\\\textbf{or}\\
Verifier \overset{TL(T, \lambda, \mu),\ N,\ x}\longrightarrow Prover\\
Verifier\ \overset{y}\longleftarrow Prover\\
 y \overset{?}= x^{2^T}\]

\hypertarget{a-note-on-trapdoor-spows-and-trapdoor-vdfs}{%
\subsubsection{A note on Trapdoor-SPoWs and
Trapdoor-VDFs}\label{a-note-on-trapdoor-spows-and-trapdoor-vdfs}}

\begin{quote}
So, whoever owns the RSA private key also knows \(\phi(N)\), and can
therefore invalidate the protocol and spoof proofs at will. This does
not preclude utility: the Prover may not be given the key anyway, or the
trapdoor may be used to generate ``time-capsules''. In ``time-capsule''
constructions the owner of the private key can take advantage of his
\textbf{Trapdoor-SPoW} to quickly calculate the output of a unique
\(X \in L\), and use it as OTP key to encrypt some secret message: all
others will have to wait before they can decipher his message.
\end{quote}

\begin{quote}
At the same time, using RSA groups necessarily requires us to generate
private keys. If we wish for others to use our SPoW in a trustless
(\emph{transparent}) fashion, what can we do to prove we did not keep
nor use the private keys? We will delve deeper into this topic in
Section \ref{sec:vdftransparency}.
\end{quote}

\hypertarget{sec:cutchoose}{%
\subsection{``Compressing'' Time}\label{sec:cutchoose}}

Now that we've build our SPoW universal time-delay reference, we can
proceed towards refining it. While the protocol does allow us to prove
time delays, it also requires the Verifier to wait the same amount of
time as the Prover (unless he has access to the RSA trapdoor, which is
not a scenario we wish to focus on). This can be a limiting factor for
computer applications, where performance is essential. If an auction
lasting 1 hour takes place between 100 participants we want the bids to
be revealed as soon as the auction ends, but our SPoW protocol requires
each participant to wait 100 hours before they can be sure of the
winning party. Research into \emph{(verifier) scalable} SPoW protocols,
also known as VDF protocols, has recently resulted in two competing
approaches:

\begin{enumerate}
\def\labelenumi{\arabic{enumi}.}
\item
  \emph{Cut-\&-Choose}:

  Wesolowski came up with a proof
  {[}\protect\hyperlink{ref-Wes18}{28}{]} which shares some similarities
  with the Schnorr \(\Sigma\text{-Protocol}\). The idea is to
  ``generate''\footnote{we don't actually generate all the possible
    problems in practice, but only the one that will be needed. For the
    purposes of this discussion, however, it does not make any
    difference.} many problems isomorphic to the original SPoW, and then
  solve the one chosen randomly by the Verifier. Concretely, the
  isomorphic problems are derived from the SPoW output, making sure that
  they still preserve the protocol's witness (i.e.~that there have been
  numerous sequential squarings starting form the input). This way, each
  available isomorphic problem will be made dependent on the original
  SPoW problem, while having the property of being much faster to check.
  When the Verifier randomly selects and validates one of these
  isomorphic problems he can be confident that, as long as the check
  succeeds, the Prover has surely waited the correct amount of time
  calculating the SPoW. The ``Cut \& Choose'' approach was perhaps first
  employed, informally, by Rabin in 1979
  {[}\protect\hyperlink{ref-Rabin79}{41}{]}.
\item
  \emph{Recursive Cut-\&-Choose}:

  Pietrzak came up with another innovative protocol
  {[}\protect\hyperlink{ref-Piet18}{29}{]} that instead shares
  similarities with the famous Graph-Isomorphism ZKIP feasibility
  result, in that soundness is improved over multiple rounds of the
  protocol. The protocol makes use of an intermediate execution trace
  derived from the SPoW computation, where each state is made dependent
  not on the input and output, but on another state that occurred just a
  little earlier within the computation. When the Verifier recursively
  validates these intermediate results, reaching the very first one, the
  proof becomes sound. The protocol is still Cut-\&-Choose, because
  during each round the Verifier can select amongst many problems
  isomorphic to the current state. While this technique requires
  multiple rounds, increasing the verification costs, the prover
  overhead is almost completely gone (unlike in the Wesolowski
  protocol). The use of an ``execution trace'' to prove computations is
  also found in STARKs, analysed in Chapter 3.
\end{enumerate}

We will be discussing only the Wesolowski VDF due to its \emph{verifier
scalability}; however, the second one is just as valid due to its
additional \emph{prover scalability} improvement. The interesting aspect
of Pietrzak's approach is that it is better suited for time-sensitive
scenarios (e.g.~PRNG beacons), where the Prover wants to submit his
result quickly and the Verifier can spare extra storage for longer
proofs. Other interesting but less successful approaches are discussed
in {[}\protect\hyperlink{ref-BBBF18}{24}{]}, and include making use of
SNARKs (see Section \ref{sec:universalconclusion}), as well as inversion
of permutation polynomials, and modular square root constructions.

\hypertarget{building-a-vdf-protocol}{%
\subsection{Building a VDF Protocol}\label{building-a-vdf-protocol}}

To build an efficient protocol we need to find a problem that is
isomorphic to our SPoW, but which is also fast to check, or
\emph{verifier scalable}. Consider the main delaying factor in the SPoW
problem, \(x^{2^T} \pmod N\): the exponent \(2^T\) is way too large to
compute and store (for values such as \(T=2^{40}\)), so we need to use
our sequential repeated squaring method. Likewise, if we choose another
large exponent, the problem will stay the same. Say we decompose our
exponent into \[\exists r \in_R Z_\lambda: 2^T = q \cdot r\] Then, if we
assume \(\lambda \approx 100\), the value \(q\) is still very
large\footnote{the symbol \(q\) is chosen in
  {[}\protect\hyperlink{ref-Wes18}{28}{]} because it the ``quotient''
  for \(2^T / r\).}, and \(x^q \pmod N\) is nearly just as hard to
compute as the original problem. So the new problem is of similar nature
as the old one, but is it isomorphic? Here's the twist, we will use a
different check function from the one in our SPoW: instead of
\[x^{2^T} \overset{?}= y \pmod N\] We'll use
\[(x^q)^r \overset{?}= y \pmod N\] The idea is to have the Verifier
randomly choose \(r\) \emph{after} \(y\) has been submitted by the
Prover, and then wait for \((x^q)\) to be submitted by the Prover. Thus,
the Verifier only needs to perform one efficient calculation, and he can
check whether the two values \(x^q\) and \(y\) are consistent with each
other. This also has the effect of suggesting that a \(T\)-time
time-puzzle was solved, as we will discuss later in the \emph{soundness}
proof, but there are still a few details to be ironed out before we can
be satisfied.

Two more security devices need to be added to this construction. The
first deals with a detail in the \emph{soundness} proof, whereby we
choose a prime random value \(r \in_R PRIME_{2\lambda}\) instead of a
normal integer, which also changes the construction slightly because
\(2^T = q \cdot r + \textit{residue}\). Intuitively, this gives the
Prover more control over the check because he is not only bound to
values that might be unrelated to the SPoW (if the Prover is malicious),
but he can also factor in the input (e.g.~\(x^\textit{residue}\), see
below) of the SPoW to force \(x^q\) to be chosen correctly. For related
reasons, {[}\protect\hyperlink{ref-Wes18}{28}{]} also requires that we
modify our RSA group to be \(RSA_\lambda/\{\pm 1\}\). The second
observation is more of a practical requirement: since inputs for the
SPoW should be unique and randomly chosen across different protocol
executions, it is better to remove the bias of the input \(x\) by
calculating a new input \(x’ \gets H(x)\), with hash function
\(H: \{0,1\}^* \to RSA_\lambda/\{\pm 1\}\).\footnote{the hash function
  can be something like
  Keccak256{[}\protect\hyperlink{ref-keccak}{23}{]}, adapted to provide
  enough \(\lambda_{RSA}\) bits, such as by iterating inputs through a
  counter in a PRNG-like construction.} If this is not performed, any
biased input \(x_2 = x_1^\alpha\) can be exploited to speed-up the
computation by using a previous SPoW output and taking advantage of the
group's commutative properties:
\(x_2^{2^T} = (x_1^\alpha)^{2^T} = (x_1^{2^T})^\alpha = y_1^\alpha \pmod N\).

Finally, a note on efficiency --- is it possible for the Prover to
generate the auxiliary value \(x^q\) faster than using again the
repeated squaring method, such as by taking advantage of the
relationship it has with \(x^{2^T}\)? In fact, it is, and
{[}\protect\hyperlink{ref-Wes18}{28}{]} takes
\(x^q \gets x^{\lfloor\frac{2^T}{r}\rfloor}\) (the flooring is due to
the prime divisor); algorithms are discussed in the \emph{completeness}
proof. Now that we've discussed how to build the protocol, here is the
construction for a \(VDF(x,y,N)\): \[Verifier \xleftarrow{y} Prover\\
Verifier \xrightarrow{r} Prover\\
Verifier \xleftarrow{\pi} Prover\\
\pi^r x’\ ^\textit{residue} \overset{?}= y
\\\textit{where}\begin{cases}
x’ = H(x)\\
y = (x’)^{2^T} \pmod N\\
r \in_R PRIME_{2\lambda}\\
\pi = x’^{\lfloor 2^T/r \rfloor}\\
\ \textit{residue} = 2^T \pmod r\\
\end{cases}\]

\begin{description}
\item[Completeness]
\[\forall (x,y,N) = X \in VDF: Pr[\pi^r x’\ ^\textit{residue} \pmod N = y] = 1\]
This is straightforward when expanding the formula:
\[\pi^r  x'\ ^\textit{residue} = (x'\ ^{\lfloor 2^T / r \rfloor})^r  x'\ ^\textit{residue} = (x'\ ^q)^r  x'\ ^\textit{residue} \\= x'\ ^{qr + \textit{residue}} = x'\ ^{2^T} = y\]
And \(x'^{\ q}\) is calculated using \[\begin{cases}
x'\ ^q = x'\ ^{\frac{2^T -\textit{residue}}{r}} = x'\ ^{\frac{2^T - (2^T \mod r)}{r}}= x'\ ^{\lfloor 2^T / r\rfloor} \gets \mathcal{A}(x', r, T) \ \textit{(order is unknown)}\\
x'\ ^q,\ q \gets (2^T - \textit{residue})r^{-1} \pmod{\phi(N)}\ \textit{(order is known)}
\end{cases}\]

\(\mathcal{A}\) is chosen to be the ``on-the-fly long division
algorithm'', with worst complexity of \(O(2T)\), but an improved
algorithm in {[}\protect\hyperlink{ref-Wes18}{28}{]} reaches
\(O(T/log(T))\), and can also be parallelised.

Just as the SPoW, this protocol can be broken if the order of the group
\(\phi(N)\) is known to the Prover. Trapdoor-VDFs are still useful, as
mentioned before, when the owner of the private RSA key is not the VDF's
Prover.
\item[Soundness]
\[\forall X \notin VDF: Pr[\pi^r x’\ ^\textit{residue} = y] = \textit{negl}(\lambda)\]
\emph{\textbf{if} cracking RSA is hard, \textbf{and} breaking the
Adaptive Prime Roots assumption is hard, \textbf{and} \(X \notin VDF\)
but the check succeeds means that either the Prover spent less than
\(\Omega(T)\) time or he chose the wrong language parameters for
\((y, \pi)\), \textbf{then} one of \textbf{two cases} holds:}

\begin{enumerate}
\def\labelenumi{\arabic{enumi}.}
\tightlist
\item
  \emph{if the problem was solved in less than \(T\) steps then RSA was
  broken, which is assumed to be improbable;}
\item
  \emph{if the problem was solved with wrong protocol parameters
  \((y, \pi)\) and they're correlated by an exponent \(r\) in the check,
  then \(\pi\) had to have been based off of \(y\)\footnote{\(\pi\) was
    necessarily chosen after \(y\) because the relationship between the
    two requires a parameter \(r\) that needs to be given by the
    Verifier, and an honest Verifier wouldn't have continued the
    protocol without having been given a \(y\) first. So \(y\) cannot be
    based off of \(\pi\) in an attack against our adaptive protocol.}
  (see below for exact relationship) and the exponent removed (i.e.~an
  exponent-root was calculated), but removing any prime exponent
  requires calculating any prime root for a value in the group, which
  breaks the Adaptive Prime Roots assumption, which is assumed to be of
  negligible probability.}
\end{enumerate}

In other words: \[{\left.\begin{aligned}
Pr[\exists \textit{"extractor"}\ E_{POLY} \in ITM: \phi(N) \gets E(N)] = negl(\lambda) 
\\ \textbf{Adaptive Prime Roots Assumption} \\\iff \forall \alpha \in RSA_\lambda(N),\ \alpha \notin \{0, \pm 1\}, r \in_R PRIME_{2\lambda}: \\Pr[\exists \textit{"Extractor"} E'_{POLY} \in ITM: \sqrt[r]{\alpha} \pmod N \gets E'(\alpha)] = negl(\lambda)
\\ \forall x \in \mathbb{Z}_N^*: \Omega_{x^{2^T}\ w/out\ \phi(N)}(T \mu_\lambda) 
\\ \Big[\forall x', y, \pi \in RSA_\lambda(N)/\{\pm 1\}, r \in PRIME_{2\lambda}: \\\pi^r x'\ ^\textit{residue} \pmod N = y \implies \pi = \sqrt[r]{y x'\ ^{-\textit{residue}}} \pmod N\Big]
\\ \Big[X \notin TL \land \pi^r x’\ ^\textit{residue} = y \implies \textbf{(c1)  } \Omega_{y}(< T \mu_\lambda) \lor \textbf{(c2)  } \lnot(y = x'\ ^{2^T} \land \pi = x'\ ^q)\Big] 
\end{aligned}\right\rbrace}\land
\\\implies\begin{cases}
\Omega_y = \Omega_{x^{2^T}\ w/\ \phi(N)}(< T \mu_\lambda) \xLeftrightarrow{\textit{same as SPoW soundness}} Pr = \textit{negl}(\lambda)\ \textit{\textbf{(case 1)}}\\
(y \neq x'\ ^{2^T} \lor \pi \neq x'\ ^q) \land \pi = \sqrt[r]{y x'\ ^{-\textit{residue}}} \pmod N\ \textit{\textbf{(case 2)}}
\\\implies \exists \alpha \in RSA_\lambda(N)/\{\pm 1\}: y \gets x^{2^T} \alpha \land \pi \gets x^q \sqrt[r]\alpha \implies \exists \textit{"Extractor"} E'_{POLY}
\\\iff negl(\lambda) = Pr[\exists E'_{POLY}] = Pr[\pi^r x’\ ^\textit{residue} = y \land (y \neq x'\ ^{2^T} \lor \pi \neq x'\ ^q)] 
\\\qquad\qquad\qquad= Pr[\pi^r x’\ ^\textit{residue} = y \land X \notin L]
\end{cases}\]

The soundness of the protocol relies on the same assumptions as the SPoW
protocol, as well as the inability to find prime roots in groups of
unknown order. In particular, two assumptions are required for this
protocol to work:

\begin{enumerate}
\def\labelenumi{\arabic{enumi}.}
\item
  Cracking RSA is hard
\item
  There is no faster way to solve the puzzle w/out either breaking the
  underlying SPoW time-lock puzzle, or breaking the Adaptive Roots
  Assumption. \emph{It is an open question whether this assumption can
  be reduced directly to the RSA hardness one, but it feels like a
  natural outcome and it would be a nice security improvement.}

  There are two main attacks that the Adaptive Prime Roots Assumption
  takes care of, when an attacker is targeting our VDF protocol. Let us
  also omit any residues for the sake of keeping things simple. First,
  the attacker could guess the expected Verifier's choice of \(r\), and
  subsequently choose a random value \(\pi\) and set \(y = \pi^r\). This
  is easily staved off by using a bruteforce-resistant security
  parameter (e.g.~\(2^{256}\)), for example based off of our RSA
  parameter \(2\lambda \approx 200\). The second attack deals with the
  reason as to why we choose \(r\) to be prime, and not just any random
  number from \(\mathbb{Z}_{2\lambda}\). The reason is that if \(r\)
  turns out to be a smooth-integer, then the attacker could choose
  \(y = \alpha^B\), for random \(\alpha\) and \(B\) the product of many
  prime powers (up to some limit); then,
  \(\pi = \sqrt[r]{y} = \alpha^{B/r} \pmod N\) with
  \(B/r \pmod {\phi(N)} = B/r\) if there is no residue, hence there is
  no need to know the group order to calculate the root because we don't
  need to work within the group. If the \(r\) is prime, then there is a
  really high probability that there will be a residue left, with an
  exception for the unlikely scenario where it is chosen to be equal to
  one of \(\phi(N)\)'s factors.
\end{enumerate}
\end{description}

\begin{description}
\item[Scalability]
This protocol is fully succinct, because it is both \emph{verifier
scalable}
\[O_V(polylog(T)) \Longleftarrow O_V(2 \cdot \lambda) = O_V(2)\] and it
has a \emph{succinct proof}
\[O_{|\pi|}(|x| = \lambda_{RSA}))  \Longleftarrow O_{|\pi|}(1 \cdot \lambda_{RSA} + 1 \cdot \lambda) = O_{|\pi|}(1 \cdot \lambda_{RSA})\]

Specifically, the checking algorithm for the Verifier \(V\) only
requires two (\(\pi^r\) and \(x'\ ^{\textit{residue}}\)) small
\(RSA_\lambda\) group exponentiations, which require respectively
\(|r|\) and \(|\textit{residue}|\) group-squarings using the
``square-and-multiply'' algorithm, and a group multiplication to put
them together for the equality check. As for the messages required to
complete the proof, only \(\pi\) and \(r\) need to be transferred, the
first one is just a group element and the second one is much smaller
(due to the fact that \(\lambda_{RSA}\) is derived from \(\lambda\))
\end{description}

\hypertarget{sec:vdftransparency}{%
\subsection{Eliminating Trust Issues}\label{sec:vdftransparency}}

Now that we've achieved such a cool protocol, let's address a final
issue: how to make multiple parties use the same VDF without fear of
cheating? Alternatively, how to achieve \textbf{transparency}?

The only construction we've mentioned so far, using RSA groups, clearly
requires someone to generate private keys which, as we've seen earlier,
can be used to break the protocol. What we really need are techniques to
prevent anyone from owning the private key, so that nobody even has the
chance to cheat. Here are a three strategies discussed by
{[}\protect\hyperlink{ref-Wes18}{28}{]}:

\begin{enumerate}
\def\labelenumi{\arabic{enumi}.}
\item
  \textbf{Alternative modulus generation}: There is an approach to
  generating RSA groups, presented by
  {[}\protect\hyperlink{ref-Sander99}{42}{]}, which aims to completely
  skip the private key generation by randomly selecting a large modulus
  which can satisfy RSA requirements with high probability. If this
  modulus is indeed large and chosen randomly, nobody should be able to
  extract \(\phi(N)\) from it. While this method is the simplest and
  most efficient way to patch our protocol, it does not always lead to
  correct RSA groups and it is believed to severely damage VDF
  sequentiality requirements, leading to more efficient \(\mu_\lambda\)
  implementations. Thus, it might break SPoW soundness assumptions and
  cannot be used reliably.
\item
  \textbf{MPC-based RSA setup}: a popular solution to solving trust
  issues is, as we've discussed already, distributing trust. As it so
  happens, secure Multi-Party Computation protocols (e.g.~Yao's garbled
  circuits {[}\protect\hyperlink{ref-twompc}{43}{]} and secret sharing
  with arithmetic circuits{[}\protect\hyperlink{ref-multimpc}{44}{]})
  would allow multiple participants to jointly generate a provably valid
  RSA modulus, without leaking the private key to anyone; a technique
  for this is presented in {[}\protect\hyperlink{ref-Boneh97}{45}{]}.
  Such MPC-based approaches are practical enough that they were also
  employed by the ZCash cryptocurrency
  {[}\protect\hyperlink{ref-Zcash}{46}{]} for its setup.

  Unfortunately, this method is secure only if at least one party in the
  computation is honest, which means that all (independent) parties
  interested in using the \emph{transparent} VDF protocol should
  participate in the setup phase, to be sure that it is trust-less.
  However, in blockchain scenarios multiple parties join the protocol
  long after the initial setup phase --- which means that some degree of
  trust is involved. As long as the number of (independent) parties
  participating in the MPC is a significant portion of the total number
  of VDF users, this method is convenient and reliable.

  Since MPC setups do involve the generation of secret random values,
  they cannot be considered strictly \emph{transparent}, according to
  the definition we gave in our VC model. However, they do provide a
  strong form of trust reduction through distribution --- which is
  commonly cited as being the core of Bitcoin's ``trustless'' design. We
  can consider them to be a weaker form of \emph{transparency}, perhaps
  called ``trustlesness''.
\item
  \textbf{Alternative Groups}: a newer approach, given by
  {[}\protect\hyperlink{ref-Wes18}{28}{]} in his VDF construction, has
  been to replace RSA with a trapdoor-free multiplicative group, also
  called ``Class Group of an Imaginary Quadratic Field''
  {[}\protect\hyperlink{ref-Buchmann88}{47}{]} (adaptation to RSA
  presented in {[}\protect\hyperlink{ref-BBHM02}{48}{]}). This approach
  promises to be uncompromising, since the group order is not known even
  to the party setting it up. However, this method still requires
  someone to generate the public parameters, so there is an assumption
  to be made: no other setup procedure for these groups needs to allow
  for a Trapdoor, so either the known setup procedure is the only one
  available, or any other procedure cannot leak the group order. Given
  that these groups have not yet been sufficiently studied by the
  cryptographic community, this method can be considered to be less
  reliable. It is, however, a very interesting topic for future
  research, and it provides an innovative solution to the problem at
  hand.
\end{enumerate}

\hypertarget{a-note-on-vdfs-as-transparency-enablers}{%
\subsubsection{A note on VDFs as transparency
enablers}\label{a-note-on-vdfs-as-transparency-enablers}}

\begin{quote}
Given that VDFs can be used to build trustless Randomness Beacons, and
that these short-term\footnote{i.e.~they cannot be used, once revealed,
  for new protocol executions; for example, in a lottery system you may
  only use numbers which have will be revealed after the bidding phase
  is over. Because of this, you will need to keep using ``fresh'' beacon
  outputs.} random numbers can be used to setup other cryptographic
protocols in a transparent fashion, it is imperative that VDFs
themselves be transparent as well. However, one does not need to go
overkill with the setup procedure --- if the protocols which make use of
the Randomness Beacon have less or equal security requirements than that
of the VDF (i.e.~the number of participants in the MPC phase is still
significant compared to the total number of users of the protocol), the
MPC setup procedure is still good enough for their purpose. In fact,
there are current plans {[}\protect\hyperlink{ref-ethereumvdfmpc}{40}{]}
to implement a blockchain-wide VDF for the Ethereum cryptocurrency,
rather than as an isolated third-party instance, such that all the
``smart contracts'' (i.e.~subprotocols) running within Ethereum already
lie within its security requirements.
\end{quote}

\hypertarget{conclusion}{%
\section{Conclusion}\label{conclusion}}

We've seen two very powerful techniques for building VC protocols: (1)
arithmetisation, used in Homomorphic Authenticators for outsourcing
small-degree computations; (2) interactivity and randomness, used in
VDFs for ``compressing'' computations and measuring time. It is my
belief that while many non-universal VC protocols have been considered
as part of separate fields, we should try to converge them under the
domain of Verifiable Computation, to compare them and understand the
most efficient designs behind specific cryptographic VC properties. Such
designs can later be abstracted and employed for building more
expressive protocols, as demonstrated by the execution trace idea found
in both Pietrzak's VDF protocol and many other Universal VC protocols
(e.g.~STARKs), discussed in the next chapter.

Unfortunately, and due to time constraints, this chapter only barely
scratches the surface of all non-universal proof protocols that have
been built over the past decades, so I leave as an open question the
analysis and unification of the remaining ones. To motivate the reader
in that direction, allow me to acknowledge other very interesting and
influential systems:

\begin{itemize}
\tightlist
\item
  \textbf{Sigma Protocols}: \(\Sigma\text{-Protocols}\) is the field
  representative of traditional Zero-Knowledge Interactive Proof
  systems, which were developed decades ago with the intent of building
  more practical constructions through the use of relaxed VC properties.
  This is an extensive field, focusing primarily on public key
  authentication, seeing the likes of the famous Schnorr Identification
  and Signature scheme. Protocols from this field share many properties
  with the universal proof systems discussed in the next chapter. The
  Ring-based Learning With Errors scheme found in
  {[}\protect\hyperlink{ref-Benhamouda15}{49}{]} is an interesting
  recent development in this field.
\item
  \textbf{Proof of Work}: this field was made popular by the deployment
  of the Bitcoin cryptocurrency, which uses it for its transactions
  (i.e.~proofs). A lot of research from the cryptocurrency communities
  has gone into extending this field with more efficient constructions,
  resulting in improved consensus solutions for decentralised-trust
  protocols. A notable evolution of this field is VDF protocols, which
  we analysed in this chapter.
\item
  \textbf{(RSA) Accumulators}\footnote{an interesting starting point for
    the reader might be
    {[}\protect\hyperlink{ref-AccumulatorSoK}{50}{]}, which offers a
    systemisation of this field, including constructions not based on
    hidden order groups.}: this interesting field, whose protocols
  implement very efficient operations for checking membership of an
  element in a set, typically makes use of hidden order groups
  (e.g.~RSA). Such constructions can also support other set operations,
  such as union and intersection. This field comes closest to
  implementing the \emph{universality} property found in Universal VC
  schemes presented in the next chapter. Fun fact: Zerocoin, the
  precursor to the Zerocash protocol that the Zcash
  {[}\protect\hyperlink{ref-Zcash}{46}{]} cryptocurrency implements, was
  actually based upon RSA Accumulators.
\item
  \textbf{Attribute Based Encryption (ABE)}: such systems take advantage
  of user identities to establish public key pairs, which offers the big
  advantage of being able to send a single message to a specific
  hierarchy of users without needing to collect many different keys. The
  use of a public authority (i.e.~Trusted Third Party) is typically
  required, and almost all such systems employ homomorphic encryption,
  used to build complex relationships between messages and identity
  attributes.
\end{itemize}

\hypertarget{universal-vc-compilers}{%
\chapter{Universal VC Compilers}\label{universal-vc-compilers}}

A revolution in the applicative world of cryptography, first with
Blockchain technology and now with Zero-Knowledge proofs, has been
developing over the last decade. The success achieved by these protocols
is starting to spark excitement, in the hopes that it could change not
just our societal functions (e.g.~cryptocurrencies vs traditional fiat
money), but also the way we interact online and develop software. The
main goal of these efforts has been that of developing innovative
cryptography to help us regain trust in a trustless world, to help us
base all our communications on verifiable statements: to build a
\textbf{``Proof of All''}.

So far we've discussed the basic building blocks for cryptographic
proofs, and how specialised protocols can be designed to handle private
or computationally-sensitive scenarios. One common characteristic, and
potential downside, of using such non-universal protocols is that
adapting them to particular use-cases requires technical know-how;
compromises (in terms of VC properties) are often also required to
retain efficiency or privacy. In this chapter we will be taking a step
towards an uncompromising solution, and the jack-of-all-trades when it
comes to Zero-Knowledge proofs, \emph{Universal} VC protocols. These
systems do not only protect the privacy of their users, but they can
also guarantee the integrity of any computation. Because such protocols
can be used to generate proofs based on any other program, automatically
and without much technical know-how, I call them \textbf{Universal Proof
Compilers}.

The focus of the chapter will be understanding and designing the
fundamentals of a protocol which marks a breakthrough in the field of VC
technology: \emph{zk-STARKs}. While this construction is fairly recent,
it is the result of many years of research and has already been well
received by the cryptographic community. This system is the first
\textbf{concretely efficient} (i.e.~suited for realistic usage)
Universal VC compiler that is also post-quantum safe, and does not
require any form of trusted setup (i.e.~it can be used out-of-the-box,
unlike zk-SNARKs). Towards the end of the chapter we will also mention
alternative systems to zk-STARKs that have been developed in recent
years.

\hypertarget{zk-starks}{%
\section{zk-STARKs}\label{zk-starks}}

What are zk-STARKs? Glad you asked:

\begin{itemize}
\tightlist
\item
  \emph{zk}, as in zero-knowledge and privacy-preserving;
\item
  \emph{Scalable}, or efficient, as proving requires little increased
  overhead, generated proofs are relatively ``small'' (or acceptable) in
  size, and verification takes exponentially less time than executing
  computations naïvely (i.e.~almost instantly, even for very heavy
  ones);
\item
  \emph{Transparent}, as in there is no requirement for a trusted setup,
  like in zk-SNARK systems;
\item
  \emph{ARgument}, as in a computationally secure cryptographic proving
  scheme achieving completeness and soundness for a specific language;
\item
  of \emph{Knowledge}, as in based on statements with relation to
  publicly known information (see more in Section \ref{sec:spec}).
\end{itemize}

But most importantly of all, (zk)STARKs are \emph{Universal} Verifiable
Computation systems. Unfortunately, these definitions are not sufficient
to build or understand such systems. The construction by Ben-Sasson et
al.~presented in {[}\protect\hyperlink{ref-STARK}{51}{]} is fairly
complex, filled with engineering-specific details (the protocol was
designed to be concretely efficient), and overall tough to digest even
for cryptography students. For these reasons, and following the goals of
the thesis, I chose to focus on design principles, foregoing formal
proofs in favour of a simplified understanding. In this section I will
break down the core concepts of zk-STARKs, showing how a general purpose
computational-integrity statement can be converted into a proof.

In Section \ref{sec:starkmain} I will give an overview of how we're
going to approach building a STARK. Each step of our design represents a
problem instance that abstracts the following step, thus providing a
useful overview for breaking down STARKs. By taking a stricter
mathematical formalisation of the initial problem statement and
following the given reduction steps, it is possible to synthesise the
protocol into a single statement, comparable to that provided within the
original paper.\footnote{the original paper in
  {[}\protect\hyperlink{ref-STARK}{51}{]} also formalises and takes care
  of multiple engineering optimisations, which I only briefly touch upon
  later on. These details can be considered to be essential for the
  implementation of a practical STARK and are an important contribution
  to the achievements of the paper, as well as the basis for the
  official open-source implementation provided by the authors in
  {[}\protect\hyperlink{ref-libSTARK}{52}{]}.}

Each later subsection will present: the main objective of a universal VC
system (Section \ref{sec:ci}); the intermediate arithmetisation process
required to break down normal computations into usable components
(Section \ref{sec:arith}); \(2POLY\), the name I give to the core
subprotocol used by STARKs to implement a VC proof (Section
\ref{sec:2poly}); concrete performance results achieved by STARKs
through interactivity, and security (i.e.~soundness) assumptions
(Section \ref{sec:starkscalable}); the privacy extension to convert
STARKs into zk-STARKs (Section \ref{sec:starkaddzk}); \(FRI\), the
subprotocol used by \(2POLY\) for probabilistic degree testing (Section
\ref{sec:fri}).

\hypertarget{sec:starkmain}{%
\subsection{The Main Design}\label{sec:starkmain}}

The single most important tool which is used by all known Universal V.C.
systems is that of \textbf{arithmetisation}. It is the process of
converting a question on the integrity of a general purpose calculation
into a mathematical statement which we can manipulate through
cryptographic means. For zk-STARKs this is the \emph{polynomial
comparison} problem, for their zk-SNARk predecessors it is
\emph{quadratic arithmetic programs}, but similar ideas arise in all the
competing universal VC protocol systems, exhibiting varying approaches:
homomorphic cryptography, multiparty computation, probabilistic
checkable proofs, and interactive proofs. The original problem statement
provided here is reduced (not without any assumptions, as we will see
later) to an algebraic statement on polynomials, for which we actually
have a working cryptographic protocol. It is the final problem of
polynomial comparison on which we will focus our cryptographic tools
deriving from PCPs and IPs, the rest is mostly arithmetisation.

First, we will informally introduce the main problem of Computational
Integrity and Privacy which we are trying to solve (Section
\ref{sec:ci}); then we will perform a few arithmetisation steps which
bring us closer to a formal statement on polynomials (Section
\ref{sec:arith}); then, we will present the core polynomial comparison
and proximity testing protocols used by zk-STARKs (Section
\ref{sec:2poly}).

Here is an overview of the problems addressed by our design, in
increasing order of specialisation:

\begin{longtable}[]{@{}llll@{}}
\toprule
\begin{minipage}[b]{0.18\columnwidth}\raggedright
Arithmetisation Step\strut
\end{minipage} & \begin{minipage}[b]{0.24\columnwidth}\raggedright
Problem\strut
\end{minipage} & \begin{minipage}[b]{0.24\columnwidth}\raggedright
Description\strut
\end{minipage} & \begin{minipage}[b]{0.24\columnwidth}\raggedright
VC Benefit\strut
\end{minipage}\tabularnewline
\midrule
\endhead
\begin{minipage}[t]{0.18\columnwidth}\raggedright
1\strut
\end{minipage} & \begin{minipage}[t]{0.24\columnwidth}\raggedright
Generic Statement\strut
\end{minipage} & \begin{minipage}[t]{0.24\columnwidth}\raggedright
\emph{Was the output of this computation, within the specified
time-frame, correct?}\strut
\end{minipage} & \begin{minipage}[t]{0.24\columnwidth}\raggedright
\emph{Universality}\strut
\end{minipage}\tabularnewline
\begin{minipage}[t]{0.18\columnwidth}\raggedright
2\strut
\end{minipage} & \begin{minipage}[t]{0.24\columnwidth}\raggedright
Computational Integrity\&Privacy Statement\strut
\end{minipage} & \begin{minipage}[t]{0.24\columnwidth}\raggedright
\emph{Is it true that Output=Program(Input) within T steps?}\strut
\end{minipage} & \begin{minipage}[t]{0.24\columnwidth}\raggedright
\emph{Universality}\strut
\end{minipage}\tabularnewline
\begin{minipage}[t]{0.18\columnwidth}\raggedright
3\strut
\end{minipage} & \begin{minipage}[t]{0.24\columnwidth}\raggedright
Algebraic Problem\strut
\end{minipage} & \begin{minipage}[t]{0.24\columnwidth}\raggedright
\(f(x) \overset{?}= y\), \(O_f(T)\)\strut
\end{minipage} & \begin{minipage}[t]{0.24\columnwidth}\raggedright
\emph{Universality}\strut
\end{minipage}\tabularnewline
\begin{minipage}[t]{0.18\columnwidth}\raggedright
4\strut
\end{minipage} & \begin{minipage}[t]{0.24\columnwidth}\raggedright
Execution Trace Algebraic Problem\strut
\end{minipage} & \begin{minipage}[t]{0.24\columnwidth}\raggedright
\(ee \overset{?}\in \mathscr{C}\), \(ee \textit{ execution trace}\),
\(\mathscr{C} \textit{ constraints}\), \(|ee| = T+1\)\strut
\end{minipage} & \begin{minipage}[t]{0.24\columnwidth}\raggedright
\emph{Soundness} (\(2POLY\) call format), \emph{Scalability}\strut
\end{minipage}\tabularnewline
\begin{minipage}[t]{0.18\columnwidth}\raggedright
5\strut
\end{minipage} & \begin{minipage}[t]{0.24\columnwidth}\raggedright
Polynomial Comparison Problem\strut
\end{minipage} & \begin{minipage}[t]{0.24\columnwidth}\raggedright
\(f(x) \overset{?}= g(x)\), \((f,g) \in \mathbb{F}[x]\),
\(deg(f) = deg(g)\)\strut
\end{minipage} & \begin{minipage}[t]{0.24\columnwidth}\raggedright
\emph{Soundness} (check), \emph{Zero-Knowledge}, \emph{Scalability}
(engineering optimisations), \emph{Transparency}\strut
\end{minipage}\tabularnewline
\bottomrule
\end{longtable}

\hypertarget{sec:ci}{%
\subsection{Original Problem Statement}\label{sec:ci}}

We're looking to build a system which can represent any VC problem,
i.e.~a ``Proof of All'' system; before building it we need to define it,
in order to state our requirements and boundaries. The researchers
behind zk-STARKs provide a language to define any trustless computation,
called \textbf{Computational Integrity and Privacy} (CIP) problem
statements. Such statements represent the state of the art of what
current cryptographic proof systems can achieve in any computational
model.

First off, let's clarify the name ``computational''. With this name, we
simply wish to allow the system's users to make statements regarding any
sort of general purpose computation. The \emph{universality} property of
our VC model suffices to satisfy this requirement. Here are the main
properties defining CIP problem statements:

\begin{enumerate}
\def\labelenumi{\arabic{enumi}.}
\item
  \emph{Integrity}

  In order to trust the output of a specific computation, we need to
  consider that a Prover may be incentivised to cheat. We can think of
  income tax statements, for example, where a citizen is trying to
  perform tax evasion by submitting false claims regarding his income.
  To prevent this, and to trust the validity of the Prover's claim, we
  need to somehow ``bind'' the computation's output to the actual
  requirements of the computation. I also informally consider this to be
  the \underline{binding property} of a CIP statement, and it can be
  accomplished through the \emph{completeness} and \emph{soundness}
  properties which we defined in our VC model.
\item
  \emph{Privacy}

  What happens if the output of a specific computation can be revealed,
  but not its input? Consider a scenario in which I'm buying drinks at a
  bar and I need to provide identification to the bartender, so that he
  may check that I am of legal age to drink alcohol, but I do not want
  to reveal anything else about my age, name, nationality, height, or
  gender. To allow a Prover to make such a privacy-friendly claim, we
  need to somehow ``hide'' our computation's inputs (i.e.~my personal
  details, in the given example) from the Verifier. I informally
  consider this to be the \underline{hiding property} of a CIP
  statement, and it can be accomplished through the VC model's
  \emph{zero knowledge}, \emph{transparency}, and \emph{post-quantum
  safety} properties.

  The transparency requirement is necessary in contexts where I want
  anybody to be able to verify my claim, at any point in time, without
  the need for a trusted setup phase. Post-quantum resistance is also
  important in any cryptographic system meant to stand the test of time,
  thus becoming a reliable standard for the protection of data many
  decades (if not centuries) down the road.\footnote{note, the main
    assumptions made by zk-STARKs are: (1) the existence of
    cryptographic One-Way hash functions; (2) the Random Oracle Model.
    These assumptions are amongst the oldest to exist in cryptography,
    and they have defied all sorts of cryptanalysis, including recent
    Quantum computer developments.}
\item
  \emph{Efficiency}

  Along with the previous two fundamental properties, Ben-Sasson et
  al.~mention this additional and more practical requirement. We are
  concerned with the realisation of concrete systems, which can be used
  under realistic and fair conditions, using hardware that is commonly
  available to any average Prover or Verifier.

  Assume you're tasked with extracting all the facial images of people
  passing through an airport on a specified date, and then matching them
  against a known-criminals' database, as part of a police
  investigation. You wish to provide the list of matches as evidence for
  a court hearing, but the court is skeptical of your work and wishes to
  double-check the results. If the computation took 20 hours to
  complete, will the court need to take just as long to verify your
  statement?

  Universal zero-knowledge proof systems of the past were actually very
  burdensome in this regard, easily requiring terabytes of memory for
  even the simplest calculations; the most efficient proof systems were
  Sigma protocols, but they were only appropriate for very specialised
  computations. To make our system actually usable, we need to somehow
  reduce its impact to be minimal for the Prover, and actually even
  convenient (i.e.~much faster than normal execution) for the Verifier.
  With regards to the communication complexity of our system, it should
  stay within acceptable levels of Internet communication.\footnote{for
    example, proving a single CIP statement typically requires the
    transfer of a few hundred bytes with zk-SNARK systems, and a few
    hundred kilobytes with zk-STARK systems. Considering that a CIP
    statement can be used to further compress other CIP statements, both
    complexities are acceptable even for repeated use in space-sensitive
    environments, such as decentralised Blockchains.} To realise this
  CIP property, we will have to implement multiple properties of our VC
  model: \emph{prover scalability}, \emph{verifier scalability},
  \emph{proof succinctness}, and \emph{non-interactivity}\footnote{the
    main strategies we will use to get all these properties are: (1)
    arithmetisation; (2) random querying. In alternative systems, such
    as zk-SNARKs, homomorphic encryption is also used, with a boost in
    efficiency but a loss in privacy (specifically, transparency).}.
  With regards to non-interactivity, it is an important efficiency
  measure because it allows using our system even when neither party can
  communicate at the same moment. The proofs can be batched in advance,
  and sent off for inspection at a later time.
\end{enumerate}

Finally, we can formalise our system's language to be

\begin{description}
\tightlist
\item[CIP]
\emph{Is it true that Output=Program(Input) within T steps?}
\[\iff \Big\{  (P, T, x, y) \bigm\vert y = P(x),\  \textit{ “Program” } P \in ITM,\ O_P(T \textit{ steps})  \Big\}\]\footnote{``steps''
  here represents a state change within the program. When comparing with
  other systems, it is useful to convert a step to the number of CPU
  cycles that are required for each state change after converting the
  program into a binary circuit. If it is hard to identify a specific
  state, then every single instruction can be interpreted as a step,
  with the state being the totality of the program's variables.}
\end{description}

A state-of-the-art system proving that \(X \in CIP\) needs to implement
the following VC properties:

\begin{itemize}
\tightlist
\item
  universality (it's intrinsic)
\item
  completeness
\item
  soundness
\item
  zero-knowledge
\item
  scalability
\end{itemize}

Additionally, zk-STARKs also implement

\begin{itemize}
\tightlist
\item
  non-interactivity
\item
  transparency
\item
  post-quantum safety
\end{itemize}

\hypertarget{sec:spec}{%
\subsubsection{A few notes on Program Specifications}\label{sec:spec}}

I would like to take a moment to debate on the utility of CIP statements
in real-life scenarios. In general, the problem we're trying to solve is
not always related to a specific program, as much as it is to a specific
computation:

\begin{description}
\tightlist
\item[Generic Statement]
\emph{was the output of this computation, within the specified
time-frame, correct?}
\end{description}

Given such a generic requirement, it may not always be necessary to
start execution of a STARK from the CIP statement of a binary program.
The requirements for the computation may, in fact, already be defined by
human generated \textbf{program specifications}, which document the
desired functionality of the program being analysed. While such
documentation is often seen underdeveloped (or lacking) even in the
biggest software projects -- since unit test-cases are a cheaper
alternative -- it can still serve as a concise and efficient definition
for the core functionality of a specific computation. In fact, it allows
us to skip the CIP's binary program conversion phase, and directly use
our program specification for the intermediate arithmetisation phases.

Here are two example scenarios, one which appeals to program
specifications and one which appeals to CIP statements on binary
programs:

\begin{itemize}
\item
  \emph{Copyright-Protected Streaming}: a cloud provider's technician is
  tasked with adding DRM to their video streaming service, in order to
  comply with a recently approved European Copyright Directive. The only
  issue is that the service specialises on streaming encrypted videos,
  as an added privacy benefit. The technician immediately thinks to use
  his favourite zero-knowledge universal VC proof system, zk-STARKs, so
  as to retain privacy of the streams and minimise the impact of the new
  feature on the service's performance. In this scenario the constraints
  are very simple: each source file needs to be checked against a list
  of blacklisted files, then it is encrypted and checked against the
  corresponding stream.

  While the technician could write a program to do this, compile it, and
  send it over to the clients so that they can generate CIP-based
  proofs, there are several downsides: (1) waste of resources, as
  developing and deploying consumer-level applications can take a lot of
  man hours; (2) bugs, as traditional testing does not typically
  guarantee that the security specifications are met by the program with
  a high degree of certainty; (3) performance hit for the clients,
  because full arithmetisation of a stateful binary program can lead to
  much more complex constraints and larger execution traces than is
  really necessary, also leading to bloated proof sizes; (4) last but
  not least is security, because the users are asked to trust that
  executing a binary program will not compromise the confidentiality of
  their files\footnote{note that even if the program was released as
    open source, it still takes a much longer time to analyse thousands
    of lines of code (also including libraries) rather than just a few
    lines of specification requirements.}.

  A much smarter solution is that of taking advantage of appropriately
  documented program specifications for the requirements of the DRM
  feature, and sending those off to the clients in a standardised
  format. The streaming service's users can then take advantage of
  trusted zk-STARK implementations to build proofs based upon a very
  small set of constraints.
\end{itemize}

\begin{itemize}
\item
  \emph{CTF Challenge}: in ``Capture The Flag'' competitions,
  participants typically take part in jeopardy-style cybersecurity games
  where they must solve multiple challenges to score points. One popular
  category of these challenges, known as ``Pwning'', requires that
  participants discover a vulnerability hidden somewhere within a given
  program; to verify that a player has exploited the program
  successfully, instead of just manually bypassing the security checks
  through binary editing, the program is uploaded to a sandboxed server
  and players are restricted to feeding it input through an internet
  socket.

  In all common recurrences of this scenario, a few issues arise: (1) a
  server with high computational and bandwidth capacities needs to be
  rented to host the vulnerable program; (2) the vulnerable program also
  needs to be sandboxed or virtualised to protect the server's
  integrity, leading to further impacts on computational requirements;
  (3) in case of oversights made during setup of the sandbox, the server
  itself may become vulnerable, leading to a potential compromise of the
  whole competition\footnote{a well justified concern when dealing with
    participants whose expertise is cybersecurity and penetration
    testing, actually!}; (4) some malicious actors may choose to carry
  out a DoS attack on the server by overloading the vulnerable program
  with continuous inputs, leading to an abrupt end of the whole
  competition.\footnote{another common recurrence in CTF
    competitions\ldots{}}

  In this specific scenario, applying STARKs using the CIP binary
  program statement makes perfect sense. There is no need to apply
  zero-knowledge (HTTPs is probably sufficient), but there is still a
  desire to verify knowledge of the vulnerability in a very short
  time-frame, and without potential compromise of the server.
  Furthermore, the requested knowledge directly relates to a specific
  stateful program execution, so it makes sense to arithmetise that same
  program along with its every nook and cranny. Thanks to STARKs, CTF
  competition maintainers could host challenges at a fraction of
  previous costs, without worrying too much about security of the
  hosting server.
\end{itemize}

\hypertarget{sec:zkstatements}{%
\subsubsection{A note on Zero-Knowledge
Statements}\label{sec:zkstatements}}

In this section, we regarded the input \(x\) of a program as part of the
CIP language statement. Truthfully, things are a little different when
we consider the need for zero-knowledge. With zero-knowledge we actually
aim to hide the input \(x\), so it cannot be part of the statement, it
will instead be part of the witness. At the same time, having a
statement of the form

\begin{description}
\tightlist
\item[zk-CIP]
\[\Big\{  (P, T, y) \bigm\vert \exists x: y = P(x),\  \textit{ “Program” } P \in ITM,\ O_P(T \textit{ "steps"})  \Big\}\]
\end{description}

does not always equate to a proof of knowledge. For example, proving
that a number is composed of prime factors and proving that these
factors are known constitute two entirely different ordeals. Because of
this, typical zero-knowledge CIP statements aim to prove knowledge of a
secret through a publicly known element (that we'll call \(h\) instead
of \(x\)). This public input should uniquely identify the secret,
without revealing any information. The most popular way to do this, with
an computationally indistinguishable amount of information revealed, is
through a cryptographic hash function \(H\):

\begin{description}
\tightlist
\item[zk-CIP of knowledge]
\[\Big\{  (P, T, h, y) \bigm\vert \exists x: y = P(x) \land h = H(x),\ \textit{ “Program” } P \in ITM,\ O_P(T \textit{ "steps"})  \Big\}\]\footnote{formally,
  it might actually be more correct to place all constraints inside the
  program, so that they can be considered to be part of the
  arithmetisation steps needed for a STARK:
  \(P'(x,h) \overset{def}= \{y=P(x) \land h=H(x)\}\).}
\end{description}

This simple solution is also useful for authenticating users based on
their public keys or other forms of public data that constitute a unique
reference to the secret.

\medskip

\emph{NOTE: when performing such proofs, it's very important to take
into consideration hash functions that are better suited to the
algebraic nature of STARKs. This is due to the complexity that arises
when arithmetising such ``functions'' which are typically optimised for
real processors. Because of this, the authors of the paper opted to make
use of a Davies-Meyer \(AES\text{-based}\) hash construction
({[}\protect\hyperlink{ref-STARK}{51}{]},
{[}\protect\hyperlink{ref-DMhash}{53}{]}), which offered better
performance compared to \(SHA2\) when used in the binary fields that
their polynomials were based upon. This concept also applies to the
\(2POLY\) and \(FRI\) protocols that we will see later, due to the
requirement of a commitment scheme.}

\hypertarget{sec:arith}{%
\subsection{Intermediate Arithmetisations}\label{sec:arith}}

The first important step to take is that of turning our problem
statement on binary inputs, outputs, and programs into a statement on
algebraic objects. In particular, the most difficult aspect of this
transition is that of converting a stateful binary program into a
function. In the original paper this is performed through a series of
complex engineering steps called APR (Algebraic Placement and Routing)
reduction, where the whole state of the program is also abstracted,
including RAM and networking\footnote{of course, applying the APR to
  real programs running on MacOS/Linux operating systems and Intel/AMD
  processors is not yet realistic, so the researchers provided a proof
  of concept in {[}\protect\hyperlink{ref-libSTARK}{52}{]} using a
  simple RISC virtual machine called TinyRAM
  {[}\protect\hyperlink{ref-TinyRAM}{54}{]}.}. For our purposes it will
suffice to assume that we have already converted, perhaps thanks to
well-defined program specifications mentioned in Section \ref{sec:spec},
CIP statements into the following Algebraic Problem:

\begin{description}
\tightlist
\item[AP1]
\[\Big\{ (x,f,y,T) \bigm\vert f(x) = y,\ O_f(T),\ f:D \to C \Big\}\]
\end{description}

We can now stop our process for a moment, to reflect on what it means to
achieve a protocol with \emph{scalability}. In the context of IP proofs,
the Prover comes up with a randomised problem that is isomorphic to the
original one, which allows revealing the witness under masked disguise.
But even if we were to forego \emph{zero-knowledge} and reveal our
witness directly (the input \(x\)), it would still take the Verifier
\(O_f(T)\) steps to check our \(AP1\) problem statement, which is just
as long as naïve execution and so it precludes \emph{verifier
scalability}.

VDFs are another family of protocols which is similar to STARKs, because
they are also trying to make really long computations become efficient
to verify. For the VDF construction by Wesolowski
{[}\protect\hyperlink{ref-Wes18}{28}{]}, the choice of an isomorphic
problem is justified because the specific algebraic properties of the
chosen problem lead to powerful relationships that are efficient to
verify, leveraging the security provided by RSA groups. But in our
scenario, we are dealing with generic computations which may not possess
such neat algebraic properties, so we cannot take advantage of such
shortcuts. The construction by Pietrzak
{[}\protect\hyperlink{ref-Piet18}{29}{]}, instead, offers an approach
that is a step closer towards the right direction. The idea is to
explicitly expand the witness of the given problem into an execution
trace, resulting in an isomorphic problem that allows the Verifier to
randomly and efficiently inspect parts of the computation. Each state
within the trace can be queried, and, with some randomness and smart
recursion, the boundaries of the trace are checked without relying on
clever isomorphism assumptions. We will take a similar approach of
expanding our witness input \(x\) into an execution trace leading up to
\(y\), and our burdensome computation will be reduced to a few
constraints with complexity much lower compared to that of the original
execution. Because of this, I call the technique \textbf{witness
expansion} and \textbf{constraint compression}. We will leave the
``randomness and smart recursion'' counterpart to the \(2POLY\)
subprotocol, which will leverage interpolation and proximity testing to
obtain prover and verifier \emph{scalability}.

The new problem can be regarded as checking an execution trace against
one or more constraints:

\begin{description}
\tightlist
\item[AP2]
\[\Big\{(ee, \mathscr{C}, T) \bigm\vert ee \in \mathscr{C},\ \mathscr{C} \textit{ "Constraints"}, |ee| = T+1, deg(\mathscr{C}) \ll T\Big\}\]
\end{description}

To define what constraints are, and for the sake of simplicity, we can
consider two case scenarios:

\begin{enumerate}
\def\labelenumi{\arabic{enumi}.}
\tightlist
\item
  \emph{Domain-based constraints}: each of the elements of the trace
  must satisfy a specific set-membership condition. This can be useful
  in very simple scenarios where we just want to check whether each
  element of a list lies within a given domain.
\item
  \emph{Polynomial-based constraints}: this scenario is more realistic,
  and it considers the requirements that a normal program would have.
  They can be represented as polynomials, taking as input one or more
  execution states.
\end{enumerate}

We will elaborate on reducing these two scenarios to a \(2POLY\) problem
in the following subsections.

\emph{NOTE: a single state of the execution trace defined above can be
composed of multiple variables, especially when extracted from a binary
program. The authors of the paper handled this case efficiently by
considering each variable separately, splitting a single execution trace
into multiple Reed-Solomon codes; this allows for notable space savings
after interpolation, and the trace evaluations can later be joined
through a linear combination.}

\hypertarget{sec:domainconstraints}{%
\subsubsection{Domain-based Constraints}\label{sec:domainconstraints}}

Let's assume to have been given the following problem:
\[\forall x \in D: f(x) \overset{?}\in \mathscr{C},\ \\with\ |\mathscr{C}| \ll deg(f),\ f: D \to C\]

Where \(\mathscr{C}\) is precisely the domain constraint, and \(f\) is
the function indexing an execution trace. Our main objective is to
reduce the original statement to a comparison between two polynomials
\((f',g')\): \[f \in \mathscr{C} \overset{?}{\iff} f' = g'\]

First, we shall convert the set membership constraint to a vanishing
polynomial, where \(True\) values for the membership relationship end up
evaluating to zero: \[\forall x\in D: C(f(x)) = 0
\\C(y) \overset{def}= (y - \mathscr{C}[0])(y-\mathscr{C}[1]) ... (y- \mathscr{C}[|\mathscr{C}|-1])\]
Unfortunately, this equation is not yet sufficient, because it is bound
to a specific domain. We will see later on that the \(2POLY\)
subprotocol needs to work on domains which can be extended, so we must
expand the domain of our inputs to span over all the integers.\footnote{as
  mentioned previously, the optimised variant of STARKs actually works
  with specific fields. More on this later.} To help us do this, we can
recall a useful theorem:

\begin{description}
\tightlist
\item[Th. Vanishing Polynomial Composition]
\emph{it is always possible to extend the domain of a univariate
vanishing polynomial through (de)composition}:
\[\forall x\in D: P(x) = 0
\\\iff \exists P': \land
\begin{cases}
P(x) = Z_D(x) P'(x) \\
deg(P) = |D| + deg(P')
\end{cases}
\\with\ Z_D(x) \overset{def}= \prod_{i \in D} (x - i)\] \emph{(note:
\(Z_D\) is also common notation to denote a polynomial vanishing on all
of the domain \(D\). The polynomial \(P'\) can be extracted by the
prover by interpolating \(P\) and calculating \(P/Z_D\).)}
\end{description}

We can now extract the full polynomial:
\[\exists P': C(f(x)) = Z_D(x)P'(x) \land deg(P') = deg(C) - |D|\] And
reduce to a \(2POLY\) problem: \[\exists P': \begin{cases}
f'(x) = g'(x)
\\deg(f') = deg(g')
\\\ 
\\f'(x) = C(f(x))
\\g'(x) = Z_D(x) P'(x)
\end{cases}\]

We've discussed reduction to polynomial comparison, and we know that the
\(2POLY\) protocol will take care of efficient comparison. However, the
the scalable version of \(2POLY\) would have the Verifier defer querying
(for a point \(i\)) \(f'(i)\) and \(g'(i)\) to the Prover, which would
only guarantee that two random polynomials given by the prover are
equivalent. To make sure that the domain (i.e.~\(Z_D\)) and codomain
(i.e.~\(C\)) constraints are respected, and to retain \textbf{constraint
soundness} of our CIP problem, the Verifier needs to call the \(2POLY\)
protocol using a special format:
\[C(\underline{f(i)}) = Z_D(i) \cdot \underline{P'(i)}\] where the
underlined parts are provided by the Prover, and the rest is calculated
by the Verifier. This type of check still retains scalability, because
\(C\) is of low-degree by assumption, and we now show an efficient
technique for evaluating \(Z_D\).

\medskip

\emph{NOTE: in the original paper, the authors actually work with
polynomials with domain taken from multiplicative field subgroups, in
order to optimise vanishing polynomial evaluations:}
\[Z_D(x) = \prod_{i \in D} (x - i)\ \land D \subseteq (\mathbb{F}, \times)
\\\overset{Th.Lagrange}{\implies} Z_D(x) = x^{|D|} - 1\ \land x \in \mathbb{F}
\\\Big( \iff \forall i \in \mathbb{Z}_{|D|}: Z_D(g^i) = 0 \  \land \langle g \rangle = D \Big)\]
\emph{with}

\begin{description}
\tightlist
\item[Th. Lagrange]
\[(\mathbb{G}, \times) \implies \forall x \in \mathbb{G}: x^{|\mathbb{G}|} = 1\]
\end{description}

\emph{This improves polynomial evaluation times from \(O(|D|)\) to
\(O(log(|D|))\) thanks to the square-and-multiply algorithm for
multiplicative field exponentiation.}

\hypertarget{polynomial-based-constraints}{%
\subsubsection{Polynomial-based
Constraints}\label{polynomial-based-constraints}}

In this more realistic scenario, we will check whether a specific
execution trace \(ee\) follows the given program constraints:
\[ee \in \mathscr{C},\ ee: D \to C
\\\mathscr{C} = \{ \mathscr{C}_{BOUNDARY},\ \mathscr{C}_{EXECUTION}\}
\\deg(\mathscr{C}) \ll deg(ee)\]

\(\mathscr{C}_{BOUNDARY}\) can be considered to be the \textbf{boundary
constraint} polynomial (or list), which identifies the value that
specific elements of the execution trace need to have; for example, the
first/last elements might correspond to specific input/output values for
a given CIP statement. \(\mathscr{C}_{EXECUTION}\) can be considered to
be one or more \textbf{execution constraint} polynomials, which define
relationships between intermediate execution trace states; typically,
they are one or more state changing functions.\footnote{I will only
  consider one state changing function, but there can be multiple; I
  will later make considerations on joining two (or more) constraint
  polynomials.} Let's start with the necessary definitions for these
constraint functions. \[\mathscr{C}_{BOUNDARY} : D_B \to C
\\ D_B \subseteq D\] \[\mathscr{C}_{EXECUTION} : C_E \to C
\\ C_E \subseteq C^{deg(C_{EXECUTION})}\] And we can now start defining
the constraint relationships:
\[\forall i \in D_B : \mathscr{C}_{BOUNDARY}(i) = ee(i)\]
\[\forall (i_{prev}, i_{next}) \in D_E : C_{EXECUTION}(i_{prev}) = ee(i_{next})
\\D_E \subseteq D^{deg(\mathscr{C}_{EXECUTION}) + 1}\] As can be noted,
the domains for these functions only apply to a subset of the execution
trace. This is evident when we consider that boundaries apply typically
only to specific elements of the trace, and state changing functions
apply to a specific pattern of elements in the trace (e.g.~subsequent
states).

Just like we did for domain-based constraints, we can convert the
relationships to vanishing polynomials: \[\forall i \in D_B : C_B(i) = 0
\\ C_B(i) \overset{def}= \mathscr{C}_{BOUNDARY}(i) - ee(i)\]
\[\forall i \in D_E: C_E(i) =0\\
\begin{aligned}
C_E(i) &\overset{def}= C_E(i_{prev}, i_{next}) 
= C_E(i_1, ..., i_{\deg(\mathscr{C}_{EXECUTION})}, i_{next})
\\&=\mathscr{C}_{EXECUTION}(ee(i_1), ..., ee(i_{\deg(\mathscr{C}_{EXECUTION})})) - ee(i_{next})
\end{aligned}\]

We can now expand the domain using the same theorem on vanishing
polynomials that we used previously:
\[\exists P': C_B(i) = Z_{D_B}(i)P'(i) \land deg(P') = deg(C_B) - |D_B|\]
\[\exists P'': C_E(i) = Z_{D_E}(i) P''(i) \land deg(P'') = deg(C_E) - |D_E|\]

Thus, obtaining two distinct \(2POLY\) problems:
\[\exists P': \begin{cases}
f'(x) = g'(x)
\\deg(f') = deg(g')
\\\ 
\\f'(x) = C_B(x)
\\g'(x) = Z_{D_B}(x) P'(x)
\end{cases}\] \[\exists P'': \begin{cases}
f''(x) = g''(x)
\\deg(f'') = deg(g'')
\\\ 
\\f''(x) = C_E(x)
\\g''(x) = Z_{D_E}(x) P''(x)
\end{cases}\]

\emph{NOTE: the authors of the paper don't actually call \(2POLY\) twice
for these two statements, but they define a randomised (by the Verifier)
linear combination to join them all together and check them at once.
This is especially useful, considering that there may be multiple
execution constraints, each detailing different conditions on successive
execution trace indexes.}

Finally, the same domain and codomain \textbf{constraint soundness}
considerations mentioned in Section \ref{sec:domainconstraints} apply
here. The call formats for \(2POLY\) are:

\[\mathscr{C}_{BOUNDARY}(i) -\underline{ee(i)} = Z_B(i)\cdot \underline{P'(i)}\]
\[\mathscr{C}_{EXECUTION}(\underline{ee(i_1)}, \underline{ee(i_2)}, ...) - \underline{ee(i_{next})} = Z_{D_E}(i) \underline{P''(i)}
\\\textit{with } i = (i_1, ..., i_{next}) \in D_E\]

\hypertarget{sec:2poly}{%
\subsection{The Polynomial Comparison Problem}\label{sec:2poly}}

All of our efforts so far can be seen as having one main goal: reducing
everything to the 2 polynomials' comparison protocol presented in this
subsection. This is because this protocol satisfies two main properties
that we're after, and that are typically harder to achieve in a
universal proof system: \emph{scalability} and \emph{zero
knowledge}\footnote{\emph{universality} is another important one, but
  that's precisely what we achieve through our problem reductions!}

How does this problem take form? Essentially, the Verifier is given (the
evaluations of) two polynomials and asked to verify whether they're
equal or not. We want to do this in the fastest way possible. Here is
our typical language definition:

\begin{description}
\tightlist
\item[2POLY(\(\mathbb{F}\))]
\[\Big\{ (f,g) \bigm\vert f(x) = g(x) \land deg(f) = deg(g) = d \land f,g \in \mathbb{F}[x] \Big\}\]\footnote{You
  will notice that we're using polynomials with coefficients taken from
  a field, this is useful for efficiency optimisations that we will
  outline later. For now, just consider all elements to be integers.} or
just
\[\Big\{ (f,g) \bigm\vert \forall i \in D: f(i) = g(i),\ f: D \to C,\  g: D \to C,\ |D|=d+1\Big\}\]
\end{description}

For now, we will assume that the polynomials (i.e.~lists) are of the
same degree, and we will focus on building a protocol to check their
equality; Section \ref{sec:fri} will take care of the degree check. The
first approach the Verifier can take to solving this problem is just
naïve comparison: \[\forall i \in D : f(i) \overset{?}= g(i)\] This
method gives us \textbf{perfect soundness}, but it also takes \(O(d+1)\)
steps to run. Assuming that each polynomial is an extremely long
execution trace, this check would force the Verifier to waste too much
time, thus precluding \emph{verifier scalability} from our final
solution.

We can do much better by \textbf{slightly increasing the soundness
error}, the same way that PCPs (Section \ref{sec:pcp}) employ
``probabilistic checks''. Let us, then, consider the error probability
of checking the polynomials against just a single element of the domain,
which I will call ``succinct query''. The error occurs on any index
which makes our check succeed, in spite of having two different
polynomials; here is an example with the errors highlighted in
red:\footnote{again, note that in this plot the polynomials map to real
  numbers, but they will be part of a field when used for real programs.}

\begin{center}
\begin{tikzpicture}
\clip (-0.5,-0.5) rectangle (3.6,3.5); 
\draw[step=1cm, gray, very thin, help lines, loosely dashed] (0,0) grid (5,5); 
\filldraw[fill=red!20, draw=red] (-0.1,-10) rectangle (0.1,10);
\filldraw[fill=red!20, draw=red] (0.9,-10) rectangle (1.1,10);
\filldraw[fill=red!20, draw=red] (2.9,-10) rectangle (3.1,10);
\draw[->] (-0.5,0) -- (3.3,0) node[right]{$x$};
\draw[->] (0,-0.5) -- (0,3.3);
\foreach \x in {1, 2,..., 3}
    \draw (\x cm, -0.5pt) -- (\x cm, 0.5pt) node[anchor=north] {$\x$};
\foreach \y in { 1, 2,..., 3}
    \draw (-0.5pt,\y cm) -- (0.5pt,\y cm) node[anchor=east] {$\y$};
\draw[thick, domain=-1:5,  samples=100, color=black!30!green] 
    plot (\x,1/2*\x^3 - 2*\x^2 + 5/2*\x);
\draw[thick, domain=-1:5,  samples=100, color=blue] 
    plot (\x,\x^3 - 4*\x^2 + 4*\x);
\node[black!30!green] at (2,2) {$f(x)$};
\node[blue] at (3,1) {$g(x)$};
\fill[red] 
    (0,0) circle [radius=2pt]
    (1,1) circle [radius=2pt] 
    (3,3) circle [radius=2pt];
\end{tikzpicture}
\end{center}

We can see by the plot that the size of the errors space is much smaller
compared to the rest of the domain, but is it really the worst case
scenario? Thankfully, there is a well-known theorem regarding
polynomials which can answer this question:

\begin{description}
\tightlist
\item[Th. Polynomial Comparison]
\emph{two differing univariate polynomials of degree \(d\) are equal in
at most \(d\) evaluation points.}
\end{description}

Now we have all the necessary information to calculate the error rate of
a succinct query: \[\begin{aligned}
Pr[error] \iff &Pr[X \notin L \land check(X) = True]
\\\iff &Pr[(f,g)\notin 2POLY \land prob\_check(f,g) = True]
\\\iff &Pr[f(x) \neq g(x) \land \exists x_0 \in_R D. f(x_0) = g(x_0)] = \frac{d}{|D|}
\end{aligned}\]

\emph{NOTE: an alternative way to visualise this problem, and leading to
the same probability, can be seen as the application of the
Schwartz--Zippel Lemma
{[}\protect\hyperlink{ref-SZLemma1}{55}{]}--{[}\protect\hyperlink{ref-SZLemma3}{57}{]}
to probabilistic polynomial identity testing.}

So, our error rate is dependent on the degree of the given polynomials,
and the size of the domain they're evaluated on. In order to decrease
this ratio, we have two available methods:

\begin{enumerate}
\def\labelenumi{\arabic{enumi}.}
\tightlist
\item
  \emph{Compress the polynomials}: to decrease \(d\), we need to replace
  our lists with equivalent alternatives of lower degree. In the given
  \(2POLY\) problem this is not possible because the lists are given as
  is, but within the STARK context the polynomials actually relate to
  execution traces. Each element of a trace can be anything, as long as
  it complies with the given constraints -- it may also contain
  irrelevant local variables, after being extracted from a complex
  program! Carefully crafting such execution traces can result in a
  reduction of their size, and of our polynomials' degrees. Another
  technique is that of carefully interpolating the execution traces: the
  authors of the paper convert an execution trace to multiple Reed
  Solomon codes, obtaining further compressions because each local
  variable is considered separately!
\item
  \emph{Add Redundancy}: to increase \(|D|\), we need to increase the
  space from which we can pick our succinct queries. To do this, we can
  simply have the Prover give polynomial evaluations over a domain that
  is much larger than their degree, and this easily be obtained through
  interpolation.
\end{enumerate}

Since the second method can always be applied to our \(2POLY\) problem,
we can always apply it to obtain any desired soundness error
\(\epsilon\):
\[\epsilon = \frac{d}{|D|} \implies |D| = \frac{d}{\epsilon}\]

This protocol will be completed with the \(FRI\) protocol in Section
\ref{sec:fri} for checking our original assumption that the two
polynomials have the same degree.

\medskip

\emph{NOTE: the approach of adding redundancy can also be applied to to
probabilistic polynomial comparison using directly coefficients instead
of evaluation points, with an even better soundness error. The basic
idea is to multiply both polynomials with a random polynomial of large
degree, thereby spreading out single coefficients across multiple ones.}

\medskip

\emph{NOTE2: it is also possible to use multiple dependent queries to
further improve the accuracy. Care must be taken to respect the
zero-knowledge requirements described in Section \ref{sec:starkaddzk}.}

\hypertarget{sec:starkscalable}{%
\subsection{Scalability through Interactivity}\label{sec:starkscalable}}

We've discussed just how we can have precise but succinct polynomial
identity tests with just a single evaluation point, but how do we
evaluate this point? The polynomials need to be interpolated to add
redundancy, how much is this going to cost us? It's time to reveal the
trick that has made so many proof protocols successful: interactivity.
Thanks to interactivity, the Verifier can ask the Prover for auxiliary
information with regards to the original problem, without compromising
the actual integrity or privacy of the statement at hand. Any good
interactive protocol makes use of a \emph{Cut\&Choose} technique (like
the one discussed in Section \ref{sec:cutchoose}), where the Prover
sends over an alternative representation of his original problem, after
the Verifier has made his choice. In the scenarios discussed within
STARKs, the original problem is typically a polynomial evaluated on a
specific domain, and the choice of the Verifier is a single point within
that polynomial. In traditional non-universal proof schemes, it is
common for the researchers to seek out an alternative representation of
the original problem that is: (1) isomorphic to the original one, (2)
randomise-able, and (3) that does not reveal anything about the original
problem's witness (whenever it is a non-deterministic secret fixed by a
public key or hash function). An example of this can be, in Schnorr
protocols, the task of finding a masked private key dependent on the
original secret, or, in the Wesolowski VDF, the adaptive prime roots
assumption used to request a randomised exponentiation strongly coupled
with the original one.

In STARKs, however, we do not have access to such isomorphic problems
for universal CIP statements that \textbf{also} inherently bind the
Verifier's choice to the original statement. Because of this, to make
sure that the Prover does not cheat based on the Verifier's selection,
we ask the Prover to make use of a Commitment Scheme to bind the
original problem statement to the Verifier's choices:

\begin{enumerate}
\def\labelenumi{\arabic{enumi}.}
\item
  \emph{Commit}

  the Prover commits to each possible evaluation of the interpolated
  polynomial on the required domain. Each evaluated point will be kept
  hidden by the Commitment Scheme (due to its ``hiding'' property),
  which is useful for the Zero-Knowledge extension discussed later. This
  step is the ``cut'' part of the \emph{Cut\&Choose} technique.
\item
  \emph{Query}

  the Verifier chooses one (or more) point(s) from the polynomial that
  he would like to query. This step is the ``choose'' part of the
  \emph{Cut\&Choose} technique.
\item
  \emph{Reveal}

  the Verifier opens the commitment for the requested points, revealing
  the requested evaluation; because a Commitment Scheme is ``binding'',
  he will not be able to change the value of the evaluations that were
  committed in the first step (as the Verifier would notice and abort
  the protocol).
\end{enumerate}

Thanks to this neat trick, the full domain of the evaluated polynomial
will be kept consistent by the Prover, otherwise either the reveal step
or the subsequent soundness check required by the protocol will
fail.\footnote{that is, as long as the check is truly sound. See the
  bottom of this subsection for a discussion on soundness assumptions
  for STARKs.}

There are still a few details to iron out:

\begin{itemize}
\item
  \emph{Who interpolates the polynomials?} The Prover interpolates the
  polynomials using their original domain (e.g.~the execution trace) and
  evaluates them on the domain defined in the \(2POLY\) subprotocol
  (which in practice can be quite large, and at least 100 times larger
  than the original domain for 99\% soundness accuracy). Interpolation
  and evaluation was combined into a single process using a
  state-of-the-art quasi-linear time algorithm for Reed-Solomon codes
  based on additive-FFT techniques, described in
  {[}\protect\hyperlink{ref-AddFFT}{58}{]}. This also produces our
  quasilinear \emph{scalable prover} VC property, having
  \(O_P(T \log^2 T)\)\footnote{\(T\) is the same as the one discussed in
    the \(CIP\) problem statement.}.
\item
  \emph{What is the communication complexity?} While the Verifier only
  needs to request a few points to be evaluated on a specific domain
  (with values at most of size \(|\mathbb{F}| = 64\ bits\)), the
  commitments made by the Prover take up as a large amount of such
  values. This can lead up to as many as \(|D'| \cdot 64\ bits\), with
  \textbar D'\textbar{} being easily x100 or more times the size of the
  original domain (such as the size of the execution trace). This is not
  practical for the Verifier to store, and imposes a huge strain on
  communications.

  Because of such issues, the authors decided to rely on the
  Kilian-Micali ({[}\protect\hyperlink{ref-Kilian92}{59}{]},
  {[}\protect\hyperlink{ref-Micali00}{60}{]}) ``argument compiler'' for
  PCPs\footnote{the improvement made by Micali was for the
    non-interactive version of the protocol.}, which basically uses a
  Merkle-Tree {[}\protect\hyperlink{ref-MerkleTrees}{61}{]} (whose
  leaves are the Prover's evaluation points) as basis for the Commitment
  Scheme, sending just a single hash value (i.e.~the tree's root) as
  commitment. However, each revealed evaluation (i.e.~leaf of the tree)
  also needs to verify the commitment using an ``authentication path'',
  which is basically a tuple of the necessary hash values required to
  traverse and validate the Merkle-Tree from the revealed leaf up to the
  tree's root. If we assume that we're using a cryptographic hash
  function
  \(H: \{0,1\}^\lambda \times \{0,1\}^\lambda \to \{0,1\}^\lambda\), the
  final proof ends up becoming a \emph{succinct proof} for its size
  complexity:
  \(O_{|\pi|}(\textit{\#queries} \cdot (\log(|\mathbb{F}|)\ bits + \textit{pathlen}\cdot \lambda\ bits)) \overset{plus\ FRI}{\approx} O_{|\pi|}(\log^2 T)\).\footnote{the
    left summand refers to the size of each queried and revealed point,
    the right summand refers to the size of each authentication path
    required to validate the revealed point (the path is logarithmic
    with relation to the total number of elements); complexity is also
    affected by the \(FRI\) subprotocol described later, yielding the
    squared \(\log T\) result.}

  Finally, in the case of the \(FRI\) protocol for low degree testing,
  there will be multiple polynomials to commit to, so multiple
  Merkle-Tree roots will have to be used and authenticated.
\item
  \emph{What Commitment Scheme to use?} Commitment Schemes are typically
  built using some randomness and a cryptographic hash function,
  \(SHA2\) or \(\textit{Keccak}\) is a typical choice. In STARKs, it
  turns out that using using \(SHA2\) was too costly for the
  arithmetisation, so they used the same Davies-Meyer
  \(AES\text{-based}\) construction
  {[}\protect\hyperlink{ref-DMhash}{53}{]} that we mentioned earlier
  (Section \ref{sec:zkstatements}).
\end{itemize}

Thanks to all of these efforts, as well the requirement on low-degree
constraints \(\mathscr{C}\) (Section \ref{sec:arith}), and taking into
consideration the \(FRI\) subprotocol discussed later, we are also able
to achieve \emph{verifier scalability} for \(O_V(\log^2 T)\). For a
concrete comparison with zk-SNARKs, we have approximately \(1/10th\)
proving time, half verification time, and \(100\) to \(1000\) times the
proof length.

On a final note, let us consider the \textbf{security assumptions} that
we require for a full zk-STARK proof, leading to a \emph{transparent}
and \emph{post-quantum safe} system:

\begin{enumerate}
\def\labelenumi{\arabic{enumi}.}
\tightlist
\item
  Existence and availability of cryptographic one-way Hash Functions
\item
  Validity of the Random Oracle Model (ROM) (only for the
  \emph{non-interactive} variant)
\item
  Existence of a ZK Argument of Knowledge Statement (Section
  \ref{sec:zkstatements})
\item
  Public Randomness Source (for \emph{transparency})
\end{enumerate}

\hypertarget{the-non-interactive-variant}{%
\subsubsection{The Non-Interactive
variant}\label{the-non-interactive-variant}}

Non-interactive STARKs were proven to exist using the ROM model, which
is performed using the traditional Fiat-Shamir heuristic
{[}\protect\hyperlink{ref-FS87}{12}{]} shown in our model ({[}Section
sec.~\ref{sec:fs}); but keep in mind that it reduces our perfect
\emph{zero-knowledge} scheme to a \textbf{computational
\emph{zero-knowledge}} scheme.

\hypertarget{sec:starkaddzk}{%
\subsection{Adding Zero-Knowledge}\label{sec:starkaddzk}}

Let us now turn the page to what is probably the most captivating
feature of zk-STARKs: \textbf{perfect \emph{zero knowledge}}!
Shockingly, and with great distinction from previous schemes based on
homomorphic cryptography (e.g.~zk-SNARKs), it is actually the easiest
property of the protocol to achieve. To see why, we need to pay respects
to our adamant use of pure polynomials, and to the shoulders of giants
on which our \(2POLY\) subprotocol stands upon: Reed-Solomon codes and
Shamir's Secret Sharing.

Reed-Solomon {[}\protect\hyperlink{ref-ReedSolomon}{62}{]} codes
originated in the 60s, with the objective of introducing error
correction functionality to error-prone communication links. Their main
success was realising that redundancy can be used to efficiently
describe polynomials and detect errors with a very small overhead, a
notion which granted us our \emph{verifier scalability}. The codes
relied on a well known theorem on polynomials:

\begin{description}
\tightlist
\item[Th. on Polynomial Interpolation]
\emph{A univariate polynomial of degree \(d\) is uniquely defined by
\(\geq d+1\) points.}\footnote{The theorem is also evident when
  considering that \(d+1\) evaluations can be put into a system of
  equations containing \(d+1\) variables for all the polynomial's
  coefficients -- solving the system leads to the correct solution.}
\end{description}

At the same time, Reed-Solomon codes offered a solid theoretical
foundation for secret sharing in the 80s, when Shamir's Secret Sharing
{[}\protect\hyperlink{ref-SSS}{63}{]} scheme was introduced. The main
idea is to hide a secret in the very first element of a list, which is
itself the evaluation of a polynomial of degree \(d\) on an arbitrarily
large domain\footnote{the evaluation typically starts from zero, so the
  coefficient of degree zero for the polynomial is simply the secret,
  and all other coefficients can be selected randomly from the same
  domain of the secret.}. Each different element of the list, except for
the first one, is distributed to a group of trusted users; when at least
\(d+1\) users come together, they're able to recover the secret through
Lagrangian interpolation.\footnote{as we discussed in Section
  \ref{sec:starkscalable} the authors of the paper actually take
  advantage of more efficient interpolation algorithms.}

The nice feature about Shamir's scheme is that it is information
theoretically secure, because not even computationally unbounded
attackers with access to \(\leq d\) shares can retrieve any information
on the secret. This serves as a foundation for our \emph{zero-knowledge}
property, here is the protocol extension:

\begin{enumerate}
\def\labelenumi{\arabic{enumi}.}
\item
  \emph{Deny Querying the Execution Trace}: the Verifier is not allowed
  to perform queries from the execution trace's original domain, as any
  of its values may contain traces of the original witness (i.e.~the
  input \(x\)). Likewise, in Shamir's scheme the first element of the
  evaluation list, typically containing the secret, is never shared.
\item
  \emph{Introduce Randomness}: the execution trace is extended with
  uniformly selected noise, equal to as many elements as the number of
  queries performed by the Verifier.

  This is performed because preventing the Verifier from querying the
  original domain of the execution trace is not sufficient to achieve
  zero-knowledge. While the same concept of Shamir's scheme applies --
  in that owning less than \(degree\ +1\) (i.e.~\(|ee|\)) evaluations is
  not always sufficient to recover the full secret (i.e.~\(ee\)) -- the
  same context does not. Specifically, in Shamir's secret sharing
  \textbf{at least} \(d\) elements of the original polynomial are
  uniformly selected random values\footnote{this is a direct result of
    the fact that at least \(d\) coefficients of the degree \(d\)
    polynomial are uniformly selected random values.}, so each possible
  interpolation of a degree \(d\) polynomial from \(d\) points is
  equally likely\footnote{if the polynomial is part of a field
    \(\mathbb{F}[x]\), then there are \(|\mathbb{F}|\) possible
    polynomials, all with the same probability of being correct}, making
  it perfectly hiding. In our scenario, instead, we cannot assume that
  at least \(\#queries\) intermediate states are uniformly random, as
  the opposite is often true because these states tend to be dependent
  on each other or take a particular shape/form with non-uniform
  probability. Because of this, some possible interpolations of a degree
  \(deg(ee)\) polynomial are more likely to occur, leading to a leakage
  of information for each query provided to the Verifier.

  In order to prevent these ``partial interpolation'' attacks, however
  unlikely they may be, we can simply append \(\#queries\) uniformly
  selected random values to the original execution trace. Not only does
  this provide perfect zero-knowledge (as in Shamir's scheme), but it
  still retains soundness with regards to the given STARK execution and
  boundary constraints. To see why, consider that the constraints only
  validate the domain of the original execution trace, so any noise
  added outside of that domain is still acceptable. This can also be
  easily deduced by considering the lack of restrictions on the contents
  of the polynomial \(P'\) found in the theorem on vanishing polynomial
  composition that was presented earlier (Section
  \ref{sec:domainconstraints}).
\end{enumerate}

\hypertarget{sec:fri}{%
\subsection{The (Low) Degree Testing Problem}\label{sec:fri}}

One important condition of the \(2POLY\) protocol, and the secret behind
its scalability, is the requirement ``\(deg(f) = deg(g) = d\)''. In
fact, knowing the degree of a specific polynomial allows for great
optimisations, such as those seen in Reed-Solomon
{[}\protect\hyperlink{ref-ReedSolomon}{62}{]} error-correction codes and
Shamir's Secret Sharing {[}\protect\hyperlink{ref-SSS}{63}{]} scheme.
The authors of the zk-STARK paper came up with a protocol for validating
a stated degree, called \emph{Fast Reed-Solomon Interactive Oracle
Proofs of Proximity}, or \(FRI\) (presented in
{[}\protect\hyperlink{ref-FRI}{64}{]}).\footnote{a variant of \(FRI\)
  with improved soundness, called \(DEEP\)-\(FRI\) and which we will not
  be discussing here, was recently published in
  {[}\protect\hyperlink{ref-DEEP-FRI}{65}{]}} The key innovation of this
protocol is providing a concrete \textbf{Proximity Testing Protocol},
made possible through the use of Interactive Oracle Proofs (IOPs) and
other engineering optimisations. In this subsection, I will discuss how
a PCP for degree testing can be made scalable through interactivity.

First things first, let's start with the complex name: (1) the protocol
is \emph{fast}, in that it is concretely efficient and can be used for
practical purposes (e.g.~STARKs); (2) the problem statement is based on
\emph{Reed-Solomon} codes; (3) the method to solve the problem is a
combination of IP and PCP proof methodologies, called \emph{IOP}; (4) we
do not test for equality with a specific degree, but for
\emph{proximity}.

The reason that the problem needs to be relaxed to proximity testing is
that it is actually quite hard to achieve a concretely efficient
protocol for checking degree equality, so we relax our assumptions a
little bit. We transform our part of the \(2POLY\) statement on degrees
from something of the form \(deg(f) = deg(g) = d\) to something like
\(deg(f) \approx deg(g) \approx d\); to be exact, through use of
Reed-Solomon codes we can make our statement become
\(deg(f) \leq d \land deg(g) \leq d\) without loss of soundness
precision for the \(2POLY\) protocol. However, the actual result
deriving from our implementation will lead us to two statements, to be
checked separately, based on proximity to \(d\):
\(deg(f) \leq d + d_0 \land deg(g) \leq d + d_0\) (for some ``small''
\(d_0\) based on \(d\)). As we will discuss later, the reason that we
have \(d_0\) proximity is because \(FRI\) gets more reliable, in cases
of malicious Provers, as the distance between the polynomial's real
degree and \(d\) gets larger; because such proximity statements can only
guarantee that the tested degree will be close to (or less than) \(d\),
they are called \emph{``low''} degree testing problems. For practical
purposes the concrete distance between \(d\) and \(d_0\) is typically
low\footnote{i.e.~\(| d_0 | \approx 1 - \rho^\frac{1}{4}\) in FRI
  {[}\protect\hyperlink{ref-FRI}{64}{]}, and
  \(\approx 1 - \rho^\frac{1}{2}\) in DEEP-FRI
  {[}\protect\hyperlink{ref-DEEP-FRI}{65}{]}, with \(\rho\) being the
  compression rate for the RS code that is found in the \(FRI\) problem
  statement, mentioned formally later.}, and, to accommodate for this
inconvenience, we can easily increase precision of the \(2POLY\) test by
increasing the degree to \(d + d_0\).

We can now move onto the formal problem language that this subprotocol
tests:

\begin{description}
\tightlist
\item[FRI]
\[\Big\{ (f, d) \bigm\vert f \in RS[\mathbb{F}, D, \rho],\ f: D \to C,\ d = \rho |D| ,\ (D,C) \subseteq \mathbb{F} \Big\}\]
\emph{\(RS[\mathbb{F}, D, \rho]\) represents all Reed-Solomon codes
mapping to a field \(\mathbb{F}\) and evaluated on a space \(D\)
(i.e.~code redundancy length is \(|D|\)) and whose compression rate is
\(\rho\). In short, we're stating that \(deg(f) < d\); one can easily
turn it into \(deg(f) \leq d',\ d' = d-1\).}
\end{description}

How do we go about tackling this problem? If we use a naïve check to get
the degree of \(f\) we must interpolate it on the full domain \(D\), but
\(O(|D|)\) complexity is far too costly. Improvements were made in the
90s {[}\protect\hyperlink{ref-RS96}{66}{]} to bring the test complexity
down to \(O(d+1)\) as long as we were testing for proximity, and further
improvements with regards to this problem were carried on in the field
of PCP proofs. We shall further improve this result to \(O(log(d))\)
with \(3\) simple steps. The core innovation of the \(FRI\) protocol is
that of improving upon the traditional PCPP (PCP of Proximity) tests
through IOPP (IOP of Proximity); the main idea is to send multiple
proofs (or oracles, when queried through the Prover) to the Verifier,
which reduce the problem to a simpler one over time. Here are our steps:

\begin{enumerate}
\def\labelenumi{\arabic{enumi}.}
\tightlist
\item
  Reduce \(f\) to a polynomial \(f'\) of degree \(deg(f') = deg(f) / 2\)
\item
  \(f \gets f'\), go back to step 1 and repeat for \(log(d)\) steps
\item
  Check that \(f\) is of degree \(0\)
\end{enumerate}

In order to reduce \(f\) to \(f'\), we take advantage of a decomposition
technique that shares similarities with the Berlekamp-Welch algorithm
{[}\protect\hyperlink{ref-Berlekamp-Welch}{67}{]} for error correction
of Reed-Solomon codes, and is exactly the same one used by the
\emph{divide-et-impera} Cooley-Tukey algorithm for the (inverse) Fast
Fourier Transform (FFT) {[}\protect\hyperlink{ref-FFT}{68}{]}. The idea
is to split the polynomial between odd and even coefficients, each
becoming its own polynomial of degree half of the original
one:\footnote{we will assume, for simplicity, that the domain \(D\) of
  \(f\) be 2-smooth (i.e.~\(\exists k \in \mathbb{N}: |D| = 2^k\)); the
  degree of \(f\) is of the same form.} \[\begin{aligned}
f(x) &= \sum_{i=0}^{deg(f)} f_i  \cdot x^i 
\\&= \sum^{deg(f)/2} f_{2i} \cdot x^{2i} + \sum^{deg(f)/2} f_{2i + 1} \cdot x^{2i+1} 
\\&= \sum^{deg(f)/2} f_{2i} \cdot x^{2i} + x \sum^{deg(f)/2} f_{2i+1} \cdot x^{2i} 
\\&= \sum^{deg(f)/2} f_{even_i} \cdot (x^2)^i + x \sum^{deg(f)/2} f_{odd_i} \cdot (x^2)^i 
\\&= f_{even}(x^2) + x f_{odd}(x^2)\end{aligned}\]

Now, let's consider an auxiliary ``composition'' polynomial \(g(x,y)\):
\[\begin{aligned}
f(x) &= \forall (x^2=y): f_{even}(y) + x f_{odd}(y) 
\\&= \forall (x^2 = y): g(x, y)\end{aligned}\]

Whenever \(x \in D\) is mapped onto \(y\) by squaring, we shall call
that domain \(D'\), such that \(y \in D'\). The polynomial \(g\) has the
important property of being derived from \(f\) and being decomposable
into smaller degrees, \(deg_x(g) \leq deg(f)/2\) and
\(deg_y(g) \leq 1\), this can easily be visualised if we abstract away
one of the variables: \[\begin{cases}
g_x = g_0 + x \cdot g_1
\\g_y = f_{even_0} \cdot y^0 + ... + f_{even_{deg(f)/2}} \cdot y^{deg(f)/2} + g_0 (f_{odd_0} \cdot y^0 + ... + f_{\deg(f)/2} \cdot y^{deg(f)/2})
\end{cases}\] Because of such considerations, \(|D'| = |D|/2\).

We can now generate all the polynomials for our 3-step plan:
\[f^{(i)} \overset{def}=\begin{cases}\begin{aligned}
f^{(0)} &\gets \forall x \in D: f(x)
\\\exists x_0 \in \mathbb{F}: f^{(1)} &\gets \forall y \in D^{(1)}: g^{(0)}(x_0, y)
\\\exists x_1 \in \mathbb{F}: f^{(2)} &\gets \forall y \in D^{(2)}: g^{(1)}(x_1, y)
\\...
\\\exists x_{\log d} \in \mathbb{F}: f^{(\log d)} &\gets \forall y \in D^{(\log d)}: g^{(\log d)}(x_{\log d}, y)
\end{aligned}\end{cases}\] \emph{(with \(g^{(i)}\) decomposition of
\(f^{(i)}\), \(|D^{(i+1)}| = |D^{(i)}|/2\)). Assuming, in the honest
case, that \(deg(f) = d\), clearly \(f^{(\log d)}\) will be be of degree
\(0\): a constant repeated up to \(|D^{(\log d)}|\) times. When we apply
the Kilian-Micali construction for interactive PCPs, we will have the
the Verifier generate (uniformly randomly) and send points \(x_i\), and
the Prover generate and commit the list of evaluations for each
\(f^{(i)}\); evaluations can be calculated either by evaluating the
decomposed polynomial on \(y\) for \(x = x_i\), or just by interpolating
the \(g_y\) shown above using \(deg(f^{(i)})/2\) points of the form
\((\alpha, f^{(i)}(\alpha))\). This process is also called the
\textbf{commit-phase} in \(FRI\).}

Now that we've seen how to validate \(deg(f)\) by checking that it
reduces to a constant function after \(log(d)\) steps, how dow we check
consistency of that final constant value? We should check that each
transition made by the Prover is actually correct, traversing through
each polynomial one-by-one in a way that is totally similar to the
Pietrzak VDF {[}\protect\hyperlink{ref-Piet18}{29}{]} technique that we
mentioned in the Intermediate Arithmetisation section. We will also be
doing so efficiently through a succinct querying of each polynomial
\(f^{(i)}\), called ``oracle'' in the IOP model that \(FRI\) is based
upon; this second part of \(FRI\) is called the \textbf{query-phase}.
The main idea is to check (for each round of our 3-step process) that a
polynomial \(f\) reduces to a polynomial \(f'\) correctly:
\[f \in RS[\mathbb{F},D,\rho] \overset{?}{\implies} f' \in RS[\mathbb{F}, D', \rho]\]
To check for consistency, let's try to reduce \(f'\) to some other
polynomial: \[f'(y) \equiv g(x_0, y)\] Unfortunately the Verifier cannot
afford to directly interpolate \(g(x,y) = f(x)\), nor can he afford to
interpolate \(g_y\), but he can afford to interpolate \(g_x\) for some
value \(y_0\):
\[\exists y_0 \in D': g(x, y_0) \iff \exists (\alpha_0, \alpha_1) \in D, y_0 \in D': Interpolate\Big[(\alpha_0,\ g(\alpha_0,\ y_0)),\ (\alpha_1,\ g(\alpha_1,\ y_0))\Big]\]
Which can be reduced to \(f\) (i.e.~\(\forall x^2 = y: g(x,y) = f(x)\))
quite easily when we query two ``related'' points from \(f\):
\[\iff \exists \alpha\in D, y_0 \in D', \alpha^2 = y_0: Interpolate\Big[(\alpha,\ g(\alpha,\ y_0)),\ (-\alpha,\ g(-\alpha,\ y_0))\Big]
\\\iff \exists \alpha \in D, y_0 \in D', \alpha^2 = y_0: Interpolate\Big[(\alpha,\ f(\alpha)),\ (-\alpha,\ f(-\alpha))\Big]\]
Therefore, we end up with the following \textbf{succinct consistency
check} by having the Verifier query from the Prover
\(f'(y_0), f(\alpha), f(-\alpha)\):
\[f'(y_0) \overset{?}= g(x_0, y_0) = Interpolate\Big[(\alpha,\ f(\alpha)),\ (-\alpha,\ f(-\alpha))\Big]\ (x_0)
\\\textit{(with $\alpha^2 = y_0$)}\]

Finally, the actual soundness analysis (i.e.~precision) for this
consistency check (and the whole \(FRI\) protocol) is the toughest part
of any IOPP or PCPP protocol, and something that we will not get into
detail. Suffice to say that as long as the Prover is honest the protocol
works just fine, and when he lies about the degree of \(f\) the protocol
works very well when the real distance of \(deg(f)\) from the claimed
degree is large (because the soundness probability is based on a
function of the distance). When this distance is small, the \(FRI\)
protocol cannot reliably detect it, but it is a very small distance
(which was already improved in the \(DEEP-FRI\)
{[}\protect\hyperlink{ref-DEEP-FRI}{65}{]} extension) and we have
mentioned above how the \(2POLY\) protocol can be easily adapted for
this issue by increasing its precision. Furthermore, the protocol can be
considered concretely efficient, with
\(O_P < 6 \cdot d,\ O_V < 21 \cdot \log(d),\ O_{|\pi|} < 2 \cdot \log(d)\).

\medskip

\emph{NOTE: degree testing in this section is fully pq-safe, but in
other proof systems (e.g.~SNARKs) it is typically based on homomorphic
encryption.}

\medskip

\emph{NOTE2: the protocol should be applied to both \(2POLY\)
polynomials to check that they are of the right degree, but the authors
of the paper take advantage of the fact that any linear combination of
the two polynomials leads to the same degree as one of them, so they
check just a single composite polynomial.}

\medskip

\emph{NOTE3: the authors of the paper actually discuss improving both
performance and soundness of the protocol by adjusting the values in the
Merkle tree of the Kilian-Micali commitment in such a way that a single
subtree will contain both points \((\alpha_i, -\alpha_i)\), and that
further down the tree we also find the other points
\((\alpha_{i+1}, -\alpha_{i+1})\) in such a way that we can re-use
\(f'(y_0)\), leading to very efficient authentication paths for the
commitment scheme. It is an open question whether better commitment
structures than Merkle trees can lead to more improvements, such as
reduced communication sizes \(O_{|\pi|}\).}

\hypertarget{sec:universalconclusion}{%
\section{Conclusion}\label{sec:universalconclusion}}

zk-STARKs are an incredibly powerful tool that can be used not only to
build privacy-friendly applications, but also to drastically reduce the
computational costs required to validate outsourced computations online.
In short, the power of such universal compilers is that of being able to
answer any sort of yes/no question:

\begin{itemize}
\tightlist
\item
  \emph{Do you have the right password?}
\item
  \emph{Do you have the right password for user John?}
\item
  \emph{Do you have the authority and balance to perform a transfer of
  EUR 100 towards John?}
\item
  \emph{Were these 1000 images analysed using the Machine Learning model
  I gave you?}
\item
  \emph{Does your result comply with the Smart Contract we uploaded to
  Ethereum?}
\item
  \emph{Does my program meet security specifications, implying that it
  is free of bugs?}
\end{itemize}

And such answers can be checked by the Verifier in time that is much,
much faster compared to simply (and naïvely) analysing all the required
data himself. When the proposed question takes this binary format, the
privacy of any needed data can be preserved by the protocol as long as
there is access to other public and binding data published by some
trusted authority (e.g.~hashed citizen identities published by the
government) or computationally inherent to the question's context
(e.g.~a known composite number uniquely defined by its prime factors).

While we've seen the current state-of-the-art in the domain of VC
systems, let's take a moment to consider constructions that have been
developed using alternative approaches. The majority of such systems
derive from the older and quite alike fields of cryptographic protocols:
IPs (Interactive Proofs) and PCPs (Probabilistically Checkable Proofs),
or alternative approaches with comparable semantics. Within the context
of constructions stemming from PCPs, there are two common solutions for
solving degree testing of arithmetic circuits: (1) multiplicatively
homomorphic encryption (e.g.~zk-SNARKs) and (2) proofs of proximity
(e.g.~zk-STARKs). While all these competing systems are part of the
cryptographic VC domain (most of them also being very recent), they can
be grouped into different fields according to their theoretical design:

\begin{itemize}
\tightlist
\item
  \textbf{MPC in the Head}: this is the only other field, apart from
  STARKs, which achieved both transparency and post-quantum safety. The
  main idea behind of such protocols is that of simulating independently
  (i.e.~``in the head'') a Multi-Party Computation (MPC), and then
  revealing it to the Verifier upon request. Amongst the most popular
  constructions there are: ZKBoo
  {[}\protect\hyperlink{ref-Giacomelli16}{69}{]}, ZKBoo++
  {[}\protect\hyperlink{ref-Chase17}{70}{]}, and Ligero
  {[}\protect\hyperlink{ref-Ames17}{71}{]};
\item
  \textbf{zk-SNARKs}: this is currently the most successful field of VC
  protocols, with multiple open-source libraries available to the public
  and even a successfully deployed privacy-friendly cryptocurrency,
  Zcash {[}\protect\hyperlink{ref-Zcash}{46}{]}. SNARKs
  {[}\protect\hyperlink{ref-SNARK}{72}{]}, traceable back to SNARGs
  {[}\protect\hyperlink{ref-Gentry11}{73}{]} and typically based on QAPs
  (Quadratic Arithmetic Programs) and homomorphic cryptography, have
  received a lot of attention from researchers, giving birth to many
  theoretical designs such as: Geppetto
  {[}\protect\hyperlink{ref-Costello15}{74}{]}, Pinocchio
  {[}\protect\hyperlink{ref-Gentry13}{75}{]}, Groth's
  {[}\protect\hyperlink{ref-Groth16}{76}{]} (most popular one), SNARKs
  for C {[}\protect\hyperlink{ref-BenSasson13}{77}{]}, and very recently
  Aurora {[}\protect\hyperlink{ref-BenSasson19}{78}{]}, Sonic
  {[}\protect\hyperlink{ref-Maller19}{79}{]}, and Libra
  {[}\protect\hyperlink{ref-Xie19}{80}{]} (most interesting due to its
  linear prover complexity);
\item
  \textbf{zk-STARKs}: this very innovative solution was groundbreaking
  due to all the VC features it implements, and for being valid even in
  realistic scenarios, it is published in
  {[}\protect\hyperlink{ref-STARK}{51}{]};
\item
  \textbf{Interactive Proofs for Muggles}: this ``older'' (compared with
  other protocols) design by
  {[}\protect\hyperlink{ref-Goldwasser15}{81}{]} is one of the few which
  is based on IPs rather than PCPs; Hyrax
  {[}\protect\hyperlink{ref-Wahby17}{82}{]} is an interesting recent
  development.
\item
  \textbf{Linear PCPs} while this DLP-based (Discrete Logarithm Problem)
  field is not strictly universal, as proofs can only guarantee that
  given inputs lie within a specific range (e.g.~boundary constraints),
  Bulletproofs {[}\protect\hyperlink{ref-Bunz18}{83}{]} have often been
  compared to other universal systems because of their applicability to
  cryptocurrencies (they were developed for the Monero
  {[}\protect\hyperlink{ref-Monero}{84}{]} cryptocurrency) and their
  (now outclassed) performance.
\end{itemize}

On a final note, most of the recent research in this field has been
published with consideration for applicative scenarios, especially
Blockchain-based solutions, by providing library implementations and
concrete performance analyses. Amongst the most successful applications
based on such research we find the privacy-friendly cryptocurrencies
Zcash {[}\protect\hyperlink{ref-Zcash}{46}{]} and Monero
{[}\protect\hyperlink{ref-Monero}{84}{]}, and privacy-friendly
smart-contract (e.g.~Ethereum programs) outsourcing and decentralised
exchanges in ZEXE {[}\protect\hyperlink{ref-ZEXE}{85}{]}.

\hypertarget{references-bibliography}{%
\chapter{References \& Bibliography}\label{references-bibliography}}

\begin{verbatim}
\setlength{\parindent}{-1.24cm}
\setlength{\leftskip}{1.24cm}
\setlength{\parskip}{8pt}
\end{verbatim}

\hypertarget{refs}{}
\leavevmode\hypertarget{ref-blockchain-forbes}{}%
{[}1{]} M. del Castillo, `Big Blockchain: The 50 Largest Public
Companies Exploring Blockchain'.
\url{https://www.forbes.com/sites/michaeldelcastillo/2018/07/03/big-blockchain-the-50-largest-public-companies-exploring-blockchain/},
2018.

\leavevmode\hypertarget{ref-bitconnect}{}%
{[}2{]} R. Hackett, `Police Nab Alleged Boss Behind Bitcoin Pyramid
Scheme Bitconnect'.
\url{http://fortune.com/2018/08/20/bitcoin-scam-bitconnect-arrest/},
2018.

\leavevmode\hypertarget{ref-ethresearch}{}%
{[}3{]} `Ethereum Research'. \url{https://ethresear.ch}.

\leavevmode\hypertarget{ref-GMR85}{}%
{[}4{]} S. Goldwasser, S. Micali, and C. Rackoff, `The knowledge
complexity of interactive proof systems', \emph{SIAM Journal on
computing}, vol. 18, no. 1, pp. 186--208, 1989.

\leavevmode\hypertarget{ref-PCPTheorem}{}%
{[}5{]} S. Arora, C. Lund, R. Motwani, M. Sudan, and M. Szegedy, `Proof
verification and the hardness of approximation problems', \emph{Journal
of the ACM (JACM)}, vol. 45, no. 3, pp. 501--555, 1998.

\leavevmode\hypertarget{ref-AS98}{}%
{[}6{]} S. Arora and S. Safra, `Probabilistic checking of proofs: A new
characterization of np', \emph{Journal of the ACM (JACM)}, vol. 45, no.
1, pp. 70--122, 1998.

\leavevmode\hypertarget{ref-Babai91first}{}%
{[}7{]} L. aszl o Babai, L. Fortnow, L. Levin, and M. Szegedy, `Checking
computations in polylogarithmic time', in \emph{Proceedings of the 23rd
annual acm symposium on theory of computing}, 1991, pp. 21--31.

\leavevmode\hypertarget{ref-Babai91second}{}%
{[}8{]} L. Babai, L. Fortnow, and C. Lund, `Non-deterministic
exponential time has two-prover interactive protocols',
\emph{Computational complexity}, vol. 1, no. 1, pp. 3--40, 1991.

\leavevmode\hypertarget{ref-IOP}{}%
{[}9{]} E. Ben-Sasson, A. Chiesa, and N. Spooner, `Interactive oracle
proofs', in \emph{Theory of cryptography}, 2016, pp. 31--60.

\leavevmode\hypertarget{ref-Quisquater90}{}%
{[}10{]} J.-J. Quisquater \emph{et al.}, `How to explain zero-knowledge
protocols to your children', in \emph{Advances in cryptology --- crypto'
89 proceedings}, 1990, pp. 628--631.

\leavevmode\hypertarget{ref-Blum86}{}%
{[}11{]} M. Blum, `How to prove a theorem so no one else can claim it',
in \emph{Proceedings of the international congress of mathematicians},
1986, vol. 1, p. 2.

\leavevmode\hypertarget{ref-FS87}{}%
{[}12{]} A. Fiat and A. Shamir, `How to prove yourself: Practical
solutions to identification and signature problems', in \emph{Conference
on the theory and application of cryptographic techniques}, 1986, pp.
186--194.

\leavevmode\hypertarget{ref-ROM}{}%
{[}13{]} M. Bellare and P. Rogaway, `Random oracles are practical: A
paradigm for designing efficient protocols', in \emph{Proceedings of the
1st acm conference on computer and communications security}, 1993, pp.
62--73.

\leavevmode\hypertarget{ref-fiatshamirisalie}{}%
{[}14{]} N. Bitansky \emph{et al.}, `Why ``fiat-shamir for proofs''
lacks a proof', in \emph{Theory of cryptography conference}, 2013, pp.
182--201.

\leavevmode\hypertarget{ref-ArthurMerlin}{}%
{[}15{]} L. Babai, `Trading group theory for randomness', in
\emph{Proceedings of the seventeenth annual acm symposium on theory of
computing}, 1985, pp. 421--429.

\leavevmode\hypertarget{ref-ArthurMerlin2}{}%
{[}16{]} L. Babai and S. Moran, `Arthur-merlin games: A randomized proof
system, and a hierarchy of complexity classes', \emph{Journal of
Computer and System Sciences}, vol. 36, no. 2, pp. 254--276, 1988.

\leavevmode\hypertarget{ref-ArthurMerlin3}{}%
{[}17{]} L. aszl o Babai, L. Fortnow, L. Levin, and M. Szegedy,
`Checking computations in polylogarithmic time', in \emph{Proceedings of
the 23rd annual acm symposium on theory of computing}, 1991, pp. 21--31.

\leavevmode\hypertarget{ref-Johnson02}{}%
{[}18{]} R. Johnson, D. Molnar, D. Song, and D. Wagner, `Homomorphic
signature schemes', in \emph{Cryptographers' track at the rsa
conference}, 2002, pp. 244--262.

\leavevmode\hypertarget{ref-Gennaro12}{}%
{[}19{]} R. Gennaro and D. Wichs, `Fully homomorphic message
authenticators'. Cryptology ePrint Archive, Report 2012/290, 2012.

\leavevmode\hypertarget{ref-Fiore16}{}%
{[}20{]} D. Fiore, A. Mitrokotsa, L. Nizzardo, and E. Pagnin, `Multi-key
homomorphic authenticators', in \emph{International conference on the
theory and application of cryptology and information security}, 2016,
pp. 499--530.

\leavevmode\hypertarget{ref-Fiore13}{}%
{[}21{]} M. Backes, D. Fiore, and R. M. Reischuk, `Verifiable delegation
of computation on outsourced data', in \emph{Proceedings of the 2013 acm
sigsac conference on computer \& communications security}, 2013, pp.
863--874.

\leavevmode\hypertarget{ref-SBB18}{}%
{[}22{]} L. Schabhüser, D. Butin, and J. Buchmann, `Context hiding
multi-key linearly homomorphic authenticators', in \emph{Cryptographers'
track at the rsa conference}, 2019, pp. 493--513.

\leavevmode\hypertarget{ref-keccak}{}%
{[}23{]} G. Bertoni, J. Daemen, M. Peeters, and G. Assche, `The keccak
reference', \emph{Submission to NIST (Round 3)}, vol. 13, pp. 14--15,
2011.

\leavevmode\hypertarget{ref-BBBF18}{}%
{[}24{]} D. Boneh, J. Bonneau, B. Bünz, and B. Fisch, `Verifiable delay
functions', in \emph{Annual international cryptology conference}, 2018,
pp. 757--788.

\leavevmode\hypertarget{ref-RSW96}{}%
{[}25{]} R. L. Rivest, A. Shamir, and D. A. Wagner, `Time-lock puzzles
and timed-release crypto', Massachusetts Institute of Technology,
Cambridge, MA, USA, 1996.

\leavevmode\hypertarget{ref-Merkle78}{}%
{[}26{]} R. C. Merkle, `Secure communications over insecure channels',
\emph{Commun. ACM}, vol. 21, no. 4, pp. 294--299, Apr. 1978.

\leavevmode\hypertarget{ref-LW15}{}%
{[}27{]} A. K. Lenstra and B. Wesolowski, `A random zoo: Sloth, unicorn,
and trx.', \emph{IACR Cryptology ePrint Archive}, vol. 2015, p. 366,
2015.

\leavevmode\hypertarget{ref-Wes18}{}%
{[}28{]} B. Wesolowski, `Efficient verifiable delay functions.',
\emph{IACR Cryptology ePrint Archive}, vol. 2018, p. 623, 2018.

\leavevmode\hypertarget{ref-Piet18}{}%
{[}29{]} K. Pietrzak, `Simple verifiable delay functions', in \emph{10th
innovations in theoretical computer science conference (itcs 2019)},
2018.

\leavevmode\hypertarget{ref-BBF18}{}%
{[}30{]} D. Boneh, B. Bünz, and B. Fisch, `A survey of two verifiable
delay functions'. Cryptology ePrint Archive, Report 2018/712, 2018.

\leavevmode\hypertarget{ref-timedkeyescrow1}{}%
{[}31{]} M. Bellare and S. Goldwasser, `Encapsulated key escrow'. MIT
Laboratory for Computer Science Technical Report, 1996.

\leavevmode\hypertarget{ref-timedkeyescrow2}{}%
{[}32{]} M. Bellare and S. Goldwasser, `Verifiable partial key escrow.',
in \emph{ACM conference on computer and communications security}, 1997,
vol. 1997, pp. 78--91.

\leavevmode\hypertarget{ref-timedcommitments}{}%
{[}33{]} D. Boneh and M. Naor, `Timed commitments', in \emph{Annual
international cryptology conference}, 2000, pp. 236--254.

\leavevmode\hypertarget{ref-BGB17}{}%
{[}34{]} B. Bünz, S. Goldfeder, and J. Bonneau, `Proofs-of-delay and
randomness beacons in ethereum', \emph{IEEE Security and Privacy on the
blockchain (IEEE S\&B)}, 2017.

\leavevmode\hypertarget{ref-BCG15}{}%
{[}35{]} J. Bonneau, J. Clark, and S. Goldfeder, `On bitcoin as a public
randomness source'. Cryptology ePrint Archive, Report 2015/1015, 2015.

\leavevmode\hypertarget{ref-traplottery}{}%
{[}36{]} @mabbamOG, `TrapLottery 0.2: Automated Lottery on the
Blockchain'. \url{https://github.com/mabbamOG/traplottery}, 2018.

\leavevmode\hypertarget{ref-SWARM}{}%
{[}37{]} V. Trón, A. Fischer, D. Nagy, Z. Felföldi, and N. Johnson,
`Swap, swear, and swindle: Incentive system for swarm'. Technical
Report, Ethersphere, 2016. Ethersphere Orange Papers 1., 2016.

\leavevmode\hypertarget{ref-scrypt}{}%
{[}38{]} C. Percival, `Stronger key derivation via sequential
memory-hard functions'. BSDCan, 2009.

\leavevmode\hypertarget{ref-ethash}{}%
{[}39{]} G. Wood and others, `Ethereum: A secure decentralised
generalised transaction ledger', \emph{Ethereum project yellow paper},
vol. 151, pp. 1--32, 2014.

\leavevmode\hypertarget{ref-ethereumvdfmpc}{}%
{[}40{]} J. Drake, `Minimal VDF randomness beacon'.
\url{https://ethresear.ch/t/minimal-vdf-randomness-beacon/}, 2018.

\leavevmode\hypertarget{ref-Rabin79}{}%
{[}41{]} M. O. Rabin, `Digitalized signatures and public-key functions
as intractable as factorization', Jan. 1979.

\leavevmode\hypertarget{ref-Sander99}{}%
{[}42{]} T. Sander, `Efficient accumulators without trapdoor extended
abstract', in \emph{International conference on information and
communications security}, 1999, pp. 252--262.

\leavevmode\hypertarget{ref-twompc}{}%
{[}43{]} A. C.-C. Yao, `How to generate and exchange secrets', in
\emph{27th annual symposium on foundations of computer science (sfcs
1986)}, 1986, pp. 162--167.

\leavevmode\hypertarget{ref-multimpc}{}%
{[}44{]} I. Damgård, M. Geisler, M. Krøigaard, and J. B. Nielsen,
`Asynchronous multiparty computation: Theory and implementation', in
\emph{International workshop on public key cryptography}, 2009, pp.
160--179.

\leavevmode\hypertarget{ref-Boneh97}{}%
{[}45{]} D. Boneh and M. Franklin, `Efficient generation of shared rsa
keys', in \emph{Annual international cryptology conference}, 1997, pp.
425--439.

\leavevmode\hypertarget{ref-Zcash}{}%
{[}46{]} D. Hopwood, S. Bowe, T. Hornby, and N. Wilcox, `Zcash protocol
specification', \emph{Tech. rep. 2016--1.10. Zerocoin Electric Coin
Company, Tech. Rep.}, 2016.

\leavevmode\hypertarget{ref-Buchmann88}{}%
{[}47{]} J. Buchmann and H. C. Williams, `A key-exchange system based on
imaginary quadratic fields', \emph{Journal of Cryptology}, vol. 1, no.
2, pp. 107--118, Jun. 1988.

\leavevmode\hypertarget{ref-BBHM02}{}%
{[}48{]} I. Biehl, J. Buchmann, S. Hamdy, and A. Meyer, `A signature
scheme based on the intractability of computing roots', \emph{Designs,
Codes and Cryptography}, vol. 25, no. 3, pp. 223--236, 2002.

\leavevmode\hypertarget{ref-Benhamouda15}{}%
{[}49{]} F. Benhamouda, S. Krenn, V. Lyubashevsky, and K. Pietrzak,
`Efficient zero-knowledge proofs for commitments from learning with
errors over rings', in \emph{European symposium on research in computer
security}, 2015, pp. 305--325.

\leavevmode\hypertarget{ref-AccumulatorSoK}{}%
{[}50{]} D. Derler, C. Hanser, and D. Slamanig, `Revisiting
cryptographic accumulators, additional properties and relations to other
primitives', in \emph{Cryptographers' track at the rsa conference},
2015, pp. 127--144.

\leavevmode\hypertarget{ref-STARK}{}%
{[}51{]} E. Ben-Sasson, I. Bentov, Y. Horesh, and M. Riabzev, `Scalable,
transparent, and post-quantum secure computational integrity'.
Cryptology ePrint Archive, Report 2018/046, 2018.

\leavevmode\hypertarget{ref-libSTARK}{}%
{[}52{]} @elibensasson, `libSTARK: a C++ library for zk-STARK systems'.
\url{https://github.com/elibensasson/libSTARK}, 2018.

\leavevmode\hypertarget{ref-DMhash}{}%
{[}53{]} B. Preneel, R. Govaerts, and J. Vandewalle, `Hash functions
based on block ciphers: A synthetic approach', in \emph{Annual
international cryptology conference}, 1993, pp. 368--378.

\leavevmode\hypertarget{ref-TinyRAM}{}%
{[}54{]} E. Ben-Sasson, A. Chiesa, D. Genkin, E. Tromer, and M. Virza,
`TinyRAM architecture specification, v0. 991'. 2013.

\leavevmode\hypertarget{ref-SZLemma1}{}%
{[}55{]} J. T. Schwartz, `Probabilistic algorithms for verification of
polynomial identities', in \emph{International symposium on symbolic and
algebraic manipulation}, 1979, pp. 200--215.

\leavevmode\hypertarget{ref-SZLemma2}{}%
{[}56{]} R. Zippel, `Probabilistic algorithms for sparse polynomials',
in \emph{International symposium on symbolic and algebraic
manipulation}, 1979, pp. 216--226.

\leavevmode\hypertarget{ref-SZLemma3}{}%
{[}57{]} R. A. DeMillo and R. J. Lipton, `A probabilistic remark on
algebraic program testing.', GEORGIA INST OF TECH ATLANTA SCHOOL OF
INFORMATION AND COMPUTER SCIENCE, 1977.

\leavevmode\hypertarget{ref-AddFFT}{}%
{[}58{]} S.-J. Lin, W.-H. Chung, and Y. S. Han, `Novel polynomial basis
and its application to reed-solomon erasure codes', in \emph{2014 ieee
55th annual symposium on foundations of computer science}, 2014, pp.
316--325.

\leavevmode\hypertarget{ref-Kilian92}{}%
{[}59{]} J. Kilian, `A note on efficient zero-knowledge proofs and
arguments', in \emph{Proceedings of the twenty-fourth annual acm
symposium on theory of computing}, 1992, pp. 723--732.

\leavevmode\hypertarget{ref-Micali00}{}%
{[}60{]} S. Micali, `Computationally sound proofs', \emph{SIAM Journal
on Computing}, vol. 30, no. 4, pp. 1253--1298, 2000.

\leavevmode\hypertarget{ref-MerkleTrees}{}%
{[}61{]} R. C. Merkle, `A digital signature based on a conventional
encryption function', in \emph{Conference on the theory and application
of cryptographic techniques}, 1987, pp. 369--378.

\leavevmode\hypertarget{ref-ReedSolomon}{}%
{[}62{]} I. S. Reed and G. Solomon, `Polynomial codes over certain
finite fields', \emph{Journal of the society for industrial and applied
mathematics}, vol. 8, no. 2, pp. 300--304, 1960.

\leavevmode\hypertarget{ref-SSS}{}%
{[}63{]} A. Shamir, `How to share a secret', \emph{Communications of the
ACM}, vol. 22, no. 11, pp. 612--613, 1979.

\leavevmode\hypertarget{ref-FRI}{}%
{[}64{]} E. Ben-Sasson, I. Bentov, Y. Horesh, and M. Riabzev, `Fast
reed-solomon interactive oracle proofs of proximity', in \emph{45th
international colloquium on automata, languages, and programming (icalp
2018)}, 2018.

\leavevmode\hypertarget{ref-DEEP-FRI}{}%
{[}65{]} E. Ben-Sasson, L. Goldberg, S. Kopparty, and S. Saraf,
`DEEP-fri: Sampling outside the box improves soundness', \emph{arXiv
preprint arXiv:1903.12243}, 2019.

\leavevmode\hypertarget{ref-RS96}{}%
{[}66{]} R. Rubinfeld and M. Sudan, `Robust characterizations of
polynomials with applications to program testing', \emph{SIAM Journal on
Computing}, vol. 25, no. 2, pp. 252--271, 1996.

\leavevmode\hypertarget{ref-Berlekamp-Welch}{}%
{[}67{]} L. R. Welch and E. R. Berlekamp, `Error correction for
algebraic block codes'. Google Patents, 1986.

\leavevmode\hypertarget{ref-FFT}{}%
{[}68{]} J. W. Cooley and J. W. Tukey, `An algorithm for the machine
calculation of complex fourier series', \emph{Mathematics of
computation}, vol. 19, no. 90, pp. 297--301, 1965.

\leavevmode\hypertarget{ref-Giacomelli16}{}%
{[}69{]} I. Giacomelli, J. Madsen, and C. Orlandi, `Zkboo: Faster
zero-knowledge for boolean circuits', in \emph{25th \(\{\)usenix\(\}\)
security symposium (\(\{\)usenix\(\}\) security 16)}, 2016, pp.
1069--1083.

\leavevmode\hypertarget{ref-Chase17}{}%
{[}70{]} M. Chase \emph{et al.}, `Post-quantum zero-knowledge and
signatures from symmetric-key primitives', in \emph{Proceedings of the
2017 acm sigsac conference on computer and communications security},
2017, pp. 1825--1842.

\leavevmode\hypertarget{ref-Ames17}{}%
{[}71{]} S. Ames, C. Hazay, Y. Ishai, and M. Venkitasubramaniam,
`Ligero: Lightweight sublinear arguments without a trusted setup', in
\emph{Proceedings of the 2017 acm sigsac conference on computer and
communications security}, 2017, pp. 2087--2104.

\leavevmode\hypertarget{ref-SNARK}{}%
{[}72{]} N. Bitansky, R. Canetti, A. Chiesa, and E. Tromer, `From
extractable collision resistance to succinct non-interactive arguments
of knowledge, and back again', in \emph{Proceedings of the 3rd
innovations in theoretical computer science conference}, 2012, pp.
326--349.

\leavevmode\hypertarget{ref-Gentry11}{}%
{[}73{]} C. Gentry and D. Wichs, `Separating succinct non-interactive
arguments from all falsifiable assumptions', in \emph{Proceedings of the
forty-third annual acm symposium on theory of computing}, 2011, pp.
99--108.

\leavevmode\hypertarget{ref-Costello15}{}%
{[}74{]} C. Costello \emph{et al.}, `Geppetto: Versatile verifiable
computation', in \emph{2015 ieee symposium on security and privacy},
2015, pp. 253--270.

\leavevmode\hypertarget{ref-Gentry13}{}%
{[}75{]} B. Parno, J. Howell, C. Gentry, and M. Raykova, `Pinocchio:
Nearly practical verifiable computation', in \emph{2013 ieee symposium
on security and privacy}, 2013, pp. 238--252.

\leavevmode\hypertarget{ref-Groth16}{}%
{[}76{]} J. Groth, `On the size of pairing-based non-interactive
arguments', in \emph{Annual international conference on the theory and
applications of cryptographic techniques}, 2016, pp. 305--326.

\leavevmode\hypertarget{ref-BenSasson13}{}%
{[}77{]} E. Ben-Sasson, A. Chiesa, D. Genkin, E. Tromer, and M. Virza,
`SNARKs for c: Verifying program executions succinctly and in zero
knowledge', in \emph{Annual cryptology conference}, 2013, pp. 90--108.

\leavevmode\hypertarget{ref-BenSasson19}{}%
{[}78{]} E. Ben-Sasson, A. Chiesa, M. Riabzev, N. Spooner, M. Virza, and
N. P. Ward, `Aurora: Transparent succinct arguments for r1cs', in
\emph{Annual international conference on the theory and applications of
cryptographic techniques}, 2019, pp. 103--128.

\leavevmode\hypertarget{ref-Maller19}{}%
{[}79{]} M. Maller, S. Bowe, M. Kohlweiss, and S. Meiklejohn, `Sonic:
Zero-knowledge snarks from linear-size universal and updateable
structured reference strings'. Cryptology ePrint Archive, Report
2019/099, 2019.

\leavevmode\hypertarget{ref-Xie19}{}%
{[}80{]} T. Xie, J. Zhang, Y. Zhang, C. Papamanthou, and D. Song,
`Libra: Succinct zero-knowledge proofs with optimal prover
computation.', \emph{IACR Cryptology ePrint Archive}, vol. 2019, p. 317,
2019.

\leavevmode\hypertarget{ref-Goldwasser15}{}%
{[}81{]} S. Goldwasser, Y. T. Kalai, and G. N. Rothblum, `Delegating
computation: Interactive proofs for muggles', \emph{Journal of the ACM
(JACM)}, vol. 62, no. 4, p. 27, 2015.

\leavevmode\hypertarget{ref-Wahby17}{}%
{[}82{]} R. S. Wahby, I. Tzialla, abhi shelat, J. Thaler, and M.
Walfish, `Doubly-efficient zkSNARKs without trusted setup'. Cryptology
ePrint Archive, Report 2017/1132, 2017.

\leavevmode\hypertarget{ref-Bunz18}{}%
{[}83{]} B. Bünz, J. Bootle, D. Boneh, A. Poelstra, P. Wuille, and G.
Maxwell, `Bulletproofs: Short proofs for confidential transactions and
more', in \emph{2018 ieee symposium on security and privacy (sp)}, 2018,
pp. 315--334.

\leavevmode\hypertarget{ref-Monero}{}%
{[}84{]} N. Van Saberhagen, `CryptoNote v 2.0'. 2013.

\leavevmode\hypertarget{ref-ZEXE}{}%
{[}85{]} S. Bowe, A. Chiesa, M. Green, I. Miers, P. Mishra, and H. Wu,
`Zexe: Enabling decentralized private computation', \emph{IACR ePrint},
vol. 962, 2018.

\end{document}